\documentclass[11pt]{article}
\pdfoutput=1

\usepackage{cancel,slashed}
\usepackage{bbm}
\usepackage{amssymb}
\usepackage{amsfonts}
\usepackage{amsmath}
\usepackage{graphicx}
\usepackage{cite}
\usepackage{physics}

\usepackage{ dsfont }
\usepackage{latexsym}
\usepackage{color}
\usepackage{cancel,slashed}
\usepackage[hyperfootnotes=false,linktocpage]{hyperref}
\usepackage{color}
\usepackage{transparent}
\usepackage[hang,flushmargin]{footmisc}

\usepackage{blkarray}
\usepackage{multirow}
\usepackage{empheq}

\usepackage{comment}

\newcommand{\bmat}{\left(\begin{array}}
\newcommand{\emat}{\end{array}\right)}

\def\Z{\mathbb{Z}}
\def\R{\mathbb{R}}

\def\CK {{\cal K}}
\def\a {\alpha}
\def\b {\beta}

\def\Tr{\text{Tr}}

\def\1{{\bf 1}}
\def\2{{\bf 2}}
\def\3{{\bf 3}}
\def\4{{\bf 4}}
\def\6{{\bf 6}}

\def\half{\frac{1}{2}}

\def\targ#1#2{\genfrac{[}{]}{0pt}{}{#1}{#2}}
\def\targ2#1#2{\genfrac{}{}{0pt}{}{#1}{#2}}
\def\half{{\textstyle\frac{1}{2}}}

\definecolor{mygr}{rgb}{0,0.6,0}
\definecolor{mygrey}{rgb}{0,0.1,0.2}
\definecolor{myblue}{rgb}{0,0.5,0.9}
\definecolor{myblue2}{rgb}{0,0.5,0.5}
\definecolor{myblue3}{rgb}{0,0.7,0.9}
\definecolor{myblue4}{rgb}{0,0.6,0.6}
\definecolor{myorange}{rgb}{1,0.5,0}
\definecolor{mypurple}{rgb}{0.6,0,1}
\definecolor{mygolden}{rgb}{1,0.8,0.2}
\definecolor{mycyan}{rgb}{0,1,1}
\definecolor{mymagenta}{rgb}{1,0,1}
\definecolor{mykiwi}{rgb}{0.8,1,0.5}
\definecolor{mybrown}{cmyk}{0.14, 0.42, 0.56, 0.2}
\definecolor{myturq}{cmyk}{0.99, 0, 0.2, 0.4}
\definecolor{myaubergine2}{cmyk}{0.4, 0.5, 0, 0.1}
\definecolor{myaubergine}{cmyk}{0.6,0.85,0,0}
\definecolor{CycleGreen}{cmyk}{0.52,0,1,0}
\definecolor{CycleBrown}{cmyk}{0, 0.4, 0.9, 0.2}

\DeclareFontFamily{U}{rcjhbltx}{}
\DeclareFontShape{U}{rcjhbltx}{m}{n}{<->rcjhbltx}{}
\DeclareSymbolFont{hebrewletters}{U}{rcjhbltx}{m}{n}

\DeclareMathSymbol{\lamed}{\mathord}{hebrewletters}{108}
\DeclareMathSymbol{\mem}{\mathord}{hebrewletters}{109}
\DeclareMathSymbol{\ayin}{\mathord}{hebrewletters}{96}
\DeclareMathSymbol{\tsadi}{\mathord}{hebrewletters}{118}
\DeclareMathSymbol{\qof}{\mathord}{hebrewletters}{113}
\DeclareMathSymbol{\resh}{\mathord}{hebrewletters}{114}
\DeclareMathSymbol{\pe}{\mathord}{hebrewletters}{112}
\DeclareMathSymbol{\pesofit}{\mathord}{hebrewletters}{80}
\DeclareMathSymbol{\samekh}{\mathord}{hebrewletters}{115}
\DeclareMathSymbol{\tav}{\mathord}{hebrewletters}{116}
\DeclareMathSymbol{\vav}{\mathord}{hebrewletters}{119}
\DeclareMathSymbol{\het}{\mathord}{hebrewletters}{120}
\DeclareMathSymbol{\yod}{\mathord}{hebrewletters}{121}
\DeclareMathSymbol{\zayin}{\mathord}{hebrewletters}{122}
\DeclareMathSymbol{\alephdot}{\mathord}{hebrewletters}{128}
\DeclareMathSymbol{\tsadisofit}{\mathord}{hebrewletters}{90}
\DeclareMathSymbol{\shin}{\mathord}{hebrewletters}{152}

\def\d{{\delta}}
\def\be{\begin{equation}}
\def\ee{\end{equation}}
\def\bea{\begin{eqnarray}}
\def\eea{\end{eqnarray}}
\def\bes{\begin{subequations}}
\def\ees{\end{subequations}}

\def\eps{{\epsilon}}
\def\oh{\frac{1}{2}}

\def\re{\mbox{Re}}
\def\im{\mbox{Im}}

\usepackage{multicol}
\usepackage{float}
\usepackage{caption}
\def\p {{\partial}}

\def\g {{\gamma}}

\newcommand{\cK}{\mathcal{K}}
\newcommand{\cM}{\mathcal{M}}

\newcommand{\cO}{\mathcal{O}}
\newcommand{\cT}{\mathcal{T}}

\newcommand*\widefbox[1]{\fbox{\hspace{2em}#1\hspace{2em}}}

\newenvironment{eqn}{\begin{equation}\begin{aligned}}{\end{aligned}\end{equation}\noindent}
\newenvironment{eqn*}{\begin{equation*}\begin{aligned}}{\end{aligned}\end{equation*}\noindent}

\makeatletter
\newsavebox\myboxA
\newsavebox\myboxB
\newlength\mylenA

\newcommand*\xoverline[2][0.75]{%
\sbox{\myboxA}{$\m@th#2$}%
\setbox\myboxB\null
\ht\myboxB=\ht\myboxA%
\dp\myboxB=\dp\myboxA%
\wd\myboxB=#1\wd\myboxA
\sbox\myboxB{$\m@th\overline{\copy\myboxB}$}
\setlength\mylenA{\the\wd\myboxA}
\addtolength\mylenA{-\the\wd\myboxB}%
\ifdim\wd\myboxB<\wd\myboxA%
   \rlap{\hskip 0.5\mylenA\usebox\myboxB}{\usebox\myboxA}%
\else
    \hskip -0.5\mylenA\rlap{\usebox\myboxA}{\hskip 0.5\mylenA\usebox\myboxB}%
\fi}
\makeatother

\begin{document}
\pagestyle{plain}

\makeatletter
\@addtoreset{equation}{section}
\makeatother
\renewcommand{\theequation}{\thesection.\arabic{equation}}

\pagestyle{empty}
\rightline{IFT-UAM/CSIC-21-58}
\vspace{0.5cm}
\begin{center}
\Huge{{F-theory flux vacua at \\ large complex structure}
\\[10mm]}
\normalsize{Fernando Marchesano,$^1$  David Prieto,$^1$ and Max Wiesner$^{1,2}$\\[15mm]}
\small{
${}^1$Instituto de F\'{\i}sica Te\'orica UAM-CSIC, Cantoblanco, 28049 Madrid, Spain \\[2mm] 
${}^2$ Departamento de F\'{\i}sica Te\'orica, 
Universidad Aut\'onoma de Madrid, 
28049 Madrid, Spain
\\[8mm]} 
\small{\bf Abstract} \\[5mm]
\end{center}
\begin{center}
\begin{minipage}[h]{15.0cm} 

We compute  the flux-induced F-term potential in 4d F-theory compactifications  at large complex structure. In this regime, each complex structure field splits as an axionic field plus its saxionic partner, and the classical F-term potential takes the form $V = Z^{AB} \rho_A\rho_B$ up to exponentially-suppressed terms, with $\rho$ depending on the fluxes and axions and  $Z$ on the saxions. We provide explicit, general expressions for $Z$ and $\rho$, and from there analyse the set of flux vacua for an arbitrary number of fields. We identify two families of vacua with all complex structure fields fixed and a flux contribution to the tadpole $N_{\rm flux}$ which is bounded. In the first and most generic one, the saxion vevs are bounded from above by a power of $N_{\rm flux}$. In the second their vevs may be unbounded and $N_{\rm flux}$ is a product of two arbitrary integers, unlike what is claimed by the Tadpole Conjecture. We specialise to type IIB orientifolds, where both families of vacua are present, and link our analysis with previous results in the literature. We illustrate our findings with several examples.

\end{minipage}
\end{center}
\newpage
\setcounter{page}{1}
\pagestyle{plain}
\renewcommand{\thefootnote}{\arabic{footnote}}
\setcounter{footnote}{0}


\tableofcontents

\section{Introduction}
\label{s:intro}

A powerful feature of F-theory compactifications is that they provide an overall picture of the set of string vacua, as they are directly connected to most string theory constructions via dualities. This trait is particularly significant in the context of compactifications to four dimensions, where they are in addition endowed with a notably simple and efficient mechanism to stabilise moduli. Indeed, complex structure moduli fixing in F-theory  through the presence of background four-form fluxes is a paradigmatic framework to remove unwanted neutral scalars from the low energy effective theory \cite{Grana:2005jc,Douglas:2006es,Blumenhagen:2006ci,Becker:2007zj,Marchesano:2007de,Denef:2008wq}. It is from this framework that we have developed our current understanding of the string Landscape. 

Since the F-theory flux landscape is quite vast, it is not obvious how to describe all the information encoded in complex structure moduli stabilisation. One possible approach is to treat the set of flux vacua as an ensemble, and apply statistical methods to extract their physical properties \cite{Denef:2007pq}. A different strategy is to assume that complex structure moduli are fully fixed at a very high scale, and so one can safely integrate out all of them to analyse the physics of K\"ahler moduli and localised degrees of freedom \cite{Ibanez:2012zz,Quevedo:2014xia,Baumann:2014nda}. The information of complex structure moduli stabilisation is then encoded in a set of parameters that appear in the effective theory below the flux scale, and which are oftentimes  assumed to be tunable in terms of an appropriate choice  of Calabi--Yau geometry and flux quanta.

It has however been pointed out that there could be more to it than this generic picture of complex structure moduli stabilisation. On the one hand, some works have questioned the idea that one can generically fix all complex structure moduli and at the same time satisfy the tadpole consistency conditions of the compactification \cite{Braun:2020jrx,Bena:2020xrh,Bena:2021wyr}. On the other hand, it has been shown that at asymptotic limits in complex structure field space the flux potential simplifies and its form can be classified in terms of robust Calabi--Yau data \cite{Grimm:2019ixq}, leading to certain no-go results and general arguments in favour of the finiteness of flux vacua \cite{Grimm:2020cda}. 

Clearly, these recent results point towards a rich structure underlying F-theory flux potentials that is yet to be unveiled. In order to uncover this structure, it is important to gain analytic control over F-theory flux potentials and its corresponding set of vacua. Ideally, given a Calabi--Yau four-fold and a choice of four-form fluxes, one would like to understand directly from these data how many complex structure moduli are stabilised by the potential, at which point in field space they are fixed, and what is their mass spectrum. 

It is the purpose of this work to take a non-trivial step in this direction, by providing an explicit, analytic description of F-theory flux potentials and their vacua. We do so by focusing on regions of large complex structure of smooth Calabi--Yau four-folds. In this regime, we are able to provide an explicit expression for the F-theory F-term potential for any four-fold $Y_4$, up to exponentially-suppressed terms. At this level of approximation, the only data that are needed to specify the potential are the flux quanta and certain topological numbers of the mirror four-fold $X_4$. This simplicity allows us to perform a general analysis of the vacua conditions for an arbitrary number of complex structure fields, and eventually uncover different families in which such vacua are arranged.   

An important ingredient of our analysis is the fact that at moderate and large complex structure the 4d K\"ahler potential displays a number of axionic shift symmetries, only broken by the exponentially-suppressed terms that we neglect. Because of this, each complex structure field splits  into an axionic and a saxionic component. Microscopically, the periodicity of the axions corresponds to the monodromies around the large complex structure point that act non-trivially both on the periods of the holomorphic  $\Omega$ and flux $G_4$ four-forms. It turns out that in terms of these real variables the scalar potential takes a very simple form, namely $V =\half Z^{AB} \rho_A \rho_B$, with $\rho_A$  monodromy-invariant combinations of fluxes and axions, and $Z^{AB}$ only depending on the saxions. Since the potential is  positive semi-definite and only yields Minkowski vacua, the on-shell equations amount to $Z^{AB}\rho_B =0$ $\forall A$, and so they can be solved algebraically.

Using these on-shell equations, one is able to rewrite the flux contribution to the D3-brane tadpole $N_{\rm flux}$ as a sum of positive terms, and from there derive that certain flux quanta must vanish at large complex structure in order to find vacua in this regime. Depending on which quanta vanish we distinguish different families of flux vacua, which we then analyse. In the most generic family, which is present in any Calabi--Yau four-fold $Y_4$, the number of stabilised moduli depends on the choice of fluxes, an effect that we characterise with explicit formulas. Remarkably, even in the most favourable case full moduli stabilisation is not that easy to observe: It is only manifest when the entries of  $Z^{AB}$ are computed to certain accuracy. In practice, one may compute them {\it i)} in the strict asymptotic limit \cite{Grimm:2019ixq}, {\it ii)} by approximating the periods of $\Omega$ with their leading behaviour (section \ref{sec:leading}) , and {\it iii)} by including all the polynomial corrections to such periods, neglecting only exponentially-suppressed terms (section \ref{sec:poly}). For this family of vacua only with this third description full complex structure moduli stabilisation is manifest. Less accurate descriptions yield potentials that typically have at least one flat direction. As a consequence, most vacua cannot exist at parametrically large complex structure. In fact, we find that the saxion vevs are bounded from above by roughly $K^{(3)} N_{\rm flux}^{p + \oh}$, where $K^{(3)}$ represents the minor polynomial correction to the potential, $N_{\rm flux}$ is the flux contribution to the D3-brane tadpole, and $p \leq h^{3,1}(Y_4)$ is bounded by the number of complex structure moduli.    

In this generic scheme, the condition to achieve full moduli stabilisation depends on those flux quanta that contribute to $N_{\rm flux}$. It is therefore possible that in some instances $N_{\rm flux}$ grows as we increase the number of moduli, as recently proposed by the Tadpole Conjecture in \cite{Bena:2020xrh}. Our framework allows us to propose a formula that tests this statement, and that can be checked in any compactification. Regardless of whether this happens or not, we find that in certain compactifications the Tadpole Conjecture is violated, due to a second family of vacua that emerges for them. This new family of vacua arises whenever a complex structure saxion appears at most linearly in $e^{-K}$ (with $K$ the K\"ahler potential) and the superpotential, a setup which we dub the {\em linear scenario}. Examples of this are Calabi--Yau four-folds $Y_4$ whose mirror $X_4$ is a fibration of a Calabi--Yau over a $\mathbb{P}^1$, and in particular type IIB orientifold compactifications. The new set of vacua appears at large values of the linear saxion, with $N_{\rm flux}$  a simple product of two flux quanta. The remaining non-vanishing flux quanta are such that they fix all complex structure moduli. Remarkably, in the particular case of  type IIB compactifications the polynomial corrections identified as $K^{(3)}$ are also needed to implement this full moduli stabilisation and, in fact, this family of vacua are mirror dual of the Minkowski type IIA flux vacua originally found in \cite{Palti:2008mg}. The necessity of polynomial corrections is however not a universal feature in other F-theory realisations of the linear scenario, as we show with an explicit example. This indicates that it is this more exotic family of vacua, and maybe new ones yet to be discovered, that dominate the landscape of F-theory vacua at regions of parametrically large complex structure.

The paper is organised as follows. In section \ref{s:potential} we compute the flux scalar potential for arbitrary four-folds, first using the leading terms of the periods of $\Omega$ and then including all polynomial terms. In section \ref{s:vacua} we analyse the resulting vacua equations, and in particular how a finite D3-brane tadpole affects the existence of vacua. From here we obtain the most generic family of flux vacua in the large complex structure regime, which nevertheless cannot exist at parametrically large complex structure. In section \ref{s:IIB} we apply our results to the special case of type IIB orientifold compactifications, matching them with the existing literature. In particular, we identify a family of flux vacua which is different from the generic one, in which the expression for $N_{\rm flux}$ is independent of the number of moduli. Section \ref{s:linear} upgrades this family of vacua to a genuine F-theory setup, which we dub linear scenario. In section \ref{s:examples} we illustrate our findings with explicit constructions of Calabi--Yau four-folds, whose mirror are smooth fibrations. We finally present our conclusions in section \ref{s:conclu}.

Several technical details have been relegated to the appendices. Appendix \ref{ap:georho} provides a geometric definition of the flux-axion polynomials $\rho_A$, and relates the $Z^{AB}$ with the Hodge star action on the space of four-forms. Appendix \ref{ap:curvature} gathers the different computations needed to include all the polynomial terms in the flux potential which, in terms of the mirror four-fold, can be seen as taking into account curvature corrections. Appendix \ref{ap:invariants} discusses the mondromy-invariant combination of fluxes that appear in our setup, which are the quantities that fix the saxions vevs. Appendix \ref{ap:elliptic} computes the most involved part of the vacua equations for the compactifications discussed in section \ref{sec:elliptic}, whose mirror four-fold is an elliptic fibration. 


\section{The F-theory potential at large complex structure}
\label{s:potential}

In a region of sufficiently large complex structure, the moduli space geometry of F-theory on a Calabi--Yau (CY) four-fold simplifies, in the sense that each complex structure field splits into an axionic and a saxionic real components. This not only constrains the form of the 4d effective K\"ahler potential, but also of the superpotential induced by background four-form fluxes. In this section we will compute both, and from there provide an explicit bilinear expression for the F-term scalar potential, on which we will base our subsequent analysis. In section \ref{sec:leading} we will consider the leading form of the potential at large saxion values, from which one can infer most of the intuition regarding the ensemble of flux vacua, and in section \ref{sec:poly} we will include the polynomial corrections to these leading terms. As we will see in section \ref{s:vacua}, such corrections turn out to be crucial to fully understand moduli stabilisation in F-theory. 

\subsection{The leading flux potential}
\label{sec:leading}

Let us consider F-theory compactified on a Calabi--Yau four-fold $Y_4$, which is a smooth elliptic fibration over a three-fold base $C_3$. In the presence of an internal background four-form flux $G_4$, a scalar potential is generated for both the complex structure and K\"ahler moduli of $Y_4$. On the one hand, the potential for K\"ahler moduli can be seen as a D-term potential $D = \oh \int_{Y_4} G_4 \wedge J \wedge J$, with $J$ the K\"ahler form of $Y_4$. On the other hand, the potential for the complex structure moduli can be understood as an F-term potential, with Gukov-Vafa-Witten superpotential \cite{Gukov:1999ya}
\be
W = \int_{Y_4} G_4 \wedge \Omega\, ,
\label{GVW}
\ee
where $\Omega$ is the holomorphic (4,0)-form of $Y_4$, in terms of which we define its complex structure moduli. At large volume the K\"ahler potential is given by $K = - 2 \log {\cal V}_3 + K_{\rm cs}$, where ${\cal V}_3$ is the volume of $C_3$ and only depends on its K\"ahler moduli, while
\be
K_{\rm cs} = - \log \int_{Y_4} \Omega \wedge \bar{\Omega} \, .
\label{kahler}
\ee
Both potentials are positive semi-definite, and select global, 4d Minkowski minima at those points in moduli space where the Hodge self-duality condition is satisfied \cite{Haack:2001jz}
\be
G_4 = * G_4\, .
\label{SDG4}
\ee
Those minima in which $G_4$ is a primitive (2,2)-form are, moreover, supersymmetric \cite{Becker:1997cp}. 

Our goal is to provide an explicit expression for the F-term scalar potential in terms of the complex structure moduli of the four-fold. To do so one must first determine a basis for the lattice $\Lambda_W$ of quantised fluxes that enters \eqref{GVW}, and then compute the corresponding periods of $\Omega$. It turns out that the first part of this problem is quite subtle. This lattice pairs up via \eqref{GVW} with the horizontal subspace of the middle cohomology of the four-fold  $H_H^4(Y_4) \subset H^4(Y_4)$, which is generated by $\Omega$ and its derivatives \cite{Strominger:1990pd,Greene:1993vm}. We have that $\dim H_H^4(Y_4) = 2 + 2 h^{3,1}(Y_4) + \dim H_H^{2,2}(Y_4)$, with the embedding $ H_H^{2,2}(Y_4) \subset  H^{2,2}(Y_4)$ being quite involved \cite{Braun:2014xka}. As a consequence, in a four-fold there is no clear link between the number of complex structure moduli, which is given by $h^{3,1}(Y_4)$, and the number of fluxes that enter the superpotential.\footnote{Recall that for type IIB  on a Calabi--Yau three-fold we have $b_3/2$ complex fields on the complex structure and axio-dilaton sectors, and a real lattice of background three-form fluxes of dimension $2b_3$. In sections \ref{s:linear} and \ref{s:examples} we will consider F-theory constructions that reproduce the same sort of relation.} 

Fortunately, one may implement the strategy of \cite{CaboBizet:2014ovf,Cota:2017aal} to overcome these difficulties and find concrete expressions for the F-term potential. The main idea in \cite{CaboBizet:2014ovf,Cota:2017aal} is to use homological mirror symmetry and consider the mirror four-fold of $Y_4$, which we denote as $X_4$. Then one may compactify type IIA on $X_4$, and identify the periods of $\Omega$ in $Y_4$ with the central charges of topological B-branes on $X_4$, which generate the mirror of the lattice $\Lambda_W$. In the large volume regime, this lattice can be understood as D$(2p)$-branes wrapping holomorphic $2p$-cycles, with $p=0,1,2,3,4$. The subtleties alluded above translate into constructing a basis of holomorphic 4-cycles, a set that can be generated by intersecting pairs of divisors of $X_4$. This basis can be constructed explicitly when $X_4$ is a smooth fibration, see \cite{Cota:2017aal} and the discussion in sections \ref{s:linear} and  \ref{s:examples}. An element of the corresponding lattice will have a central charge of the form $\int_{X_4} e^{J_c} \wedge F_{RR}$, where $F_{RR}$ is a closed even polyform and $J_c = B + iJ$ is the complexified K\"ahler form of $X_4$. It follows that, under these assumptions, the F-theory superpotential \eqref{GVW} can be identified with a 2d analogue of the 4d type IIA RR flux superpotential \cite{Taylor:1999ii}. 

The leading order term for the central charge $\Pi_{2p}$ of a D$(2p)$-brane wrapping a holomorphic $2p$-cycle on $X_4$ in the large volume limit is 
\begin{subequations}
\label{eq:periods}
\begin{align}
        \Pi_0&=1\, , \\
        \Pi_2^i&=-T^i\, ,\\
        \Pi_{4\, \mu} &=\oh \eta_{\mu\nu} \zeta^\nu_{ij}T^iT^j\, ,\\
        \label{period4}
        \Pi_{6\, i}&=-\frac{1}{6}\mathcal{K}_{ijkl}T^jT^kT^l\, ,\\ 
        \Pi_8&=\frac{1}{24}\mathcal{K}_{ijkl}T^iT^jT^kT^l\, ,
\end{align}
\end{subequations}
where $T^i = b^i + i t^i$, $i = 1, \dots, h^{1,1}(X_4)$ stand for the complexified K\"ahler moduli of $X_4$, and $\CK_{ijkl}$ for its quadruple intersection numbers. The index $\mu$ in $\Pi_{4\, \mu}$ runs over a basis of four-cycles generating all the intersections of a basis of Nef divisor classes $[D_i]$ on $X_4$. As a result we can write the class of their intersection as $[\g_{ij}] =  [D_i . D_j] = \zeta^\mu_{ij} [\sigma_\mu]$ for some set of integral four-form classes $[\sigma_\mu]$ and some $\zeta^\mu_{ij} \in \mathbb{Z}$. Finally $\eta_{\mu\nu} = [\sigma_\mu] \cdot [\sigma_\nu]$ is the intersection matrix of this sector, which must satisfy
\be
\CK_{ijkl} = \zeta_{ij}^\mu \eta_{\mu\nu}\zeta^\nu_{kl}  = \zeta_{ij}^\mu\zeta_{\mu, kl}\, .
\label{interrel}
\ee
where in the second equality we have defined $\zeta_{\mu, kl} \equiv [\sigma_\mu] \cdot [D_k] \cdot [D_l]$.

Applying the mirror symmetry map, the $\{T^i\}$ become the complex structure moduli of $Y_4$, where now $i = 1, \dots, h^{3,1}(Y_4)$. The set of holomorphic $2p$-cycles classes of $X_4$ becomes a lattice of horizontal four-cycles in $Y_4$, such that $[\sigma_\mu] \mapsto [\sigma_\mu^Y]$. The central charges \eqref{eq:periods} become the leading terms for the periods of the four-form $\Omega$ in the large complex structure limit, where it admits an expansion of the form
\begin{equation}
    \Omega=\alpha \pi_0+\alpha_i\pi_2^i+\sigma_\mu^Y \pi^\mu_4 +\beta^i\pi_{6i}+\beta\pi_8\, .
    \label{Omega}
\end{equation}
Here $\{ \a, \a_i, \sigma_\mu^Y, \b^i, \b\}$ represent a set of harmonic four-forms which is also an integral basis for $H^4_H(Y_4)$. Their moduli-dependent coefficients are given by
\begin{equation}
\label{eq:coeff}
        \pi_0=1\, , \quad
        \pi_2^i=T^i\, ,\quad
        \pi_{4}^\mu =\oh  \zeta^\mu_{ij}T^iT^j\, ,\quad
        \pi_{6i}=\frac{1}{6}\mathcal{K}_{ijkl}T^jT^kT^l\, ,\quad
        \pi_8=\frac{1}{24}\mathcal{K}_{ijkl}T^iT^jT^kT^l\, .
\end{equation}
These coefficients are the periods of the holomorphic form in the homological representation. One can relate these with the periods in the cohomology basis \eqref{eq:periods} using the intersection matrix for the Poincaré dual cycles to the four form basis $\{ \a, \a_i, \sigma_\mu^Y, \b^i, \b\}$. The classical intersection numbers for such cycles are 
\begin{equation}
    \int_{Y_4}\alpha \wedge \beta=1,\hspace{1cm} \int_{Y_4}\alpha_i\wedge \beta^j=-\delta_{i}^j,\hspace{1cm}
\int_{Y_4}\sigma_\mu^Y\wedge \sigma_\nu^Y=\eta_{\mu\nu}\, .
\label{intersection}
\end{equation}
In fact the intersection matrix for $\{ \a, \a_i, \sigma_\mu^Y, \b^i, \b\}$ is more involved, as \eqref{intersection} receive corrections that destroy its block-anti-diagonal form and which, in the mirror four-fold $X_4$, arise due to curvature terms. We discuss such corrections in subsection \ref{sec:poly}, where we show that they can be absorbed in a redefinition of the $G_4$-flux quanta. Thus, for the purpose of providing an explicit expression for the F-term potential, one may still work with these naive intersection numbers.

To compute the flux superpotential we only need to expand the flux $G_4$ in the same basis of four-forms
\be
G_4 =  m \a  - m^i \a_i  +\hat{m}^{\mu} \sigma_\mu^Y - e_i \b^i  + e \b\, ,
\label{G4}
\ee
where $m, m^i, \hat{m}^\mu, e_i, e \in \Z$ represent the flux quanta. Using \eqref{intersection} we obtain that the superpotential takes the form
\be
W  =  e +  e_iT^i + \frac{1}{2}\, \hat{m}^{\mu} \zeta_{\mu,kl}   T^k T^l  + \frac{1}{6}\, {\cal K}_{ijkl}\, m^i T^j T^k T^l +  \frac{m}{24}\, {\cal K}_{ijkl}\, T^i T^j T^k T^l \, .
\label{supo}
\ee
One can obtain a more symmetric expression by considering a set of integers $m^{ij}$ that satisfy
\be
\hat{m}^{\mu} = \oh  \zeta_{ij}^\mu m^{ij}\, ,
\label{hatm}
\ee
so that the superpotential becomes
\be
W  =  e +  e_iT^i + \frac{1}{4}\, {\cal K}_{ijkl}  m^{ij} T^k T^l  + \frac{1}{6}\, {\cal K}_{ijkl}\, m^i T^j T^k T^l +  \frac{m}{24}\, {\cal K}_{ijkl}\, T^i T^j T^k T^l \, .
\label{supalt}
\ee
In general the choice of $m^{ij}$ is not unique, but it is easy to see that any choice will yield the same final expression. We will predominantly use the form of the superpotential \eqref{supo}, although in some instances it will be more convenient to use the auxiliary expression \eqref{supalt} that involves the redundant set of fluxes $m^{ij}$.

Notice that this superpotential is nothing but a linear combination of the central charges $\Pi_{2p}$ in \eqref{eq:periods} which, upon mirror symmetry becomes a linear combination of the periods of $\Omega$. Indeed, we have that
\be
W = \vec{q}^{\, t} \Sigma \vec{\Pi} = e \Pi_0 - e_i \Pi_2^i + \hat{m}^\mu \Pi_{4\, \mu} - m^i \Pi_{6\, i} + m \Pi_8\, , 
\label{supoprod}
\ee
which clearly reproduces \eqref{supo}. Here we have defined the vector of fluxes $\vec{q}^{\, t} = (m, m^i,  \hat{m}^\mu, e_i, e)$, the vector of periods $\vec{\Pi}^{\, t} = (\Pi_8, \Pi_{6\, i}, \Pi_{4\, \mu}, \Pi_{2}^i, \Pi_0)$ and the pairing matrix
\begin{align}
    \Sigma = \left(\begin{matrix} 1&0&0&0&0 \\ 0&-\delta_j^i&0&0 &0 \\ 0&0&\d_{\mu \nu} &0&0\\ 0&0 &0&-\delta_i^j&0 \\ 0&0&0&0&1   \end{matrix}\right)\, .
    \label{pairingm}
\end{align}

We can also use \eqref{eq:periods}, \eqref{Omega} to compute the piece of the K\"ahler potential \eqref{kahler}. We have that
\be
    K_{\rm cs}=-\log\left[2\re (\pi_0\bar{\pi}^8)\int_{Y_4}\alpha\wedge {\beta}+2\re (\pi_2^i\bar{\pi}^6_j)\int_{Y_4}\alpha_i\wedge {\beta}^j
 +  \pi_4^\mu\bar{\pi}_4^\nu \int_{Y_4}\sigma_\mu^Y\wedge \sigma_\nu^Y \right]\, ,
  \label{Kcscomp}
\ee
from where we obtain
\begin{equation}
    K_{\rm cs}=-\log(\frac{2}{3}\mathcal{K}_{ijkl}t^it^jt^kt^l)\, .
    \label{Kcs}
\end{equation}
As expected, in this large complex structure limit the leading term of the K\"ahler potential only depends on $t^i \equiv \im T^i$, and so the field space metric displays abundant continuous shift symmetries. As we will see below, polynomial corrections to the periods \eqref{eq:periods} do modify \eqref{Kcs}, but they do not introduce a dependence on $b^i \equiv \re T^i$. This can be expected from considering type IIA compactified on the mirror manifold $X_4$, where the $b^i$ correspond to integrals of the B-field. In the large volume limit these fields can be considered as axions, since the only terms breaking the continuous shift symmetry are generated by world-sheet instanton effects and are therefore suppressed as $e^{2\pi i T^in_i  }$, $n_i \in \Z$. The same statement applies to our setup, where the periodic nature of the fields $b^i$ translates into a the familiar set of monodromies $\cT_i$ around the large  complex structure point, which act non-trivially on the basis $\{ \a_0, \a_i, \sigma_\mu, \b^i, \b^0\}$, the periods $\Pi_{2p}$ and the flux quanta, but leave $\Omega$ and $G_4$ invariant. 

This large set of axionic variables allows us to derive a simple, analytic expression for the F-term scalar potential. The main observation is that one should express the scalar potential in terms of a set of axion polynomials $\rho_A$ linear on the flux quanta, which are invariant under the action of the monodromies $\cT_i$. Because at the two-derivative level the scalar potential is quadratic in the fluxes, one recovers an expression of the form
\be
V = \oh Z^{AB} \rho_A \rho_B \, ,
\label{bilinear}
\ee
where $\rho_A \equiv \rho_A(b)$ are independent of the saxions $t^i$. The matrix entries $Z^{AB}$ do not depend on the fluxes, and so they can only depend on the axions through periodic functions. However, such periodic functions necessarily enter the periods of $\Omega$ through terms of the form $e^{2\pi i T^in_i  }$, which are exponentially suppressed in the large complex structure regime. Therefore under our assumptions we have that $Z^{AB} \equiv Z^{AB}(t)$ only depends on the saxions of the compactification, providing a simple, factorised bilinear structure for the F-term scalar potential. This same strategy was applied for type IIA 4d flux compactifications in \cite{Bielleman:2015ina,Carta:2016ynn,Herraez:2018vae,Marchesano:2020uqz}, where a potential with the structure \eqref{bilinear} was obtained, in agreement with general EFT considerations \cite{Farakos:2017jme,Bandos:2018gjp,Lanza:2019xxg}. As shown in \cite{Escobar:2018tiu,Escobar:2018rna,Marchesano:2019hfb,Marchesano:2020uqz}, this bilinear structure allows one to characterise the set of vacua in a simple, systematic manner, and even to determine the behaviour of the system away from them \cite{Valenzuela:2016yny,Grimm:2019ixq,Grimm:2020ouv}.  In section \ref{s:vacua} we will use the form \eqref{bilinear} of the F-theory F-term potential to classify the set of flux vacua at large complex structure. Finally, as pointed out in \cite{Grimm:2019ixq}, the same bilinear expression \eqref{bilinear} holds near other points at infinite distance in complex structure field space, and so in principle our strategy could be extended to these regions as well. 

To find the bilinear expression \eqref{bilinear} one must use the well-known no-scale properties of F-theory compactifications to simplify the Cremmer et al. \cite{Cremmer:1982en} formula for the F-term potential. In particular, the fact that the K\"ahler moduli do not appear in the superpotential translates into the following simplified expression \cite{Haack:2001jz,Giddings:2001yu}
\begin{equation}
    V=e^{K}\sum_{i,j} K^{i\bar{j}}D_{i}W D_{\bar{j}}\overline{W}\, ,
    \label{Cremmer}
\end{equation}
where $i, j = 1, \dots , h^{3,1}(Y_4)$ run over the complex structure moduli of $Y_4$. Here $D_i = \p_i + (\p_i K)$ stands for the supergravity covariant derivative, while $K^{i\bar{j}}$ is the inverse of the K\"ahler metric $K_{i\bar{j}} \equiv \p_i\p_{\bar{j}} K$. Because the K\"ahler potential is independent of the complex structure axions, it is more convenient to express both in terms of tensors with real indices $g_{ij} \equiv \frac{1}{4} \p_{t^i}\p_{t^j} K =  K_{i\bar{j}}$. These read
\be
g_{ij} = 4\frac{\mathcal{K}_i\mathcal{K}_j}{\mathcal{K}^2}-3\frac{\mathcal{K}_{ij}}{\mathcal{K}}\, \qquad g^{ij} =\frac{4}{3}t^it^j-\frac{1}{3}\mathcal{K}\mathcal{K}^{ij} \, ,
  \label{metric}
\ee
with $\cK^{ij}$ the inverse of $\cK_{ij}$, and we have defined the contractions
\be
 \mathcal{K}\equiv\mathcal{K}_{ijkl}t^it^jt^kt^l\, , \quad \mathcal{K}_i\equiv \mathcal{K}_{ijkl}t^jt^kt^l\, , \quad \mathcal{K}_{ij}\equiv \mathcal{K}_{ijkl}t^kt^l\, , \quad \mathcal{K}_{ijk}\equiv \mathcal{K}_{ijkl}t^l\, .
\ee

The expression \eqref{Cremmer} is already positive semi-definite and  bilinear, but still not of the form \eqref{bilinear}. To make explicit the factorisation between axions and saxions, one must define the flux-axion polynomials $\rho_A$, which capture the discrete symmetries of the superpotential, and whose geometric interpretation and general definition is given in appendix \ref{ap:georho}. In our setup they read
\begin{subequations}
\label{rhos}
\begin{align}
    \rho=&\ e+e_ib^i+\frac{1}{2}\hat{m}^\mu\zeta_{\mu,kl} b^k b^l +\frac{1}{6}\mathcal{K}_{ijkl}m^ib^jb^kb^l+\frac{1}{24}m \mathcal{K}_{ijkl}b^ib^jb^kb^l\, , \\
    \rho_i=&\ e_i+\hat{m}^\mu\zeta_{\mu,il} b^l+\frac{1}{2}\mathcal{K}_{ijkl}m^jb^kb^l+\frac{1}{6}m\mathcal{K}_{ijkl}b^jb^kb^l\\
    \hat{\rho}^\mu=&\ \hat{m}^\mu+\zeta_{ij}^\mu b^im^j+\frac{1}{2}\zeta_{ij}^\mu b^ib^j\ , \\
    \tilde{\rho}^i=&\ m^i+mb^i\, ,\\
    \tilde{\rho}=&\ m\, .
    \end{align}
    \end{subequations}
As pointed out in \cite{Herraez:2018vae}, these polynomials are related to each other via derivatives, leading to a convenient way to express for the superpotential and F-terms. For the case at hand we have
\begin{subequations}
    \label{eq: superpotential and deriv}
\begin{align}
\label{eq:suporho}
    W=&\rho+i\rho_it^i-\frac{1}{2}\zeta_\mu\hat{\rho}^\mu-\frac{i}{6}\mathcal{K}_i\tilde{\rho}^i+\frac{\mathcal{K}}{24}\tilde{\rho}\, ,\\
    \partial_{i}W=&\rho_i+i\zeta_{\mu i}  \hat{\rho}^\mu-\frac{1}{2}\mathcal{K}_{ij}\tilde{\rho}^j-\frac{i}{6}\mathcal{K}_i\tilde{\rho}\, ,
\end{align}
\end{subequations}
together with $\partial_j K = 2 i \cK_j/\cK$, and where we have defined the contractions $\zeta_\mu \equiv \zeta_{\mu,ij} t^it^j$ and $\zeta_{\mu i} \equiv \zeta_{\mu,ij} t^j$. Plugging these expressions into \eqref{Cremmer} and using the properties of the metrics \eqref{metric} one finds the following expression for the F-theory flux potential
\begin{align}
    V= e^{K} \left[4\left(\rho-\frac{\mathcal{K}}{24}\tilde{\rho}\right)^2+g^{ij}\left(\rho_i+\frac{\mathcal{K}}{6}g_{ik}\tilde{\rho}^k\right)\left(\rho_j+\frac{\mathcal{K}}{6}g_{jl}\tilde{\rho}^l\right)+
    g^{ij}_{P}\zeta_{\mu i}\zeta_{\nu j} \hat{\rho}^\mu  \hat{\rho}^\nu
    \right]\, ,
    \label{scalarpot}
\end{align}
where $g_P^{ij}$ is the primitive component of the inverse metric, i.e. $g_P^{ij}=\frac{1}{3}(t^it^j-\mathcal{K}\mathcal{K}^{ij})$. This expression for the potential is one of the main results of this section. It reproduces the bilinear, factorised structure in \eqref{bilinear} as a sum of three positive semi-definite terms, that correspond to a block-diagonal structure for the saxion-dependent matrix $Z$. Indeed, if we arrange the flux-axion polynomials in a vector of the form
\be
\vec{\rho}^{\, t} = \left(\tilde{\rho}, \tilde{\rho}^i,   \hat{\rho}^{\mu}, \rho_i,   \rho   \right) \, ,
\label{vecrho}
\ee
then the said matrix reads
\begin{align}
\label{ZAB}
Z^{AB} =
\frac{e^K\cK}{3}
\begin{pmatrix}
\frac{\cK}{24} & & & & -1 \\
&  \frac{\cK}{6} g_{ij} & &  \d^i_j &  \\
& & \frac{6}{\cK}  g^{ij}_{P}\zeta_{\mu i}\zeta_{\nu j} & &  \\
&  \d^i_j & &  \frac{6}{\cK} g^{ij} &  \\
-1 & & & &  \frac{24}{\cK} \\
\end{pmatrix} \, ,
\end{align}
which can be easily taken to a block-diagonal form. Notice that each block is singular, and that their ranks add up to $2h^{3,1}(Y_4)$. Therefore, generically the vacua equations $Z^{AB}\rho_B =0$ amount to impose $2h^{3,1}(Y_4)$ conditions on the same amount of unknowns, namely the complex structure real fields. Finally, note that we can rewrite this expression as 
\begin{align}
\label{ZABdiag}
2{\cal V}_3^2 Z =  {\rm diag} \left(\frac{\cK}{24},  \frac{\cK}{6} g_{ij},  g_{\mu\nu},  \frac{6}{\cK}g^{ij},  \frac{24}{\cK} \right)
- 
\chi_0\, ,
\end{align}
where $g_{\mu\nu} \equiv \eta_{\mu\nu} -2 (\cK^{ij}-\cK^{-1} t^it^j) \zeta_{\mu i}\zeta_{\nu j}$ and 
\begin{align}\label{eq:chi0}
    \chi_0 = \left(\begin{matrix} 0&0&0&0&1 \\ 0&0&0&-\delta_j^i &0 \\ 0&0&\eta_{\mu\nu} &0&0\\ 0&-\delta_i^j &0&0&0 \\ 1&0&0&0&0   \end{matrix}\right)\, ,
\end{align}
encodes the intersection numbers  \eqref{intersection}.
 As it follows from the results of appendix \ref{ap:georho}, splitting $Z$ in these two terms corresponds to the well-known expression for the scalar potential
\be
V = \frac{1}{4{\cal V}_3^2} \left[ \int_{Y_4} G_4 \wedge * G_4 -   \int_{Y_4} G_4 \wedge G_4\right]\, ,
\ee
at this level of approximation. As we will see below, the polynomial corrections to the scalar potential will respect the factorisation between axions and saxions, and therefore the bilinear structure \eqref{bilinear}. On the one hand, the corrections to the intersection numbers \eqref{intersection} will modify $\vec{\rho}$ but not $Z$. On the other hand, the corrections to the K\"ahler potential  \eqref{Kcs}  will leave $\vec{\rho}$ invariant but destroy the block-diagonal structure of $Z$.

It is instructive to compare the above results with previous analysis in the literature. For instance, one would recover the F-theory flux potential analysed in \cite{Marsh:2015zoa} by setting $m^i = \hat{m}^\mu = e_i = e =0$ and keeping only $m$ as a non-vanishing quantum of flux. The scalar potential would still look the same, but the axion dependence in \eqref{rhos} would become very simple. As we will see in section \ref{s:vacua}, vacua with $m \neq 0$ are not allowed at sufficiently large complex structure, in agreement with the result of \cite{Marsh:2015zoa}. Including the remaining flux quanta does a priori allow us to find non-trivial extrema of the potential, as we will also study in the next section.

One may also compare \eqref{scalarpot} with the asymptotic potentials analysed in \cite{Grimm:2019ixq} restricted to the particular case of the  large complex structure limit. In the language of \cite{Grimm:2019ixq}, the approximation that leads to the expression \eqref{scalarpot} lies in between those that result in the asymptotic form of the potential and its strictly asymptotic form. To achieve the latter one must take the expression \eqref{ZABdiag} and replace each of the entries in $ {\rm diag} \left(\frac{\cK}{24},  \frac{\cK}{6} g_{ij},  g_{\mu\nu},  \frac{6}{\cK}g^{ij},  \frac{24}{\cK} \right)$ by its leading term on the complex structure saxions $t^a$, which amounts to replace the Hodge star operator by its strictly asymptotic approximation $C_{\rm sl(2)}$. The plain asymptotic form of the potential (that is, replacing $*$ by $C_{\rm nil}$) is achieved by adding further polynomial corrections to \eqref{scalarpot}, which we now turn to analyse. As we will see, full moduli stabilisation is only achieved when these corrections are taken into account. Moreover, their presence leads to important restriction on the space of flux vacua, which remain undetected if only the strictly asymptotic form of the potential is used.

\subsection{Polynomial corrections}
\label{sec:poly}

The leading form of the potential \eqref{scalarpot} receives several corrections of different nature, which can be classified in terms of corrections to the superpotential and K\"ahler potential. In the following we will address those that depend on the complex structure sector and are polynomial corrections to $W$ and $e^{-K}$. 
These can be treated like perturbative corrections to the leading potential, as opposed to exponentially-suppressed corrections. 
Taking these polynomial corrections into account permits to extend our analysis to regions where the complex structure saxions are only moderately large, so that the exponential corrections of the form $e^{2\pi i T^in_i }$ can still be neglected. The reader not interested in the details of the following derivation may only focus on the results \eqref{supofinal} and \eqref{Kcscorr}, that summarise the polynomial corrections for the superpotential and K\"ahler potential, and proceed to the next section.

To compute the said corrections let us again consider type IIA compactified in the mirror four-fold $X_4$. Here the polynomial corrections that arise in the K\"ahler sector are due to curvature corrections, while the exponential corrections that we will neglect arise from world-sheet instanton effects. The polynomial corrections are encoded in the asymptotic expression for the D$(2p)$-brane central charges, as computed in \cite{Gerhardus:2016iot} and reviewed in appendix \ref{sap:corrper}. They correct the leading terms in  \eqref{eq:periods} as
\begin{subequations}
\label{eq:corrper}
\begin{align}
    \Pi_0^{\rm corr}=&1\, ,\\
     \Pi_{2}^{i\, {\rm corr}}=&-T^i\, ,\\
     \Pi_{4\, ij}^{\rm corr} =&\frac{1}{2}\cK_{ijkl}T^kT^l+\frac{1}{2}\left(\CK_{iijk}+\CK_{ijjk} \right) T^k + \frac{1}{12}\left(2\CK_{iiij} + 3\CK_{iijj} + 2\CK_{ijjj} \right) +  K_{ij}^{(2)} \, ,\\
    \Pi_{6\, i}^{\rm corr} =&-\frac{1}{6}\cK_{ijkl}T^jT^kT^l-\frac{1}{4}\cK_{iijk}T^jT^k-\frac{1}{6}\cK_{iiij}T^j - K^{(2)}_{ij}T^j \nonumber\\
    &- \half K^{(2)}_{ii} -\frac{1}{24}\cK_{iiii}+i K_i^{(3)} \, ,\\
    \Pi_8^{\rm corr}=&\frac{1}{24}\cK_{ijkl}T^iT^jT^kT^l+ \half K_{ij}^{(2)} T^iT^i - iK_i^{(3)} T^i + K^{(0)}\, , 
\end{align}
\end{subequations}
where we have defined
\be
K^{(2)}_{ij} = \frac{1}{24}\int_{X_4} c_2(X_4) \wedge D_i \wedge D_j\, , \qquad K_i^{(3)} = \frac{\zeta(3)}{8\pi^3}\int_{X_4} c_3(X_4)\wedge D_i \, ,
\label{K23}
\ee
and
\be
K^{(0)} = \frac{1}{5760}\int_{X_4} 7c_2(X_4)^2-4c_4(X_4)\, .
\label{K0}
\ee
Notice that here we are working with the redundant set of four-cycles $\g_{ij}= D_i .  D_j$. 

From these expressions it is possible to compute how the corrected version of the F-theory superpotential \eqref{supo} looks like. Indeed, mirror symmetry translates \eqref{eq:corrper} into the corrected periods of $\Omega$ in $Y_4$, and so one simply needs to multiply them by the $G_4$ flux quanta, as in \eqref{supoprod}. We will then obtain an expression of the form
\be
W = \vec{q}^{\, t} \Sigma \vec{\Pi}^{\rm corr} = e \Pi_0^{\rm corr} - e_i \Pi_2^{i,{\rm corr}}+ \hat{m}^\mu \Pi_{4\, \mu}^{\rm corr} - m^i \Pi_{6\, i}^{\rm corr} + m \Pi_8^{\rm corr}\, , 
\label{supoprodcor1}
\ee
The main subtlety boils down to relating the periods associated to the central charges of the mirror B-branes wrapping the basis of 4-cycles, i.e. $\Pi_{4\, \mu}^{\rm corr}$, with the periods associated to the B-branes wrapping the (potentially degenerated) system that contains all intersections of divisors, i.e. $\Pi_{4 \, ij}^{\rm corr}$. Since we know the expressions for the latter, it will be more convenient to express \eqref{supoprodcor1} in terms of them and the flux quanta $m^{ij}$ defined in \eqref{hatm}. We thus start by rewriting 
\be
W  = e \Pi_0^{\rm corr} - e_i \Pi_2^{i,{\rm corr}}+ \frac{1}{2}m^{ij}\zeta^\mu_{ij} \Pi_{4\, \mu}^{\rm corr} - m^i \Pi_{6\, i}^{\rm corr} + m \Pi_8^{\rm corr}\, . 
\label{supoprodcor2}
\ee
The representations $\Pi_{4\, \mu}^{\rm corr}$ and $\Pi_{4\, ij}^{\rm corr}$  are related by a linear combination of periods as follows
\begin{eqn}
    \Pi_{4 \, \mu}^{\rm corr}=\mathcal{M}^{0}_\mu \Pi^{\rm corr}_0+\mathcal{M}_{\mu, i} \Pi_2^{i, {\rm corr}}+\mathcal{M}^{ij}_\mu \Pi^{\rm corr}_{4 \, ij}\,.
    \label{eq: Pimu transformation}
\end{eqn}
Upon contracting with the tensor $\zeta^\mu_{ij}$, the linearity of the $H^4(X_4)$ central charge components  with respect to the intersections $[D_i.D_j]$ allows us to find and inverse relation of the form\footnote{Note this linearity is not preserved once we include the contributions D2 and D0 contributions to the central charge, motivating the need of the additional terms in the transformations \eqref{eq: Pimu transformation} and \eqref{eq: Piij transformation}. To be more precise the central charges are linear over the K-theory classes of sheaves $\mathcal{O}_{D_i\cap D_j}$ not on $[D_j.D_j]$ themselves.} 
\begin{eqn}
    \Pi^{\rm corr}_{4 \, ij}=\hat{\mathcal{M}}^{0}_{ij} \Pi^{\rm corr}_0+\hat{\mathcal{M}}_{ij, k} \Pi_2^{k, {\rm corr}}+\zeta^\mu_{ij}\Pi_{4 \, \mu}^{\rm corr}\,,
    \label{eq: Piij transformation}
\end{eqn}
where we define new (integral) matrix components $\hat{\mathcal{M}}^0_{ij}$ and $\hat{\mathcal{M}}_{ij,k}$ appropriately. Substituting $\zeta^\mu_{ij}\Pi_{4 \, \mu}^{\rm corr}$ back into \eqref{supoprodcor2} we obtain
\begin{eqn}
    W = \left(e-\frac{1}{2}\hat{\mathcal{M}}_{ij}^0m^{ij}\right) \Pi_0^{\rm corr} - \left(e_i+\frac{1}{2}\hat{\mathcal{M}}_{jk,i}m^{jk}\right) \Pi_2^{i,{\rm corr}}+ \frac{1}{2}m^{ij} \Pi_{4\, ij}^{\rm corr} - m^i \Pi_{6\, i}^{\rm corr} + m \Pi_8^{\rm corr}\, .
    \label{supoprodcor3}
\end{eqn}
 This is a rather involved expression, but it becomes more manageable if one distinguishes between two classes of corrections that appear in the periods of $\Omega$. The first one corresponds to corrections to the intersection numbers \eqref{intersection}, and the second one to the K\"ahler potential \eqref{Kcs}. As we will see, each of these corrections has a different effect on the F-term scalar potential, which becomes more transparent when it is written in the bilinear form \eqref{bilinear}.

To compute the corrections to the intersection numbers \eqref{intersection}, one may again consider type IIA compactified on the mirror manifold $X_4$. There, two D$(2p)$-branes wrapping holomorphic cycles on $X_4$ of complementary dimension have a natural topological intersection number, that can be thought of as the mirror dual to \eqref{intersection}. Then, on a D-brane wrapping a $2p$-cycle with $p \geq 2$, a non-trivial curvature may induce lower-dimensional D-brane charges. This affects the index that counts the open strings stretching between the two D-branes, and which in the absence of induced charges amounts to the intersection number between cycles. The curvature-corrected open string index between two B-branes $\mathcal{E}$ and $\mathcal{F}$ reads
\begin{align}\label{intersectionmatrix}
    \chi(\mathcal{E}, \mathcal{F}) = \int_{X_4} \text{Td}(X_4)\left(\text{ch}\, \mathcal{E}\right)^\vee \left(\text{ch}\, \mathcal{F}\right)\,,
\end{align}
where $\text{ch}\, \mathcal{E}$ is the Chern character of $\mathcal{E}$, and the Todd class for a Calabi--Yau four-fold is
\begin{align}\label{eq: Todd}
    \text{Td}(X_4) = 1 + \frac{c_2}{12} + \frac{3c_2^2 - c_4}{720} \,. 
\end{align}
Finally, for an element $\beta \in H^{2k}(Y,\mathbb{Z})$ we define $\beta^{\vee} = (-1)^k \beta$. It is the topological index \eqref{intersectionmatrix} that is well-behaved under the mirror map, and gives the actual intersection numbers of the four-forms that appear in \eqref{Omega}, instead of \eqref{intersection}. Nevertheless, it turns out that, upon applying the proper redefinitions, one can still use the intersection matrix \eqref{intersection}.

Indeed, the open string index for holomorphic $2p$-cycles on $X_4$ is computed in appendix \ref{sap:corrper}, with the result
\begin{align}\label{eq:chicorr}
    \chi = \Lambda^T \chi_0 \Lambda \,,
\end{align}
where $\chi_0$ is defined as in \eqref{eq:chi0} and
\begin{align}\label{eq:Lambda}
    \Lambda = \left(\begin{matrix} 1&0&0&0&0 \\ 0&\delta_i^j & 0&0&0 \\ \frac{1}{24} c_2^\mu&-\frac{1}{2}\zeta^\mu_{ii}& \delta^\mu_\nu & 0&0 \\ 0 & \frac{1}{6}\cK_{jiii} + K^{(2)}_{ji} & -\frac{1}{2}\left(\cK_{jkkl} +\cK_{jkll}\right)\mathcal{M}_\nu^{kl}+\mathcal{M}_{\nu,j} &\delta_j^i&0 \\ 
   K^{(0)} & -\frac{1}{24} \cK_{iiii} - \half K^{(2)}_{ii} & \lambda_{kl}\mathcal{M}_\nu^{kl} +\mathcal{M}^0_\nu&0&1 \end{matrix}\right)\,,
\end{align}
contains the corrections induced by the curvature. Here we have defined $c_2(X_4) = c_2^\mu \sigma_\mu$ and $ \lambda_{kl}= \frac{1}{12}\left(2\cK_{kkkl} + 3\cK_{kkll} +2 \cK_{klll}\right) + K^{(2)}_{kl}$. Notice that $\Lambda$ is independent of $K_i^{(3)}$.

With these expressions at hand, it is easy to see that the superpotential \eqref{supoprodcor1} can be rewritten in terms of the homological periods $\vec{\pi}^{\rm corr}=\chi^{-1}\vec{\Pi}^{\rm corr}$ as
\be
W^{\rm corr} = \left(\Lambda  \Sigma \vec{q}\right)^{\, t}\cdot  \chi_0 \cdot \left(\Lambda \vec{\pi}^{\rm corr}\right) \, ,
\label{suporot}
\ee
where 
\be
\Lambda \vec{\pi}^{\rm corr} = \left(\begin{matrix}1&0&0&0&0 \\ 0&\d^j_i&0&0&0 \\0&0&\d_\mu^\nu&0&0 \\ -iK_i^{(3)}&0&0&\d_j^i&0 \\0&-iK_i^{(3)} &0&0&1 \end{matrix}\right) \left(\begin{matrix} \pi_0 \\ \pi^i_2 \\ \pi^{\mu}_4 \\ \pi_{6i} \\ \pi_8 \end{matrix} \right) \, ,
\ee
amd $\vec{q}$ and the sign matrix $\mathcal{S}$ are the same objects introduced below \eqref{G4}. The components of $\vec{\pi}^{\rm corr}$ can be interpreted as the corrected moduli-dependent coefficients of $\Omega$ in the expansion \eqref{Omega}. Here we will not need the precise expression of such components, because the quantities of interest only depend on $\Lambda \vec{\pi}^{\rm corr}$. The expression \eqref{suporot} implies that, when taking into account the polynomial corrections in our F-theory setup, one can still use the classical intersection numbers \eqref{intersection} if one makes the replacements
\be
\label{eq: correction map}
\vec{q} \to \Sigma \Lambda \Sigma \vec{q} \, , \qquad \vec{\pi} \to \Lambda \vec{\pi}^{\rm corr}\, ,
\ee
in all the computations of the previous subsection. That is, in \eqref{Omega} we perform the replacements
\be
\pi_{6i} \to \pi_{6i} -i K_i^{(3)} \, , \qquad \pi_8 \to \pi_8 -i  K_i^{(3)} T^i\, ,
\label{picorr2}
\ee
and in \eqref{G4} we replace the flux quanta by
\begin{subequations}\label{eq:fluxshift}
\begin{align}
    \bar{m}^\mu &=  \hat{m}^{\mu} + m^i\frac{\zeta^\mu_{ii}}{2}  +m\frac{c_2^\mu}{24}  \,,\\
    \bar{e}_j &= e_j+\hat{m}^\mu\left[\frac{1}{2} \left(\cK_{jkkl}+\cK_{jkll}\right) \mathcal{M}_\mu^{kl}-\mathcal{M}_{\mu,j}\right] +m^i \left(\frac{1}{6} \cK_{jiii} + K_{ij}^{(2)}\right) \,,\\
    \bar{e} &= e + \hat{m}^\mu \left(\mathcal{M}_\mu^{ij}\lambda_{ij}+\mathcal{M}^0_\mu\right) +m^i\left(\frac{1}{24} \cK_{iiii} + \oh K_{ii}^{(2)}\right) + m K^{(0)}\,. 
\end{align}
\end{subequations}

Alternatively, one can make use of the flux quanta $m^{ij}$ and the relations \eqref{eq: Pimu transformation} and \eqref{supoprodcor3} to find
\begin{subequations}\label{eq:fluxshift2}
\begin{align}
       \bar{m}^{ij} &=  \hat{m}^{\mu} + \frac{m^k}
       {2}\delta^{i}_k\delta^{j}_k  +m\frac{c_2^{ij}}{24}  \,,\\
    \bar{e}_j &= e_j+\frac{m^{kl}}{4} \left(\cK_{jkkl}+\cK_{jkll}+2\hat{\mathcal{M}}_{kl,j}\right)+m^i \left(\frac{1}{6} \cK_{jiii} + K_{ij}^{(2)}\right) \,,\\
    \bar{e} &= e +\frac{m^{ij}}{2} \left(\lambda_{ij}-\hat{\mathcal{M}}^0_{ij}\right) +m^i\left(\frac{1}{24} \cK_{iiii} + \oh K_{ii}^{(2)}\right) + m K^{(0)}\,.  
\end{align}
\end{subequations}
The matrices $\mathcal{M}$ and $\hat{\mathcal{M}}$ are useful to make the invariance under a choice of basis of B-branes explicit. In practice, the most convenient choice of integral basis will generically be given by a subset of the intersection of divisors $\{[\sigma_\mu]\}=\{[D_i.D_j]\}|_{(i,j)\in J}$. In those cases, the  entries of the matrix $\mathcal{M}$ simplify notably, since we can take $\mathcal{M}_{\mu}^{ij}=1$ for $(i,j)\in J$ and $\mathcal{M}_{\mu}^{ij}=0$ otherwise.  Then trivially, $\mathcal{M}_\mu^0=\mathcal{M}_{\mu,i}=0$. In \eqref{eq:fluxshift2} we would still need the terms in $\hat{\mathcal{M}}_{kl}$ to account for the freedom to define the flux quanta $m^{ij}$. We can similarly remove them if we choose the convention $m^{ij}=0$ for $(i,j)\notin J$.

To sum up, idependently of the choice of basis of B-branes, the corrected expression for the GVW superpotential takes the form
\be
W^{\rm corr}  =  \bar{e} +  \bar{e}_iT^i + \frac{1}{2}\,  \bar{m}^{\mu} \zeta_{\mu,kl}  T^k T^l  + \frac{1}{6}\, {\cal K}_{ijkl}\, m^i T^j T^k T^l +  \frac{m}{24}\, {\cal K}_{ijkl}\, T^i T^j T^k T^l -i  K_i^{(3)} \left(m^i + m T^i\right)  \, . 
\label{supofinal}
\ee

This strategy to rewrite the superpotential not only gives a more manageable expression, but also yields the corrected K\"ahler potential as a byproduct. Indeed, it follows that the corrections to \eqref{Kcs} can be computed from the expression \eqref{Kcscomp}, by performing the replacements \eqref{picorr2}. One then finds that
\be
K_{\rm cs}^{\rm corr}=-\log\left(\frac{2}{3}\mathcal{K}_{ijkl}t^it^jt^kt^l + 4 K_i^{(3)} t^i \right)\, .
\label{Kcscorr}
\ee
Notice that this expression still respects the continuous shift symmetry of the axionic fields $b^i$ and it only depends on the type IIA $\alpha'^3$-corrections that correspond to the third Chern class of $X_4$. It is also a natural generalisation of the $\alpha'^{3}$-correction to the K\"ahler potential in type IIA compactifications in Calabi--Yau three-folds, see e.g. \cite{Palti:2008mg,Escobar:2018rna}. In appendix \ref{sap:corrka} we rederive the same expression using a different method, as a cross-check of our results.

From these expressions one can derive the corrections to the F-term scalar potential \eqref{scalarpot}. For this, it is useful to write the superpotential and its derivatives in terms of shifted axion polynomials. We have that 
\begin{subequations}
    \label{eq:supo and der corr}
\begin{align}
    W^{\rm corr}&=\, \bar{\rho}+i\bar{\rho}_it^i-\frac{1}{2}\zeta_\mu \bar{\rho}^{\mu}-i \left(\frac{1}{6}\mathcal{K}_i + K_i^{(3)}\right)\tilde{\rho}^i + \left(\frac{\mathcal{K}}{24} +  K_i^{(3)}t^i \right)\tilde{\rho}  \, ,\\
    \partial_{i}W^{\rm corr}&=\bar{\rho}_i+i\zeta_{\mu i}  \bar{\rho}^\mu-\frac{1}{2}\mathcal{K}_{ij}\tilde{\rho}^j -i\left(\frac{\mathcal{K}_i}{6}+K_i^{(3)}\right)\tilde{\rho}\, \, ,
\end{align}
\end{subequations}
where
\begin{subequations}
\label{corrhos}
\begin{align}
     \bar{\rho}=&\ \bar{e}+\bar{e}_ib^i+\frac{1}{2}\bar{m}^\mu\zeta_{\mu,kl} b^k b^l +\frac{1}{6}\mathcal{K}_{ijkl}m^ib^jb^kb^l+\frac{1}{24}m \mathcal{K}_{ijkl}b^ib^jb^kb^l\, , \\
    \bar{\rho}_i=&\ \bar{e}_i+\bar{m}^\mu\zeta_{\mu,il} b^l+\frac{1}{2}\mathcal{K}_{ijkl}m^jb^kb^l+\frac{1}{6}m\mathcal{K}_{ijkl}b^jb^kb^l\\
    \bar{\rho}^\mu=&\ \bar{m}^\mu+\zeta_{ij}^\mu b^im^j+\frac{1}{2}m\zeta_{ij}^\mu b^ib^j\ , \\
        \tilde{\rho}^i=&\ m^i+mb^i\, ,\\
    \tilde{\rho}=&\ m\, .
    \end{align}
    \end{subequations}
Notice that if we take $K^{(3)}_i \to 0$ the corrected scalar potential reduces to \eqref{scalarpot}, except for the flux redefinition \eqref{eq:fluxshift} that only replaces the components of \eqref{vecrho} by \eqref{corrhos}. As we show in appendix \ref{sap:corrFterm}, the effect of a non-vanishing $K^{(3)}_i$ is to modify the matrix \eqref{ZAB}, inducing  new non-vanishing entries that destroy its block-diagonal structure. Due to its complicated form, it is easier to characterise the corrections to the vacua equations in terms of the vanishing conditions for the corrected F-terms, as we do in appendix \ref{sap:corrvac}.

\subsubsection*{Monodromies}

The above expressions allow us to connect the definition of $\vec{\rho}$ with the monodromies that act on the periods of $\Omega$. We start by defining the axion-dependent rotation matrix $R(b)$ and its associated generators $P_i$ as follows
\begin{align}
\label{eq: axion matrix}
    R(b) &= \,  \begin{pmatrix}
    1&0&0&0&0 \\ b^i&\delta_k^i & 0&0&0 \\  \frac{1}{2}\zeta^\mu_{ij} b^ib^j & \zeta^\mu_{kj}b^j & \delta^\mu_\nu  & 0&0 \\ \frac{1}{6}\mathcal{K}_{ijkl}b^jb^kb^l & \frac{1}{2}\mathcal{K}_{ijkl}b^jb^l & \zeta_{\nu,ij} b^j &\delta_i^k&0 \\ 
   \frac{1}{24} \mathcal{K}_{ijkl}b^ib^jb^kb^l & \frac{1}{6}\mathcal{K}_{ijkl}b^ib^jb^l & \frac{1}{2}\zeta_{\nu,ij}b^ib^j &b^k&1
    \end{pmatrix}\, = e^{b^i P_i} , \\
  P_n  & = \,   
   \begin{pmatrix}
    0&0&0&0&0 \\ \delta^i_n &0 & 0&0&0 \\   0 & \zeta^\mu_{in}  & 0 & 0&0 \\ 0 & 0 & \zeta_{\mu,in}  & 0 &0 \\ 
  0 &0 & 0 & \delta^i_n &0
    \end{pmatrix} .
\end{align}
Here $R(b)$ transforms the flux vector $\vec{q}^{\, t} = (m, m^i,  m^{\mu}, e_i, e_0)$ into the vector of flux-axion polynomials as  $\vec{\rho}=R\vec{q} $, where $\vec{\rho}^{\,  t} = \left(\tilde{\rho}, \tilde{\rho}^i,   {\rho}^{\mu}, \rho_i,   \rho   \right)$ as defined in \eqref{vecrho}. The matrices $P_i$ are the generators of such a rotation. One can account for the polynomial corrections using the matrix $\Lambda$ \eqref{eq:Lambda} and the map \eqref{eq: correction map}. The corrected flux-axion polynomials of \eqref{corrhos} can then be similarly written as $\vec{\rho}^{\, \rm corr}(b)=R(b) \Sigma\Lambda \Sigma \vec{q}$. To proceed, it is useful to consider the superpotential in terms of the vector of  fluxes  and the central charges of B-branes as in \eqref{supoprodcor1} and rerwite it in terms of the newly introduced objects
\begin{equation}
    W=\vec{q}^{\,t}\Sigma\vec{\Pi}^{\rm corr}=(\vec{\rho}^{\, \rm corr})^t R^{t-1}(b) \Sigma \Lambda^{t-1} \vec{\Pi}^{\rm corr}\,,
    \label{eq: superpot axion-saxion split}
\end{equation}
where in the last step we have used that $\Sigma^2=I$. One can check that $R^{t-1}(b) \Sigma \Lambda^{t-1} \vec{\Pi}^{\rm corr}$ does not depend on the axions $b^i$, and so that \eqref{eq: superpot axion-saxion split} expresses the superpotential as a product of an axion-dependent and a saxion-dependent vector. Defining the monodromy matrix in terms of its action on the periods
\begin{align}
    \vec{\Pi}^{\rm corr}(T^j+1) = {\cal T}_j \cdot \vec{\Pi}^{\rm corr}(T^j)\,,
    \label{eq: monodromy}
\end{align}
and working with the saxion-dependent vector in \eqref{eq: superpot axion-saxion split} we find
\begin{eqn}
    \mathcal{T}_j=e^{-(P^{\rm corr}_j)^t}\,,
    \label{eq: monodromy P}
\end{eqn}
where $P^{\rm corr}_j \equiv \Lambda^{-1} P_j \Lambda$ and we have used that $\Sigma P_j \Sigma=-P_j$.

Alternatively, one can derive more explicit expressions by considering \eqref{eq:corrper} and substituting directly in the definition of the monodromy action \eqref{eq: monodromy} 

\begin{align}
\label{eq: monodromy central charges}
   \hat{\cal{T}}_j = \left(\begin{matrix} 1 & -\delta_{n}^j &0&0&0\\ 0 & \delta_{l}^n &-\delta^{i}_l\delta^{k}_j&0&0\\ 0 & 0  &\delta_{m}^i\delta_{n}^k &-\mathcal{K}_{ijmn}&\frac{1}{2}(\mathcal{K}_{mmnj}+\mathcal{K}_{mnnj}+\mathcal{K}_{mnjj})\\ 0&0& 0 & \delta_{i}^l &0\\ 0 &0& 0 &0 &1 \end{matrix}\right)\,.
\end{align}

Note this is the monodromy matrix represented in the (degenerated) system spanned by the intersection of divisors $[D_i.D_j]$. To relate to the original monodromy matrix in the basis containing $[\sigma_\mu]$, i.e.  $\Pi^{\rm corr}=(\Pi_8^{\rm corr}, \Pi_{6,i}^{\rm corr}, \Pi_{
4\,\mu}^{\rm corr}, \Pi_{2}^{i\, {\rm corr}}, \Pi_0^{\rm corr})$, we can use the matrices defined in \eqref{eq: Pimu transformation} and \eqref{eq: Piij transformation} to obtain
\begin{eqn}
    \mathcal{T}_j= \mathcal{M}\hat{\mathcal{T}}_j \hat{\mathcal{M}}\,,
\end{eqn}
where we have extended the matrices $\mathcal{M}$ and $\hat{\mathcal{M}}$ trivially to the full system of B-branes in both representations so that the transformation leaves the remaining central charges outside the four-cycle sector invariant as in \eqref{eq: M explicit}.


\section{Tadpoles and vacua}
\label{s:vacua}

With an explicit form for the F-term scalar potential in the large complex structure regime one may characterise the set of vacua in that region. We will pay particular attention to the fact that the flux contribution to the tadpole $N_{\rm flux}$ is bounded from above, something that forbids the presence of certain flux vacua at arbitrarily large complex structure. As we will see, this tadpole constraint leads to different moduli stabilisation scenarios, classified by which flux components are turned on. In this section we will analyse the most generic of these scenarios, in which one can clearly see that the corrections $K_i^{(3)}$ to the K\"ahler potential are crucial to stabilise all moduli. As a direct consequence, one finds an upper bound for the vev of the complex structure saxions, that depends both on $K_i^{(3)}$ and $N_{\rm flux}$. One can also consider a quite different setup in which such a bound is absent, whose general discussion we leave for section \ref{s:linear}.

\subsection{General flux vacua}

Armed with the explicit form of the potential at large complex structure, one may now analyse its set of vacua. Let us first consider the leading flux potential \eqref{scalarpot}. Since it is a sum of three positive semi-definite terms and its dependence on the K\"ahler moduli only enters through the overall factor $e^{K} \cK \propto {\cal V}_3^{-2}$, its minima correspond to Minkowski vacua where these three terms vanish. In other words, we must impose the following set of on-shell conditions
\begin{subequations}
\label{eq:Mink}
\begin{empheq}[box=\widefbox]{align}
    \rho&=\frac{1}{24}\mathcal{K}\tilde{\rho} \label{eq:Mink 0}\\
    \rho_i&=-\frac{1}{6}\mathcal{K}g_{ij}\tilde{\rho}^j \label{eq:Mink i} \\
    0&= \left(  \cK \zeta_{\mu i}- \mathcal{K}_i\zeta_{\mu}\right) \hat{\rho}^\mu 
    \label{eq:Mink mu}
\end{empheq}
\end{subequations}
where the general solution for \eqref{eq:Mink mu} reads
\be
\hat{\rho}^\mu = A \zeta^\mu + C^\mu\, , \qquad \zeta_{\mu i} C^\mu =0\quad \forall i\, , 
\label{splitmu}
\ee
with $A$, $C^\mu$ moduli-dependent quantities. For those vacua that preserve supersymmetry, we need to impose that $W=0$ on-shell. From \eqref{eq:suporho} we see that this implies two additional conditions:
\be
t^i\rho_i = 0\, , \qquad  \zeta_\mu\hat{\rho}^\mu = \frac{\CK}{6}\tilde{\rho} \, .
\label{eq:SUSY}
\ee

From our discussion in the previous section it follows that, in order to implement the polynomial corrections that correspond to $K^{(0)}$ and $K_{ij}^{(2)}$, we only need to perform the replacement 
\be
(\rho, \rho_i, \hat{\rho}^\mu) \to (\bar\rho, \bar\rho_i, \bar{\rho}^\mu)
\label{reprho}
\ee 
in \eqref{eq:Mink} and \eqref{eq:SUSY}, with the new quantities given by \eqref{corrhos}. Therefore the above equations essentially hold whenever it is a good approximation to neglect the correction due to  $K_i^{(3)}$ in the K\"ahler potential \eqref{Kcscorr}. The vacua equations that follow from including  such a  correction to the K\"ahler potential are discussed in appendix \ref{sap:corrvac}. In here we simply collect the result, approximated to linear order in $\eps_i = 6K^{(3)}_i/\CK$:
\begin{subequations}
    \label{eq:Minkcorr}
\begin{align}
 \bar\rho- \frac{1}{24}\mathcal{K}\tilde{\rho}&  = - \frac{1}{48}\eps_it^i \left[ \cK\tilde{\rho} + 18 \zeta_\mu \bar\rho^{\mu} \right] \, , \\
 \label{eq:Minkcorrrhoi}
  \bar{\rho}_i+\frac{1}{6}\mathcal{K}g_{ij}\tilde{\rho}^j & =  \frac{1}{3}  \CK_i  \left( \eps_j -  \epsilon_k t^k \frac{ \cK_j}{\cK} \right)\tilde{\rho}^j   - \frac{1}{6}\epsilon_i \cK_j \tilde \rho^j   \, , \\
  \left(\zeta_{\mu i}- \frac{\mathcal{K}_i}{\CK} \zeta_{\mu}\right) \bar{\rho}^\mu& = \frac{1}{8} \left(\eps_i - \epsilon_k t^k \frac{\cK_i }{\cK} \right)   \left( \cK \tilde \rho +  2 \zeta_\mu \bar\rho^{\mu}\right)  \, .
 \end{align}
\end{subequations}
Finally, those vacua that are supersymmetric will satisfy the additional conditions
\bea
t^i \bar{\rho}_i = \frac{1}{4} \left( \cK \eps_i\tilde{\rho}^i  -  \epsilon_k t^k \cK_j\tilde{\rho}^j \right) \, ,
\qquad 
\zeta_\mu \bar{\rho}^{\mu} =\frac{\cK}{6}\left(1 + \eps_it^i \right) \tilde{\rho} \, ,
\label{eq:SUSYcorr}
\eea
up to quadratic terms in $\eps_i$. 

\subsection{The tadpole constraint}
\label{sec:tadpole}

In any consistent F-theory compactification on a four-fold $Y_4$ one must satisfy the D3-brane tadpole condition
\be
N_{\rm flux} = \oh \int_{Y_4} G_4 \wedge G_4 = \frac{\chi(Y_4)}{24} - N_{\rm D3}\, ,
\label{tadpole}
\ee
where $\chi(Y_4)$ is the Euler characteristic of $Y_4$, and $N_{\rm D3}$ is the number of space-time filling D3-branes. The number $\chi(Y_4)$ can take a range of values depending on the four-fold, but since stability of Minkowski vacua requires $N_{\rm D3}>0$, \eqref{tadpole} sets an upper bound for $N_{\rm flux}$. Notice that we also need to impose $N_{\rm flux} > 0$ in order to find a vacuum, due to the on-shell constraint \eqref{SDG4}. We therefore have the allowed range $0\leq N_{\rm flux} \leq \chi(Y_4)/24$ for any Minkowski flux vacuum. To understand what this implies in our setup, one may easily compute the value of $N_{\rm flux}$ in terms of the expressions of section \ref{s:potential}. Starting from \eqref{G4} one finds 
\be
N_{\rm flux} \equiv \bar{e} m - \bar{e}_i m^i + \frac{1}{2} \eta_{\mu\nu} \bar{m}^{\mu} \bar{m}^{\nu}\, ,
\label{Nflux}
\ee
where the barred flux quanta are defined in \eqref{eq:fluxshift} and their presence arises from the corrections to the naive intersection numbers \eqref{intersection}. 

The interesting observation is that this expression for $N_{\rm flux}$ equals a bilinear of flux-axion polynomials, namely
\be
N_{\rm flux} = \bar{\rho} \tilde{\rho} - \bar{\rho}_i \tilde{\rho}^i + \frac{1}{2} \eta_{\mu\nu} \bar{\rho}^{\mu} \bar{\rho}^{\nu}\, .
\label{Nfluxrho}
\ee
One can check this identity directly, or by realising that the flux contribution to the tadpole \eqref{Nflux} is one of the flux monodromy-invariants that constrain the orbit of values that $\vec{\rho}$ can take. In fact, since the entries of $\vec{\rho}$ are invariant under monodromies as well, their on-shell value can only depend on such flux invariants and, because of \eqref{eq:Mink}, the same holds for the saxion vevs. The invariants that arise in generic F-theory flux  compactifications are listed in appendix \ref{ap:invariants}.

This last expression for $N_{\rm flux}$ can be evaluated at each vacuum via the on-shell conditions derived above. For simplicity, let us assume that we are in a sufficiently large complex structure regime such that the K\"ahler potential correction term $K^{(3)}_it^i$ in \eqref{Kcscorr} can be neglected. Then one may use \eqref{eq:Mink} with the replacement \eqref{reprho} to obtain
\be
N_{\rm flux} \stackrel{\rm vac}{=} \frac{\CK}{24} \left(\tilde{\rho}^2 + 4 g_{ij} \tilde{\rho}^i\tilde{\rho}^j \right)  +  \frac{1}{2} g_{\mu\nu} \bar{\rho}^{\mu} \bar{\rho}^{\nu}\, ,
\label{Nfluxvac}
\ee
where $g_{\mu\nu}$ is defined as in \eqref{ZABdiag}, and we have used that for a vector of the form \eqref{splitmu} we have that  $\eta_{\mu\nu} \hat{\rho}^\nu =  g_{\mu\nu} \hat{\rho}^\nu$, see appendix \ref{ap:georho} for details.

Along any limit of large complex structure we have that $\CK \to \infty$, because otherwise $\CK_i \to 0$ for at least some $i$, which takes us away from the regime of validity of our analysis. Then the question is if along these limits all terms on the rhs of \eqref{Nfluxvac} remain bounded from above. If they did not, no vacua would be found at sufficiently large complex structure, for any value of $\chi(Y_4)$. Since all terms are positive definite, they need to be bounded separately. 

The first term on the rhs of \eqref{Nfluxvac} is clearly unbounded, so we must impose $\tilde{\rho} = m = 0$, which then implies $\tilde{\rho}^i = m^i$. For the second term, the question is whether $\CK g_{ij} m^i m^j = (4 \CK_i\CK_j/\CK -3 \CK_{ij})m^im^j$ remains bounded or not along the different large complex structure limits. Those choices of $m^i$ where it is not bounded should be set to zero in order to find a consistent vacuum. This depends crucially on the topology of $Y_4$ through the quadruple intersection numbers $\CK_{ijkl}$ of its mirror $X_4$. A full classification of all possibilities should follow from the techniques developed in \cite{Grimm:2019ixq} applied to the special case of large complex structure limits. Here, we take a simplified approach by asking whether $\CK g_{jj}$ remains bounded or not in the case that we blow up a single modulus $t^i\rightarrow \infty$. If it does not, one should set $m^j=0$ to find vacua in that regime.

We can distinguish four different cases: 
\begin{itemize}
    \item[$(i)$] The modulus $t^i$ appears with a quartic term in the K\"ahler potential, i.e. $\cK_{iiii}\ne 0$. In this case the component $\CK g_{ii}$ is not bounded since 
    \begin{subequations}\label{Kgii1}
    \begin{align}
        {\cK} g_{ii}  \sim (t^i)^2 \rightarrow \infty\,.
    \end{align}
    In addition, for those indices $j \neq i$ such that $\cK_{iiij}\ne 0$, the diagonal term $\cK g_{jj}$ scales as
    \begin{align}
         \cK g_{jj} = \frac{4\cK_j\cK_j}{\cK}-3\cK_{jj}  \sim (t^i)^2 \rightarrow \infty\,, 
    \end{align}
    \end{subequations}
     and it is therefore also unbounded.  
    \item[$(ii)$] The modulus $t^i$ appears only cubic in the K\"ahler potential, i.e. $\cK_{iiii}=0$ but $\cK_{iiik}\ne 0$ for some $k\ne i$. In this case the component $\cK g_{ii}$ is unbounded as 
    \begin{subequations}\label{Kgii2}
    \begin{align}
        {\cK} g_{ii}  \sim \cK_{iiik} t^i t^k \rightarrow \infty\,,
    \end{align}
    with no summation involved. 
    If in addition $\cK_{iijk}\ne 0$ for some $k$, also the component $\cK g_{jj}$ is unbounded, as it scales at least as 
    \begin{align}
        \cK g_{jj} \sim t^i \rightarrow \infty \,. 
    \end{align}
    \end{subequations}
    \item[$(iii)$] The K\"ahler potential depends quadratically on the modulus $t^i$ which corresponds to $\cK_{iiij}=0, \forall j$ but $\cK_{iikl}\ne 0$ for some $k,l\ne i$. In this case the metric component $\cK g_{ii}$ does not scale:
    \begin{subequations}\label{Kgii3}
    \begin{align}
        \cK g_{ii} \sim \cK_{iikl} t^k t^l\sim \text{const.}
    \end{align}
    But the components $\cK g_{jj}$ are still unbounded, since generically they scale as 
    \begin{align}
        \cK g_{jj}\sim (t^i)^2 \rightarrow \infty\,,
    \end{align}
    \end{subequations}
    as long as $\cK_{iijk}\ne 0$ for some $k$. 
    \item[$(iv)$] Finally, if the K\"ahler potential is only linear in $t^i$, i.e. $\cK_{iikl}=0, \forall k,l$, but $\cK_{ijkl}\ne0$ for $j,k,l\ne i$ the diagonal component $\cK g_{ii}$ vanishes asymptotically as
    \begin{subequations}\label{Kgii4}
    \begin{align}
        \cK g_{ii}  \sim \frac{\cK_{ijkl} t^j t^k t^l}{t^i} \rightarrow 0 \,.
    \end{align}
    The other components $\cK g_{jj}$ are nevertheless unbounded as, generically 
    \begin{align}
         \cK g_{jj}\sim t^i \rightarrow \infty\,.
    \end{align}
    \end{subequations}
\end{itemize}

Given this behaviour of the tensor $\cK g_{ij}$, one would expect to find very few vacua in which $m^i \neq 0$ for some $i$ in regions where $t^i \gtrsim \oh \sqrt{\chi(Y_4)}, \forall i$. Exceptions to this rule may for instance happen if the index $i$ appears only linearly in the quadruple intersection numbers $\cK_{ijkl}$, and if we consider the regime $t^i \gg t^j, \forall j \neq i$. In that case one may satisfy the tadpole constraint for $m^i$ arbitrary and $m^j =0, \forall j \neq i$. A clear setup where this happens is when we consider a factorised geometry like $Y_ 4 = Y_3 \times \mathbb{T}^2$, that can be interpreted as a type IIB flux compactification, and identify $T^i$ with the complex structure of $\mathbb{T}^2$. The type IIB setup will be analysed in section  \ref{s:IIB}, while the more general linear setup will be discussed in section \ref{s:linear}. In the next subsection we will consider the more generic case in which we need to set $m = m^i =0, \forall i$ in order to find vacua in the region $t^i \gtrsim \oh \sqrt{\chi(Y_4)}$, keeping in mind that in some special cases this constraint could be stronger than necessary. For smaller saxion values these restricted flux quanta will also give rise of vacua, but there they will coexist with vacua with other flux patterns, see e.g. \cite{Denef:2005mm,Honma:2017uzn}.

\subsection{Moduli stabilisation}
\label{sec:moduli}

Motivated by the above discussion, let us restrict our attention to flux vacua at large complex structure such that 
\begin{align}
   \vec{q}^{\, t} = (m, m^i, \bar{m}^\mu, \bar e_i, \bar e) = (0,0, \hat m^\mu, \bar e_i, \bar e) \, ,
    \label{truncflux}
\end{align}
which implies that $\tilde{\rho} = \tilde{\rho}^i = 0$ and that $\bar{\rho}^\mu = \hat{m}^\mu$. In this case the flux contribution to the D3-brane tadpole reads
\be
N_{\rm flux} =  \frac{1}{2} \eta_{\mu\nu} \hat{m}^{\mu} \hat{m}^{\nu}\, .
\label{Nfluxgen}
\ee

Plugged into \eqref{eq:Mink}, the restricted fluxes \eqref{truncflux} imply
\begin{align}
\label{eq: vacuum equations m=ma=0}
\begin{cases}
    \bar{\rho}=0\\
    \bar{\rho}_i=0\\
   \cK \zeta_{\mu i} \hat{m}^\mu  = \mathcal{K}_i\zeta_{\mu}\hat{m}^\mu
\end{cases}
\end{align}
where we recall that the last equation is equivalent to the decomposition \eqref{splitmu} for $\hat{m}^\mu$. This system has the simplifying property that the  equations for axions and saxions decouple. From the first two equations we obtain
\begin{subequations}
\label{barrhoeqs}
\begin{align}
    \bar\rho&=0 \implies \bar{e}+\bar{e}_ib^i+\frac{1}{2}\hat{m}^\mu\zeta_{\mu,kl} b^k b^l=0 \stackrel{\eqref{barrhoieq}}{\implies} \bar{e}=-\frac{1}{2}\bar{e}_ib^i\, ,
\label{barrhoeq}\\
   \bar\rho_i&=0 \implies \hat{m}^\mu\zeta_{\mu,ij} b^j =-\bar{e}_i\, .
   \label{barrhoieq}
    \end{align}
 \end{subequations}

To analyse the implication of these two equations let us define the matrix $M_{ij} \equiv \hat{m}^\mu\zeta_{\mu,ij}$, and let $r$ be its rank. From \eqref{barrhoieq} we obtain a system of $r$ equations with $h^{3,1}(Y_4)$ unknowns. This system will only have a solution if the vector $\bar{e}_i$ lies in the image of $M$, which will impose $h^{3,1}(Y_4) - r$ constraints on these fluxes. Only when these constraints are met we will be able to find a vacuum, and in this case only $r$ axions will be stabilised. In particular, notice that then only $r$ complex structure fields appear in the superpotential \eqref{supofinal}. This suggests that several saxionic directions will not be stabilised either, as one can see from the third equation in \eqref{eq: vacuum equations m=ma=0}. Indeed, in general we have that $\zeta_\mu \neq 0$, as this corresponds to the volume of a holomorphic four-cycle in the mirror four-fold $X_4$, but also that it only depends on $r$ saxionic directions, and so the remaining ones are unfixed by the vacuum equations. Moreover this third equation is such that contracted with $t^i$ becomes trivial and so, in fact, it only stabilises $r-1$ saxions. Therefore at least one saxionic direction is left unconstrained, even in the case of maximal rank. 

Coming back to \eqref{barrhoeqs}, we see that only those axions $b^i$ that are fixed by \eqref{barrhoieq} will appear in \eqref{barrhoeq}, which translates into an additional constraint that must be imposed on the fluxes in order to achieve a vacuum.  This time, however, the constraint is removed when corrections to the K\"ahler potential are taken into account, similarly to the effect observed  in \cite{Palti:2008mg,Escobar:2018tiu,Escobar:2018rna} in the context of Minkowski type II flux compactifications on three-folds. Indeed, including the corrections to the K\"ahler potential couples the equations for axions and saxions, which in turn changes the counting of stabilised moduli. This can already be seen from the vacua equations corrected at linear order in the parameter $\eps_i = 6K^{(3)}_i/\CK$, see \eqref{eq:Minkcorr}, which adapted to the present case read
\begin{subequations}
    \label{eq:Minkcorrmodu}
\begin{align}
\label{corr1}
 \bar\rho&  = - \frac{3}{8}\eps_it^i  \zeta_\mu \hat m^{\mu}  \, , \\
 \label{corr2}
  \bar{\rho}_i & = 0   \, , \\
  \label{corr3}
   \left(\CK\zeta_{\mu i}-\mathcal{K}_i \zeta_{\mu}\right) \hat{m}^\mu& = \frac{1}{4} \left(\CK\eps_i - \epsilon_k t^k \cK_i  \right)   \zeta_\mu \hat m^{\mu}  \, .
 \end{align}
\end{subequations}

Notice that \eqref{corr2} is the same as before, and therefore gives $r$ equations on the axions. Similarly, \eqref{corr3} becomes trivial when contracted with $t^i$ and so, even if modified, still yields $r-1$ equations for the saxions. The main difference comes from \eqref{corr1}, which couples axions and saxions and using \eqref{corr2} becomes
\be\label{corr1+corr2}
 \bar{e}+\frac{1}{2}\bar{e}_ib^i = - \frac{3}{8}\eps_it^i  \zeta_\mu \hat m^{\mu} \, .
\ee
On the one hand, this equation no longer sets a constraint for the flux $\bar{e}$. On the other hand, plugging in the value for $b^i$ obtained from \eqref{barrhoieq} one obtains an additional equation for the saxions which, together with \eqref{corr3}, fixes the vev for $r$ of them. Using the results of appendix \ref{sap:corrvac}, one can check that this structure is in fact preserved at all orders in the correction parameter $\eps_i$, and so the counting holds at the level of polynomial terms in the scalar potential.  

To sum up, we obtain a system with only $r = \rank (M)$ complex structure fields fixed by the above vacua equations. Fixing the remaining ones would necessarily imply taking into account the exponentially-suppressed corrections that we are neglecting in our analysis. It is beyond the scope of our work to determine whether full moduli stabilisation would then be achieved or not, although in any event such fields would be extremely light in this regime. 

In general we will consider those cases in which the rank of  $M_{ij} \equiv \hat{m}^\mu\zeta_{\mu,ij}$ equals $h^{3,1}(Y_4)$, which a priori can be achieved by choosing an appropriate flux $\hat{m}^\mu$. Since in this scheme $N_{\rm flux} = \oh \eta_{\mu\nu} \hat{m}^\mu \hat{m}^\nu$, one may wonder if such flux choices restrict the possible values of $N_{\rm flux}$. Let us for instance consider the case in which the choice of $\hat{m}^\mu$ is  such that $r = h^{3,1}(Y_4)$ implies
\begin{eqn}\label{originTC}
 \zeta_{\mu, ij}M^{ij}  = \frac{1}{2\gamma}  \eta_{\mu\nu} \hat{m}^\nu + \beta_\mu \, ,
\end{eqn}
where $M^{ik}M_{kj} = \delta^i_j$, $\gamma$ is a real  function of the fluxes with a lower bound $\alpha>0$ and $\hat{m}^\mu \beta_\mu \leq 0$. Then we have that $N_{\rm flux} \geq \alpha h^{3,1}(Y_4)$, which is the sort of behaviour proposed by the Tadpole Conjecture in \cite{Bena:2020xrh}. Whenever \eqref{originTC} holds, and depending on the precise value for $\alpha$, a large number of moduli could be in tension with satisfying the upper bound for $N_{\rm flux}$, as pointed out in \cite{Bena:2020xrh}. It would be thus interesting  to determine in which cases \eqref{originTC} occurs.

We can go a step further in our analysis and impose bounds on the saxion vevs by recalling the leading solution for $\hat{\rho}^\mu$, see \eqref{splitmu}. Since now $m^i=m=0$ we have
\begin{equation}
    \hat{m}^\mu=A\zeta^\mu+C^\mu +\cO(\epsilon_i)\, ,
\end{equation}
with $C^\mu\zeta_{\mu i}=0$. Therefore, the tadpole is given by
\begin{equation}
\label{eq: tadpole constraint anstaz}
    N_{\rm flux}=\frac{1}{2} g_{\mu\nu}\hat{m}^\mu \hat{m}^\nu=\frac{1}{2}A^2\cK +\frac{1}{2}C^\mu C^\nu g_{\mu\nu}+\cO(\epsilon_i)\geq \frac{1}{2} A^2\cK+\cO(\epsilon_i)\, .
\end{equation}
On the other hand, substituting in \eqref{corr1} we obtain
\begin{equation}
    A=-\frac{4\bar{\rho}}{9 K^{(3)}_i t^i}\, .
    \label{astiA}
\end{equation}
Looking now at the equation \eqref{corr2}
\begin{equation}
    \bar{e_i}=-\hat{m}^\mu \zeta_{\mu, il}b^l\equiv - M_{il} b^l\, ,
    \label{Meq}
\end{equation}
we can infer that $\bar{\rho}$ behaves like $\bar{\rho}\sim q/P(\hat{m}^\mu)$ for some  integer $q$ and some polynomial $P(\hat{m}^\mu)$ of degree $r = \rank M$ in the $\hat{m}^\mu$. For instance, when $M$ is invertible and so $r=h^{3,1}(X_4)$, the matrix $M_{il}$ has integer combinations of the $\hat m^\mu$ as coefficients, and thus its inverse
\begin{align}
    M^{-1} = \frac{1}{\text{det}M} \sum_{s=0}^{h^{3,1}-1}M^s \sum_{k_1,\dots k_{h^{3,1}-1}} \prod_{l=1}^{h^{3,1}} \frac{(-1)^{k_l+1}}{l^{k_l}k_l!}\left(\Tr{M^l}\right)^{k_l},\quad s+\sum_{l=1}^{h^{3,1}-1} l k_l=h^{3,1}-1\,,
\end{align}
depends inversely on det $M$, which is a degree $h^{3,1}$ polynomial on the fluxes $\hat{m}^\mu$. The remaining terms appearing in $M^{-1}$ are polynomials of the integers $\hat{m}^\mu$, up to combinatoric factors.  Because in this case
\begin{eqn}
\bar{\rho} = \bar{e} - \oh M^{ij} \bar{e}_i  \bar{e}_j \, ,
\end{eqn}
with $M^{ij}$ the inverse of $M_{ij}$, we can estimate that there exists an integer $p\leq h^{3,1}(X_4)$ satisfying $N_{\rm flux}^p \bar{\rho}\gtrsim d^{2p-1}$, with $d \equiv {\rm g.c.d} \{m^\mu\}$. When $M$ is not invertible, we instead have that $p \leq r = \rank M$. Finally, using \eqref{eq: tadpole constraint anstaz}, we conclude that
\begin{equation}
\label{eq: K3 saxion bound}
   \cK <  (N_{\rm flux})^{2p+1} d^{2-4p} (K_i^{(3)}t^i)^2 \, .
\end{equation}
For a given choice of fluxes, this relation sets an upper bound on the possible values of the complex structure saxions. The details of this constraint will  heavily depend  on the topology of the mirror four-fold, through its intersection numbers and the $\alpha'^3$-correction terms $K_i^{(3)}$. For instance, notice that for a saxionic direction $t^i$ along which $\cK$ grows linearly \eqref{eq: K3 saxion bound} does not really set a bound, in agreement with our results of section \ref{s:linear}. As a very rough estimate, \eqref{eq: K3 saxion bound}  sets an overall bound for the complex structure saxion vevs of the form
\be
t^i \lesssim N_{\rm flux}^{p+\frac{1}{2}} d^{1-2p} | K_i^{(3)}|\, .
\label{boundgen}
\ee
Remarkably, our reasoning applies also when some fields are not fixed at the polynomial level.

Finally note that, even when $M$ has maximal rank, this moduli stabilisation scheme suggests that there is a saxionic field direction whose mass is suppressed by $\eps_it^i$ compared to the other ones, as it is only stabilised when the corrections to the K\"ahler potential are taken into account. To check whether the scalar mass spectrum is hierarchical or not  one should describe the potential in terms of canonically normalised fields, which we will not attempt to do in this work. Nevertheless, we already see that the key ingredient for such a potential hierarchy is the mixing between different blocks in the saxion-dependent matrix \eqref{ZAB}, which only appears due to $K_i^{(3)}$ corrections, and so by construction it is suppressed in the large complex structure regime.


\section{The type IIB limit}
\label{s:IIB}

A celebrated moduli stabilisation setup corresponds to type IIB orientifold compactifications with background three-form fluxes. In this section we specify our results to this case, neglecting the presence of D7-brane moduli and worldvolume fluxes. As we will see, our findings  imply not only a simple form for the scalar potential at large complex structure and weak coupling, but also two different moduli stabilisation schemes with an upper bound for the complex structure vevs. One of these schemes provides counterexamples to the Tadpole Conjecture of \cite{Bena:2020xrh}, that proposes a tension between full moduli stabilisation and the tadpole constraint. Such a scheme will be generalised to genuine F-theory compactifications in section \ref{s:linear}.

\subsection{The flux potential}

Type IIB compactifications with background three-form fluxes can be understood as F-theory on $(C_3 \times \mathbb{T}^2)/\Z_2$, with $C_3$ a Calabi--Yau three-fold, provided that the presence of D7-branes can be neglected for the bulk dynamics. We can then apply the results of the previous two sections by splitting the index counting complex structure moduli as $i = \{0, a\}$, where $T^0$ represents the complex structure of $\mathbb{T}^2$ and $T^a$, $a = 1, \dots, h^{2,1}(C_3)$ the complex structure moduli of the three-fold. We also impose
\be
\CK_{0abc} = \kappa_{abc}\, ,
\label{interT2}
\ee
where $\kappa_{abc}$ are the triple intersection numbers of the mirror three-fold $B_3$. From \eqref{supalt} we obtain a leading-order superpotential of the form 
\begin{align}\nonumber
 W =&\ e +  e_0 T^0 + e_a T^a + \frac{1}{2} m^{ab} \kappa_{abc} T^c T^0 + \frac{1}{2} m^{0a} \kappa_{abc} T^bT^c \\
   & +\frac{m^0}{6} \kappa_{abc} T^aT^bT^c + \frac{1}{2} m^a  \kappa_{abc} T^b T^cT^0  + \frac{m}{6}  \kappa_{abc}T^aT^b T^cT^0 \, .
   \label{supoIIBF}
\end{align}
This expression does not fully correspond to the superpotential of type IIB flux compactifications, due to the redundancy associated to the quanta $m^{ij}$. A one-to-one correspondence between flux quanta is achieved when we consider an expression of the form \eqref{supo}, which involves specifying a basis of holomorphic four-cycles classes $\{[\sigma_\mu]\}$ in the mirror four-fold $X_4 = (B_3 \times \mathbb{T}^2)/\Z_2$. 

In this case the basis $\{[\sigma_\mu]\}$ can be constructed explicitly, as follows. We first consider the $B_3$ Mori cone generators $[\mathcal{C}'^a]$,  $a = 1, \dots, h^{1,1}(B_3)$, and the divisor classes $[D_a']$, that generate its K\"ahler cone and specify its triple intersection numbers as $\kappa_{abc} = [D_a']\cdot [D_b']\cdot [D_c']$. The K\"ahler cone of $X_4$ is generated by $[D_a] = [D_a' \times \mathbb{T}^2]$, and by the class of $B_3$, which we denote as $[D_0]$. Following section \ref{s:potential}, we consider the following set of holomorphic four-cycles
\be
\gamma_{ij}  = D_i . D_j \, , \quad i = \{0, a\}\, ,
\label{gamma}
\ee
that correspond to the quanta $m^{ij}$ in \eqref{supoIIBF}. The elements of this set are not independent in homology, as opposed to the following ones
\be
H_a = D_a' \,, \qquad H_{\hat{a}} = \mathcal{C}^{\hat{a}} \times \mathbb{T}^2\, ,
\label{4formbasis}
\ee
which form the holomorphic four-form basis $\{[\sigma_\mu]\} = \{[H_a], [H_{\hat{a}}]\}$. In other words, the index $\mu$ in  \eqref{supo} splits as $\mu = \{a, \hat{a}\}$, with $a, \hat{a} = 1, \dots, h^{1,1}(B_3)$. The intersection matrix for \eqref{4formbasis} is
\be
\eta_{a \hat{a}} = [H_a] \cdot [H_{\hat{a}}] = \delta_{a \hat{a}}\, ,
\ee
with the remaining entries vanishing. The relation with the redundant set \eqref{gamma} is given by
\be
\zeta^a_{0b} = \zeta^a_{b0}  = \delta_{ab}\, , \qquad   \zeta_{a, bc}\equiv \zeta^{\hat{a}}_{bc} \eta_{\hat{a}a}= \kappa_{abc}\, ,
\label{Koszul}
\ee
with vanishing remaining entries. One can then easily check that  \eqref{interT2} is recovered from \eqref{interrel}. 

Having fixed $\{[\sigma_\mu]\}$, the superpotential for the mirror four-fold $Y_4 = (C_3 \times \mathbb{T}^2)/\Z_2$ reads
\begin{align}\nonumber
 W =&\ \bar{e} +  \bar{e}_0 T^0 + \bar{e}_a T^a +  \bar{m}_a T^a T^0 + \frac{1}{2} \hat{m}^{a} \kappa_{abc} T^bT^c  +\frac{m^0}{6} \kappa_{abc} T^aT^bT^c \\
   & + \frac{1}{2} m^a  \kappa_{abc} T^b T^cT^0  + \frac{m}{6}  \kappa_{abc}T^aT^b T^cT^0 -i  K_0^{(3)} \left(m^0 + m T^0\right) \, .
   \label{supoIIBF2}
\end{align}
where we have applied \eqref{hatm}, defined $m_a \equiv \delta_{a\hat{a}} \hat{m}^{\hat{a}}$ and already taken into account the polynomial corrections of  section \ref{sec:poly}. Notice that for the case of $Y_4 = (C_3 \times \mathbb{T}^2)/\Z_2$ we have that $K^{(0)} = K^{(2)}_{00} = K^{(2)}_{ab} = K^{(3)}_a = 0$. We similarly obtain the corrected K\"ahler potential 
\be
K_{\rm cs}^{\rm corr}=- \log (2t^0) -\log\left(\frac{4}{3}\kappa_{abc}t^at^bt^c+ 2 K_0^{(3)} \right)\, .
\label{KcscorrIIB}
\ee

One may now connect these expressions with the more standard formulation of type IIB flux compactifications on Calabi--Yau orientifolds. We start with the superpotential  \cite{Gukov:1999ya}
\begin{align}
W_{\rm IIB}=\int_{C_3} \Omega_3 \wedge G_3\,,
\end{align}
where $G_3=F_3-\tau H_3$ is the complexified three-form flux, with $\tau = C_0 - ig_s^{-1}$ the axio-dilaton. The holomorphic three-form $\Omega_3$ of the Calabi--Yau $C_3$, can be expanded as 
\begin{align}
\Omega_3= \alpha_I X^I - \beta^I \partial_I \mathcal{F}\,,
\end{align}
with $(\alpha_I, \beta^I)$ a symplectic basis of harmonic three-forms on $C_3$. The prepotential in the large complex structure limit is given by
\begin{align}
\mathcal{F} = - \frac{1}{6} \frac{\kappa_{abc}X^aX^bX^c}{X^0} + \frac{1}{2} K_{ab}^{(1)} X^aX^b + K_a^{(2)} X^aX^0 - \frac{i}{2} K^{(3)} (X^0)^2 + \dots
\end{align}
where exponential corrections have been neglected. 
By introducing the projective coordinates $z^a =X^a/X^0$ we can write the holomorphic three-form as 
\begin{align}
\Omega_3= X^0\left[ \alpha_0 + z^a \alpha_a +\left(\frac{1}{2}\kappa_{abc}z^bz^c - K_{ab}^{(1)} z^b- K_a^{(2)}  \right) \beta^a- \left( \frac{1}{6} \kappa_{abc}z^a z^bz^c + K_a^{(2)} z^a - i K^{(3)} \right) \b^0\right]\, ,
\label{O3exp}
\end{align}
where an overall $X^0$ factor has been dropped.  Similarly, we can expand the $G_3$ flux as 
\begin{align}
G_3=(F^0 -\tau H^0) \alpha_0 - (F^a-\tau H^a) \alpha_a + (F_a-\tau H_a)\beta^a + (F_0 - \tau H_0) \beta^0\, ,
\end{align}
and arrive to the following expression for the superpotential
\begin{align}\label{IIBsuperpot}
 W_{\rm IIB}=\bar G_0  +\bar G_a z^a +  \frac{1}{2} \kappa_{abc} G^a z^bz^c + \frac{1}{6} G^0 \kappa_{abc} z^az^bz^c  - i K^{(3)}  G^0\,,
\end{align}
where we have defined $G_I \equiv F_I -\tau H_I$ and $\bar G_a= G_a -K_{ab}^{(1)} G^a + K_a^{(2)} G^0$, $\bar G_0= G_0 - K^{(2)}_a G^a$. One can see that this expression matches \eqref{supoIIBF2} upon performing the identifications
\be
T^0 = \tau\, , \qquad T^a = z^a\, ,
\ee
as well as $K^{(3)}_0 = K^{(3)}$, $K^{(2)}_{0a} =  K^{(2)}_{a}$, $K^{(1)}_{ab} = \half \kappa_{aab}$ and 
\begin{align}\label{identification}
 H^0=-m\,,\qquad F^0&=m^0 \,,\qquad H^a=-m^a\,,\qquad F^a=\hat{m}^{a}\,,\\
 \bar{H}_a= -\bar{m}_a\,,\qquad \bar{F}_a&=\bar{e}_a \,,\;\;\qquad \bar{H}_0=-\bar{e}_0\,,\;\;\qquad \bar{F}_0=\bar{e}\,. \nonumber
\end{align}
Additionally, from \eqref{O3exp} one also reproduces \eqref{KcscorrIIB}, as already shown in \cite{Palti:2008mg,Escobar:2018rna}.

Using the results of section \ref{s:potential} one may give a compact expression for the resulting F-term scalar potential. The flux-axion polynomials are\footnote{For altenative definitions of flux-axion invariants in the type IIB compactifications see \cite{Bielleman:2015ina,Shukla:2019wfo,Blanco-Pillado:2020wjn}.}
\begin{align}
\nonumber \bar{\rho} &= \bar{e} + \bar{e}_0 b^0+  \bar{e}_ab^a+ \bar{m}_ab^ab^0 +  \kappa_{abc} \left(\frac{1}{2} \hat{m}^{a} b^bb^c  +\frac{1}{2} m^ab^bb^cb^0 +\frac{1}{6}m^0 b^ab^bb^c +\frac{1}{6}m b^ab^bb^c b^0\right)\,,\\
\nonumber\bar{\rho}_0&= \bar{e}_0+\bar{m}_ab^a +\kappa_{abc} \left(\frac{1}{2}m^a b^bb^c +\frac{1}{6} m b^a b^bb^c\right)\,,\\
\nonumber\bar{\rho}_a&= \bar{e}_a + \bar{m}_ab^0 +  \kappa_{abc}\left( \hat{m}^{b}b^c  + m^b b^0 b^c +\frac{1}{2}m^0 b^bb^c + \frac{1}{2} m b^b b^c b^0\right)\,,\\
\bar{\rho}_a^\prime&= \bar{m}_a +  \kappa_{abc} \left(m^b b^c +\frac{1}{2} m b^b b^c\right) \,,\\
\nonumber\hat{\rho}^{a}&= \hat{m}^{a} + m^a b^0 +m^0 b^a + m b^0 b^a \,,\\
\nonumber\tilde \rho^a&= m^a + m b^a \,,\\ 
\nonumber \tilde \rho^0&=m^0 + mb^0 \,,\\ 
\nonumber\tilde \rho&=m\, ,
\end{align}
in terms of which the potential takes the form \eqref{bilinear}.  At leading order, the saxion-dependent matrix $Z$ reads
\begin{align}
\label{ZABIIB}
Z^{AB} =
\frac{4}{3}e^K t^0\kappa
\begin{pmatrix}
\frac{1}{6}t^0\kappa & & & & & & & -1 \\
&  \frac{1}{6}\frac{\kappa}{t^0} & & & & & 1  &  \\ 
& & \frac{2}{3} t^0 \kappa g_{ab}^\kappa  & & & \delta^a_b & & \\
& & &\frac{2}{3} \frac{\kappa}{t^0} g_{ab}^\kappa  & - \delta^a_b & & & \\
& & &  -  \delta^a_b & \frac{3}{2}\frac{t^0}{\kappa} g^{ab}_\kappa & & 
\\
& &  \delta^a_b   & & &  \frac{3}{2}\frac{1}{t^0\kappa}g^{ab}_\kappa & &  \\
&1  & & & & & 6\frac{t^0}{\kappa} &  \\
-1 & & & & & & & \frac{6}{t^0\kappa} \\
\end{pmatrix} \, ,
\end{align}
where $\vec{\rho}^{\, t} = \left(\tilde{\rho}, \tilde{\rho}^0, \tilde{\rho}^a,   \hat{\rho}^{a},  \bar{\rho}_a^\prime, \bar{\rho}_a, \bar{\rho}_0,   \bar\rho   \right)$ and we have defined
\be
\kappa \equiv \kappa_{abc}t^at^bt^c\, ,\qquad \kappa_a \equiv \kappa_{abc}t^bt^c\, , \qquad \kappa_{ab} \equiv \kappa_{abc}t^c\, ,
\ee
and
\be
g_{ab}^\kappa = \frac{3}{2\kappa} \left(\frac{3\kappa_a\kappa_b}{2\kappa} - \kappa_{ab}\right)\, , \qquad  g^{ab}_\kappa =  2t^at^b - \frac{2}{3} \kappa \kappa^{ab}\, .
\ee
Notice that in this case the matrix $Z$ has the structure
\be
Z = 
\begin{pmatrix}
A & B \\ B^t & B^t A^{-1} B \\
\end{pmatrix} \, , \qquad A = A^t\, ,
\label{blockZ}
\ee
with $A$, $B$ non-singular $2h^{3,1} \times 2h^{3,1}$ matrices. This form is  preserved by polynomial corrections.

\subsection{Tadpoles and moduli stabilisation}
\label{sec:moduliIIB}

Let us analyse the conditions for Minkowski vacua and the implications of the tadpole constraint in the type IIB orientifold limit. If we consider a large complex structure regime such that the effect of the correction $K^{(3)}$ can be neglected, the vacua conditions read
\begin{subequations}
\label{eq:MinkIIB}
\begin{empheq}[box=\widefbox]{align}
    \bar\rho&=\frac{1}{6}t^0 \kappa\tilde{\rho} \label{eq:MinkIIB 0}\\
    \bar\rho_0&=-\frac{1}{6}\frac{\kappa}{t^0}\tilde{\rho}^0 \label{eq:MinkIIB 00}\\
    \bar{\rho}_a&=-\frac{2}{3}t^0 \kappa g_{ab}^\kappa\tilde{\rho}^b \label{eq:MinkIIB i} \\
    \bar{\rho}'_a &= \frac{2}{3} \frac{\kappa}{t^0} g_{ab}^\kappa \hat \rho^b
    \label{eq:MinkIIB mu}
\end{empheq}
\end{subequations}
All these equations are a straightforward application of the general result \eqref{eq:Mink} to the type IIB limit, except perhaps \eqref{eq:MinkIIB mu}. To see how it arises from  \eqref{eq:Mink mu} notice that
\begin{equation}
\label{vacua mu strategy}
    \begin{cases}
4\left(  \zeta_{\mu 0}- \frac{\mathcal{K}_0}{ \cK} \zeta_{\mu}\right) \bar{\rho}^\mu =  2  t^a \bar{\rho}_a^\prime - \frac{ \kappa_a \hat{\rho}^a}{t^0}  = 0\, ,\\
4\left(  \zeta_{\mu a}- \frac{\mathcal{K}_a}{ \cK} \zeta_{\mu}\right) \bar{\rho}^\mu = 4t^0 \bar{\rho}_a^\prime + 4\kappa_{ab}\hat{\rho}^b - \frac{3\kappa_a}{\kappa} \left(2t^0t^b  \bar{\rho}_b^\prime + \kappa_b \hat{\rho}^b\right) = 0\, ,
    \end{cases}
\end{equation}
where we used that $\zeta_{\hat{a}, 0b} = \delta_{\hat{a}b}$ and $\zeta_{a, bc} = \kappa_{abc}$. Together, these two conditions imply  \eqref{eq:MinkIIB mu}.

As in the general case, when turning on the correction  $K^{(3)}$ the above vacuum equations are corrected. In particular, for the type IIB limit we find that the equations \eqref{eq:Minkcorr} yield 
\begin{subequations}
 \label{eq:MinkcorrIIB}
\begin{align}
    \bar \rho -\frac{t^0\kappa }{6} \tilde \rho &= -\eps\, \frac{\kappa t^0 }{12} \left[\tilde \rho + \frac{9}{\kappa t^0}  \kappa_a \hat \rho^a \right]
    \,,\\
    \bar \rho_0 +\frac{1}{6}\frac{\kappa}{t^0}\tilde \rho^0 &= \eps\, \frac{1}{12} \left(\frac{\kappa}{t^0} \tilde\rho^0 - 9  \kappa_a \tilde\rho^a \right)\,,\\
    \bar \rho_a +\frac{2}{3}t^0 \kappa g_{ab}^\kappa \tilde \rho^a &= \eps\, \frac{3}{4}  \kappa_a \left({\tilde \rho^0} - \frac{t^0\kappa_b}{\kappa}\tilde \rho^b \right) \,,\\
    \bar{\rho}'_a - \frac{2}{3} \frac{\kappa}{t^0} g_{ab}^\kappa \hat \rho^b & = \eps \frac{3}{8}  \frac{\kappa_a}{\kappa}  \left(2\kappa \tilde{\rho} + \frac{\kappa_b}{t^0} \hat \rho^b + 2 t^b \bar{\rho}'_b \right)\,,
\end{align}
\end{subequations}
where we have defined  $\epsilon \equiv \frac{3 K^{(3)}}{2 \kappa}$. 

In terms of the vacua equations, one can give a more explicit expression for the tadpole condition in this setup. One begins with the topological quantity
\be
N_{\rm flux} =  \bar{e} m - \bar{e}_i m^i + \bar{m}_a \hat{m}^a = \bar{\rho} \tilde{\rho} - \bar{\rho}_i \tilde{\rho}^i +  \bar{\rho}_a^\prime \hat{\rho}^{a}\, ,
\label{NfluxrhoIIB}
\ee
which at vacua can be expressed as
\begin{eqn}
N_{\rm flux} &\stackrel{\rm vac}{=} \frac{t^0 \kappa}{6} \left(\tilde{\rho}^2 +  \frac{(\tilde{\rho}^0)^2}{(t^0)^2} + \frac{2}{3} g_{ab}^\kappa  \tilde{\rho}^a\tilde{\rho}^b  +
\frac{3}{2\kappa^2} g^{ab}_\kappa
\rho_a^\prime \rho_b^\prime \right)
\, ,
\label{NfluxvacIIB}
\end{eqn}
where we have used the conditions \eqref{eq:MinkIIB} and therefore neglected the effect of $K^{(3)}$. This approximation is justified if we aim to obtain the restriction on the fluxes  that arise in the different weak coupling, large complex structure limits $\kappa, t^0 \to \infty$, as done in section \ref{sec:tadpole}. As in there (see also \cite[appendix D]{Blanco-Pillado:2020wjn}),  we must set $\tilde{\rho} = 0$ when $t^0, \kappa/6 > \sqrt{N_{\rm flux}}$ in order to find vacua, and therefore in this regime $\tilde{\rho}^0 = m^0$, $\tilde{\rho}^a = m^a$. The remaining fluxes will then be constrained depending on the different limits that we take, which we can classify in a slightly more explicit manner as compared to the general case.

Indeed, let us  consider a scaling of the form $t^0 \sim \kappa^r \to \infty$, with $r \in \R$. If $r \geq 1$ then $t^0 \kappa g_{ab}^\kappa$ will diverge, and we will have to set $m^a = \tilde{\rho}^a$ to zero. We will also have that $t^0  g^{ab}_\kappa/\kappa$ diverges, and so $m_a = \bar{\rho}_a^\prime$ must vanish as well.  We then recover a simplified flux lattice such that $\vec{q}^{\, t} = (0,m^0,0, \hat{m}^a, 0, \bar{e}_a, \bar{e}_0, \bar{e})$, and the tadpole is given by $N_{\rm flux} = - m^0\bar{e}_0$. Alternatively, if $r<1$ the $m^0 = \tilde{\rho}^0$ must be set to zero and, generically, the same applies for $m^a = \tilde{\rho}^a$. The question is then whether $\hat{m}^a = \hat{\rho}^a$ and $m_a = \bar{\rho}_a^\prime$ must vanish or not. In fact, to have a non-trivial tadpole we need that $N_{\rm flux} = \sum_a \hat{m}^a m_a \neq 0$, and one can convince oneself that this is only possible if $t^0$ scales like $\kappa g_{aa}^\kappa$, for at least some $a$. All these are cases in which $r<1$ and $\vec{q}^{\, t} = (0,0,0, \hat{m}^a, m_a, \bar{e}_a, \bar{e}_0, \bar{e})$, which we will consider as another subset of vacua. Finally, one can check that this classification is unchanged if we add to \eqref{NfluxvacIIB} the corrections that arise from imposing \eqref{eq:MinkcorrIIB}. Let us now analyse the moduli stabilisation of both classes of vacua:

\subsubsection*{IIB1: $\vec{q}^{\, t} = (0,0,0, \hat{m}^a, {m}_a, \bar{e}_a, \bar{e}_0, \bar{e})$}

This case falls into the generic class of vacua discussed in section \ref{sec:moduli}. We have that the vacua equations \eqref{eq:MinkcorrIIB} reduce to 
\begin{subequations}
 \label{eq:MinkcorrIIB1}
\begin{align}
\label{eq:MinkcorrIIB1 0}
    \bar \rho &=  -\frac{3\eps}{4}   \kappa_a \hat m^a \,,\\
       \label{eq:MinkcorrIIB1 i}
    \bar \rho_i  &=0 \,, \qquad i= 0, a\, ,\\
    {m}_b - \frac{2}{3} \frac{\kappa}{t^0} g_{bc}^\kappa \hat m^c & = \eps \frac{3}{4t^0}  \frac{\kappa_b}{\kappa}  \kappa_a \hat m^a \, ,
     \label{eq:MinkcorrIIB1 mu}
\end{align}
\end{subequations}
to first order in $\eps$. We may now apply the general discussion in section \ref{sec:moduli} to set bounds on the saxion vevs. Using \eqref{eq:MinkcorrIIB1 mu} we find that
\be
{m}_a = A\kappa_a + C_a + \cO(\eps) \, , \qquad \hat{m}^a = 2At^0t^a + C^a + \cO(\eps)\, ,
\label{solIIB1mu}
\ee
where $C_at^a=0$, $C^a\kappa_a=0$ and $C^a\kappa_{ab}=-t^0C_b$. We therefore obtain the inequality 
\be
N_{\rm flux} = 2A^2 t^0 \kappa + C_a C^a+\cO(\eps) \geq 2A^2 t^0 \kappa +\cO(\eps)\, ,
\ee
while from \eqref{eq:MinkcorrIIB1 0} one can see that
\be
\label{eq: IIB1 A}
A = -\frac{4}{9} \frac{\bar{\rho}}{t^0 K^{(3)}}  \, . 
\ee
From here we find the following bound for the complex structure saxions,
\be
\frac{\kappa}{t^0} < N_{\rm flux}^{2p+1} d^{2-4p} (K^{(3)})^2 \, , 
\label{boundIIB1}
\ee
where $p \leq h^{2,1}(C_3)+1$ is bounded by the number of complex structure plus dilaton fields, and $d ={\rm g.c.d} (\{\hat{m}^a, {m}_a\})$. Here we have used a reasoning similar to the one below \eqref{Meq} to arrive to the inequality $N_{\rm flux}^p \bar{\rho} \gtrsim d^{2p-1}$. Finally, notice that taking into account that in this scheme $t^0 \sim \kappa g_{aa}^\kappa$, we end up with a bound for the saxions which is, again, roughly of the form \eqref{boundgen}.

To obtain a more concrete scheme one may consider that the matrix $M_{ij}\equiv \hat{m}^\mu\zeta_{\mu,ij}$, introduced below \eqref{barrhoeqs}, is invertible. In the type IIB limit and with our particular choice of fluxes this matrix is given by
\begin{equation}
 M= 
\begin{pmatrix}
0 & m_a \\ m_b & S_{ab} \\
\end{pmatrix} \, , \qquad S_{ab}\equiv \kappa_{abc}\hat{m}^{c}\, .
\end{equation}
Then, choosing the fluxes in such a way that $M_{ij}$ has an inverse requires $S_{ab}$ to be invertible as well. If that is the case, the inverse matrix  $M^{ij}$ has the form
\begin{equation}
 M^{-1}= S^{-1}
\begin{pmatrix}
-1 & S^{ac}m_c \\ S^{bc}m_{c} & SS^{ab}-S^{ac}S^{bd}m_cm_d \\
\end{pmatrix} \, , \qquad S \equiv S^{ab}m_am_b\, ,
\end{equation}
where $S^{ab}$ is the inverse of $S_{ab}$. Notice that for $M$ to be invertible we have to further ensure $S \ne 0$. If this last condition is not satisfied, the kernel of $M$ is given by 
\begin{align}
    \text{ker}(M) = \langle (1, -S^{ab} m_b)^t \rangle\,,
\end{align}
such that we have a flat direction along $T^i = (\tau, -\tau S^{ab} m_b)^t$. Given the identification \eqref{identification} this precisely reproduces  the flat direction found in \cite{Demirtas:2019sip} for  $S_{ab}$ invertible but $S =0$.

Using these results we can achieve stabilisation of all the moduli of the system. Starting with the axions, from \eqref{eq:MinkcorrIIB1 i} we have 
\begin{align}
    b^0&=-S^{-1}S^{ab}\bar{e}_am_b+\bar{e}_0S^{-1}\, ,\\
    b^a&=-\bar{e}_0 S^{-1}S^{ab}m_b-\bar{e}_bS^{ab}+S^{-1} S^{ac}m_c S^{bd}m_d\bar{e}_b=-S^{ab}(\bar{e}_b+b^0m_b)\, .
\end{align}

Regarding the saxions,  the expression \eqref{eq:MinkcorrIIB1 mu} at leading order provides us with a system of $h^{2,1}(C_3)$ independent equations of order $4$ in the set of $h^{2,1}(C_3)$ $\{t^a\}$. Hence, we can use it to express all the saxions in terms of the saxionic direction $t^0$. We can then  substitute our results in \eqref{eq:MinkcorrIIB1 0} and employ the first order corrections in $\epsilon$ to stabilise the remaining direction $t^0$. Note that we are able to ignore the corrections in \eqref{eq:MinkcorrIIB1 mu} because the first leading contribution of the saxions in \eqref{eq:MinkcorrIIB1 0} is already linear in the parameter $\epsilon$.

Looking to the shape of $M^{ij}$  we observe a very straightforward flux choice for which the matrix $M_{ij}$ is invertible, and which is related to the ansatz taken in \cite{Blanco-Pillado:2020hbw} for different reasons. Indeed, let us consider that $\hat{m}^a\neq0$ $\forall a$ and take
 $m_a=(r/q) S_{ab}\hat{m}^b$, with $q\equiv {\rm gcd}(\kappa_{abc}\hat{m}^b\hat{m}^c)$ and $r \in \Z$. Then the matrix $M^{ij}$ simplifies to
\begin{equation}
    M^{-1}= S^{-1}
    \begin{pmatrix}
    -1 & \frac{r}{q} m^a \\ \frac{r}{q} m^b & S S^{ab}-\frac{r^2}{q^2}\hat{m}^a\hat{m}^b \\
    \end{pmatrix} \, ,
\end{equation}
where in this case we have $S=(r/q)^2 (S_{ab}\hat{m}^a\hat{m}^b)$. 
Let us study the implications of our ansatz for the relations derived from the equations of motion. First we evaluate $S_{ab}\hat{m}^b=\kappa_{abc}\hat{m}^b\hat{m}^c$ using the second relation of \eqref{solIIB1mu}:
\begin{align}
    \kappa_{abc}\hat{m}^b\hat{m}^c&=4A^2(t^0)^2\kappa_a+4\kappa_{ab}AC^bt^0+\kappa_{abc}C^bC^c\nonumber\\
    &=4A^2(t^0)^2\kappa_a-4(t^0)^2AC_a+\kappa_{abc}C^bC^c\, ,
\end{align}
where in the last step we have used the relation $C^a\kappa_{ab}=-C_b t^0$. Now we combine our ansatz $m_a=(r/q)S_{ab}\hat{m}^b$ with the first relation of \eqref{solIIB1mu}. This leads to
\begin{align}
    \begin{cases}
    r=\frac{q}{4A(t^0)^2}\, ,\\
    C_a=\frac{1}{2}\kappa_{abc}C^bC^c\, .
    \end{cases}
\end{align}
Contracting the last expression with $t^a$ and taking into account that $C_at^a=0$ we conclude
\begin{equation}
    \kappa_{ab}C^aC^b=0\Rightarrow g_{ab}^\kappa C^aC^b=0\Rightarrow C^a=0\, .
\end{equation}
Hence, the ansatz $m_a=(r/q) S_{ab}\hat{m}^b$ implies that $C^a = C_a =0$ in \eqref{solIIB1mu}. Then $\hat{m}^a \propto t^a$ and $S_{ab}\propto\kappa_{ab}$, as in \cite{Blanco-Pillado:2020hbw}. Moreover the ratio $t^a/t^0$ is easily fixed at leading order, since \eqref{solIIB1mu} gives
\begin{equation}
    \frac{t^a}{t^0}=\frac{2r\hat{m}^a}{q}\, .
\end{equation}
Working now with \eqref{eq:MinkcorrIIB1 i} we have
\begin{align}
    b^0&=\frac{q^2}{r^2S_{ab}\hat{m}^a\hat{m}^b}\left(\bar{e}_0-\frac{r}{q}\hat{m}^a\bar{e}_a\right)\, ,\\
    b^a&=-S^{ab}\bar{e}_b-\frac{r\hat{m}^ab^0}{q}\, .
\end{align}
Finally, \eqref{eq: IIB1 A} determines the vev for the saxion $t^0$.

Note that in this particular setup the total tadpole $N_{\rm flux} = \sum_a {m}_a\hat{m}^a$ is a sum of positive terms and so it exceeds in value to $h^{2,1}(C_3)$. As pointed out in \cite{Bena:2020xrh} this kind of behaviour leads to a significant tension between tadpole cancellation and full moduli stabilisation for a large number of moduli. From our perspective, this would favour vacua where $C^a \neq 0$. In that case, one should apply \eqref{originTC} to see whether $N_{\rm flux}$ is bounded from below by $h^{2,1}(C_3)$ or not.

\subsubsection*{IIB2: $\vec{q}^{\, t} = (0,m^0,0, \hat{m}^a, 0, \bar{e}_a, \bar{e}_0, \bar{e})$}

This case is dual, via mirror symmetry, to the type IIA non-supersymmetric Minkowski flux vacua constructed in \cite{Palti:2008mg} and analysed from the viewpoint of the bilinear potential \eqref{bilinear} in \cite{Escobar:2018rna}. As shown in there, in this case one can solve for the vev of each field in terms of the flux vacua and the correction $\eps$. One starts with the following vacua equations 
\begin{subequations}
 \label{eq:MinkcorrIIB2}
\begin{align}
    \bar \rho  &=  0 \,,\\
    t^0\bar{e}_0 +\frac{1}{6}\kappa m^0 &= \eps\, \frac{\kappa}{6} \frac{1+ 4\eps}{2 - \eps} m^0 \,,\\
    \bar \rho_a  &= \eps\, \frac{3}{2}  \frac{\kappa_a}{2 - \eps} m^0  \,,\\
   \hat \rho^a & = 0 \,,
\end{align}
\end{subequations}
which at first order in $\eps$ are equivalent to \eqref{eq:MinkcorrIIB}, restricted to this choice of fluxes.
Borrowing the results from \cite[section 4.1]{Escobar:2018rna} and adapting them to our notation we obtain the solution
\begin{eqn}
b^0 & = - \frac{1}{3\bar e_0(m^0)^2}\left(\kappa_{abc}\hat{m}^{a}\hat{m}^{b}\hat{m}^{c}-3e_a\hat{m}^{a}m^0\right)-\frac{\bar{e}}{\bar e_0}\, , \\
b^a & = - \frac{\hat{m}^a}{m^0}\, ,
\label{axionsIIB2}
\end{eqn}
for the axions and 
\begin{eqn}
t^0 & = - \frac{1}{6}\frac{m^0}{\bar{e}_0} \kappa \left(1 - \eps \frac{1 + 4\eps}{2- \eps}\right) 
\, , \\
\kappa_a & = \frac{2- \eps }{3 (m^0)^2 \eps }  \left(2m^0 \bar{e}_a -\kappa_{abc} \hat{m}^{b}\hat{m}^{c} \right)\, ,
\label{saxionsIIB2}
\end{eqn}
for the saxions. Note that the $\kappa_a$ are determined implicitly, and that acceptable vacua correspond to saxion vevs within $\{t^a > 0 | \epsilon \ll 1\}$, which imposes a constraint on the flux quanta.\footnote{Explicit solutions to the equations for $\kappa_a$ have been proposed in \cite{Palti:2008mg}, assuming  homogeneous vevs for all $t^a$. Additionally, these equations are similar to those determining the K\"ahler moduli vevs in type IIA AdS$_4$ CY  orientifolds \cite{DeWolfe:2005uu,Marchesano:2019hfb}, and so explicit solutions for such a setup will translate into Minkowski vacua in this context.}

Notice that $t^0 \sim \kappa/6$, as could have been guessed from the leading order equation  \eqref{eq:MinkIIB 00} and the fact that $\bar{\rho}_0 = \bar{e}_0$. Also
\be
\frac{\kappa}{\kappa_a}  \lesssim (m^0)^2 |K^{(3)}| < N_{\rm flux}^2 d^{-2} |K^{(3)}|  \, , 
\label{boundIIB2}
\ee
with $d = {\rm g.c.d.} (m^0,\bar{e}_0)$. This results in an upper bound on the value of the complex structure saxions which is roughly of the form \eqref{boundgen} with $p =\frac{3}{2}$, even if the moduli stabilisation scheme under discussion is different from the one in section \ref{sec:moduli}. Note also that this bound is consistent with the regime in which $\eps \ll 1$, whenever $(m^0)^2 |K^{(3)}|$ is moderately larger than 1.  

This class of vacua constitute a counterexample to the Tadpole Conjecture proposed in \cite{Bena:2020xrh}, since the flux contribution to the tadpole $N_{\rm flux} = - m^0\bar{e}_0$ is independent of the number of complex structure moduli. There is therefore no tension between achieving full moduli stabilisation and having an $N_{\rm flux}$ that it is bounded. A key property for this to happen is the fact that most of the RR flux quanta that implement moduli fixing do not contribute to the tadpole because they do not pair up with $\bar{e}_0$ in the intersection matrix.  What is true is that all complex structure saxions $t^a$ are stabilised only when the effect of the correction $K^{(3)}$ is taken into account \cite{Palti:2008mg} which suggests that, in this case, decoupling the expression for $N_{\rm flux}$ from the number of complex structure moduli comes at the cost of having several light fields. In the next section we will generalise this scheme to genuine F-theory setups. We will see that most of the features of the type IIB case will be realised except for the bound \eqref{boundIIB2}, which may or may not be present.

\section{The linear scenario}
\label{s:linear}

The moduli stabilisation scheme \textbf{IIB2} for type IIB orientifolds provides a class of compactifications in which the flux contribution to the D3-brane tadpole $N_{\rm flux}$ is independent from the number of complex structure moduli $h^{3,1}(Y_4)$, and nevertheless one can achieve full moduli stabilisation. Therefore it is quite simple to stabilise all complex structure moduli and at the same time satisfy the bound $24 N_{\rm flux} \leq \chi(Y_4)$, a scenario whose realisation has been recently doubted \cite{Braun:2020jrx,Bena:2020xrh,Bena:2021wyr}, see also \cite{Betzler:2019kon}. In the following we would like to generalise the key features of scheme \textbf{IIB2} to more general F-theory compactifications, providing further counterexamples to the Tadpole Conjecture of \cite{Bena:2020xrh}. 

We will dub this more general setup the {\em linear scenario}, because the key ingredient will be a four-fold $Y_4$ such that at least one complex structure saxion $t_L$ only appears linearly on $\CK = \frac{3}{2}e^{-K_{\rm cs}}$ and in the superpotential. This means that $\CK$ takes the form
\begin{eqn}\label{cKlin}
 \CK = 4 \CK_L t_L + f \, ,
\end{eqn}
with $\CK_L \equiv  \cK_{L abc}t^a t^b t^c$, and $f \equiv f(t^a)$ a function independent of $t_L$ and homogeneous of degree four on the remaining saxions $t^a$. This kind of K\"ahler potential is found when the mirror four-fold $X_4$ is a  smooth three-fold fibration over $\mathbb{P}^1$,\footnote{Note that in order for $t_L$ to appear only linearly in $\cK$ the mirror $X_4$ needs to have a nef effective divisor $D_L$ such that $D_L^2=0$. The normal bundle $\mathcal{O}_{X_4}(D_L)|_{D_L}$ of $D_L$ is then trivial and by adjunction it  follows that $c_1(D_L)$ vanishes. This is satisfied whenever $X_4$ is a fibration of a CY three-fold, $K3\times \mathbb{T}^2$ or an abelian variety over $\mathbb{P}^1$, in which case $D_L$ corresponds to the class of the generic fibre. See \cite{Lee:2019oct} for a related discussion for CY three-folds.}  see section \ref{sec:counter} for an explicit example. In this case the leading saxion-dependent matrix \eqref{ZAB} is 
\be
\label{ZABlin}
 Z  = \frac{1}{2{\cal V}_3^2}
 \begin{pmatrix}
\frac{\cK}{24} & &  &  & & & -1 \\
& \frac{\cK}{6} g_{LL} & \frac{\cK}{6} g_{LL} \varepsilon_a  & &  &  1 &   \\ 
& \frac{\cK}{6} g_{LL} \varepsilon_a & \frac{\cK}{6} g_{ab}   &  & \delta^a_b &  & \\
 & & &   g_{\mu\nu} -\eta_{\mu\nu} & & &  \\
& & \delta^b_a   & &   \frac{6}{\cK}\tilde{g}^{ab} &   -\frac{6}{\cK}\tilde{g}^{ab}\varepsilon_a &  \\
& 1 &   & &  -\frac{6}{\cK}\tilde{g}^{ab}\varepsilon_a & \frac{6}{\cK}g^{-1}_{LL} + \frac{6}{\cK}\tilde{g}^{ab}\varepsilon_a \varepsilon_b&  \\
-1  & & &    & & &  \frac{24}{\cK} \\
\end{pmatrix} \, ,
\ee

where
\begin{eqn}
\frac{\cK}{6} g_{LL} = \frac{1}{6}\frac{\cK_L}{t_L\left(1 +\frac{1}{4}\frac{f}{\cK_Lt_L} \right)}\, , \qquad \varepsilon_a = \p_a \left( \frac{f}{4\CK_L}\right) ,
\end{eqn}
and $\tilde{g}^{ac}g_{cb} = \delta^a_b$. We now consider a limit which takes one or several of the saxions $t^a$ to infinity such that
 \begin{align}\label{limitL}
    t_L \sim \cK_L \to \infty \, .
\end{align}
We also assume that $t_L$ grows faster than any of the other saxions $t^a$, so that we realise the hierarchy $t_L \gg t^a$. Along such a limit $\cK \to \infty$ and $\cK g_{ab} \to \infty$. This implies that in order to find vacua we need to set  $\tilde{\rho} = \tilde{\rho}^a =0$, which translates into the flux constraint $m = m^a =0$. We also have that
\begin{eqn}
\frac{\cK}{6} g_{LL} \stackrel{\eqref{limitL}}{\to} \frac{1}{6}\frac{\cK_L}{t_L}\, ,
\label{limgLL}
\end{eqn}
and one may find vacua with $m^L \neq 0$ in this regime. Finally, one describes the fluxes $\hat{m}^\mu$ by constructing the set of four-forms $\sigma_\mu$ in the mirror four-fold $X_4$. As mentioned, we assume that $X_4$ is a three-fold fibration over $\mathbb{P}^1$. Due to this fibration structure, a basis of holomorphic four-cycles on $X_4$ can be generated from the K\"ahler cone generators $D_a$ of the fibre $X_3$ 
\begin{align}\label{linearscenarioH1}
   H_a = D_a . D_L \,,  
\end{align}
as well as by fibering the Mori cone generators $\mathcal{C}^a$ of $X_3$ over the base $\mathbb{P}^1$
\begin{align}\label{linearscenarioH2}
    H_{\hat a} = \mathcal{C}^a \rightarrow \mathbb{P}^1 \,. 
\end{align}
This last set of basis elements is related to the holomorphic four-cycles $\gamma_{ij}=D_i . D_j$ as 
\begin{align}
    [\gamma_{ab}] = \cK_{Labc}\delta^{c\hat c} [H_{\hat c}]\,. 
\end{align}
The integral basis of four-form classes $[\sigma_\mu]$ is then $\{[\sigma_\mu]\}=\{[H_a],[H_{\hat a}]\}$, and so the four-cycle index splits as $\mu = \{a, \hat{a}\}$, like in the type IIB case, and then \eqref{ZABlin} takes the form \eqref{blockZ}. The intersection matrix $\eta_{\mu\nu}$ is given by 
\begin{align}
    \eta_{a \hat b}=\delta_{a \hat b}\,,\qquad \eta_{a b}=0\,,
\end{align}
plus $\eta_{\hat a \hat b}$ in general non-vanishing and quite involved (see \eqref{intersectionexampleIII} for its form in an explicit example). The form of $g_{\mu\nu}$ will then in general be quite complicated, but given the non-vanishing entries of the intersection matrix $\eta_{\mu\nu}$, setting $\hat{m}^{\hat a}=0$ leads to $\eta_{\mu\nu} \hat{m}^\mu\hat{m}^\nu =0$ and the only contribution to the tadpole is
\be
N_{\rm flux} = - m^L\bar{e}_L\, , 
\ee
which, just as in the \textbf{IIB2} scheme, is independent from $h^{3,1}(Y_4)$. We then recover the flux vector
\begin{align}\label{eq:fluxchoiceL}
    \vec q^{\, t} = (0, m^L, 0, \hat m^a,0,   \bar e_a, \bar e_L,  \bar e)\,,
\end{align}
and we find that $m^L$ has the same role as $m^0$ in the \textbf{IIB2} setup of section \ref{sec:moduliIIB}. It moreover follows that the flux-axion polynomials \eqref{corrhos} reduce to 
\begin{align}
    \nonumber \bar \rho &= \bar e +e_Lb^L+\bar e_a b^a + \cK_{L abc}\left(\frac{1}{2}\hat m^a b^b b^c+\frac{1}{6} m^L   b^a b^b b^c\right) \,, \\
    \nonumber \bar \rho_a &= \bar e_a +  \cK_{Labc}\left(\hat m^b b^c + \frac{1}{2} m^L  b^bb^c\right)\,,\\
    \label{rhoAexample} \bar \rho_L  &= \bar e_L  \,,\qquad \bar \rho_a^\prime = 0 \,,\qquad \hat \rho^a = \hat m^a + m^L  b^a \,,\\
    \nonumber\tilde \rho^a &=0 \,,\qquad \tilde \rho^L  = m^L \,,\qquad \tilde\rho=0\, ,
\end{align}
which we recognise as the flux-axion polynomials in the \textbf{IIB2} setup upon the identifying $\cK_{Labc}$ with $\kappa_{abc}$. The leading-order vacua equations read
\begin{subequations}
    \label{vaclin}
\begin{align}
\bar{\rho}  &= 0\, ,\\
\label{vaclin1}
\bar{e}_L + \frac{\cK}{6}g_{LL} m^L  & = 0 \, ,\\
\label{vaclin2}
\bar{\rho}_a - \varepsilon_a  \bar{e}_L & =  0\, ,\\
\hat{\rho}^a  & =  0\, ,
 \end{align}
\end{subequations}
and can be solved like for the \textbf{IIB2} scheme. Indeed, the first and fourth equations fix the vev for the axions as
\begin{eqn}
b^L & = - \frac{1}{3\bar e_L(m^L)^2}\left(\cK_{L abc}\hat{m}^{a}\hat{m}^{b}\hat{m}^{c}-3e_a\hat{m}^{a}m^L\right)-\frac{\bar{e}}{\bar e_L}\, , \\
b^a & = - \frac{\hat{m}^a}{m^L}\, ,
\label{axionsL}
\end{eqn}
and the remaining ones the vev for the saxions. In particular we find that $\cK g_{LL}/6 \simeq \cK_L/6t_L$ must lie in the range $(N_{\rm flux}^{-1}, N_{\rm flux})$, and that 
\begin{equation}
    t^L=-\frac{1}{6}\frac{\mathcal{K}_Lm_L}{e_L}-\frac{f}{4\mathcal{K}_L}\, ,
\end{equation}
\begin{eqn}
\varepsilon_a  = \frac{m^L \bar{e}_a -\oh \cK_{Labc} \hat{m}^{b}\hat{m}^{c}}{ m^L\bar{e}_L}  = - \frac{N_a}{N_{\rm flux}} \quad \implies \quad N_{\rm flux} |\varepsilon_a| \gtrsim 1\, .
\label{Nfluxineqlin}
\end{eqn}
Here we have defined $N_a \equiv  m^L \bar{e}_a - \frac{1}{2} \cK_{L abc} \hat m^b \hat m^c$ as a monodromy-invariant flux combination in the present setup, more precisely the analogue of the third invariant listed in appendix \ref{ap:invariants}. We also obtain the inequality
\begin{eqn}
\frac{\cK}{6}g_{LL} |\varepsilon_a| \gtrsim N_{\rm flux}^{-2}\, .
\label{Nfluxineqlin2}
\end{eqn}

Notice that in the present setup the leading vacua equations  in principle suffice to find a set of vacua with full moduli fixing, unlike in the \textbf{IIB2} scheme. This is due to the fact that $\varepsilon_a$ appears at leading order. Nevertheless, further corrections will also contribute to the above equations, and in some cases they are needed to understand the implications of the inequality \eqref{Nfluxineqlin}. We can read off such corrections from \eqref{apeq:vacuumrho_i}, focusing on those corrections that involve $K^{(3)}_L$, which are the leading ones. To leading order  in $\epsilon_L \equiv \frac{3K^{(3)}_L}{2\CK_L}$ one can also extract them from \eqref{eq:Minkcorrrhoi}, obtaining that in \eqref{vaclin2} now we have 
\begin{eqn}
 \varepsilon_a  =\p_a \left( \frac{f}{4\CK_L}\right) - 6\, \eps_L\cK_L\frac{\cK_a}{g_{LL}\cK^2}  \stackrel{\eqref{limitL}}{\to} g_a^\infty - \frac{27}{4} K^{(3)}_L  \frac{t_L}{\cK_L} \frac{\cK_{L a}}{\cK_L} \, .
\end{eqn}
Here we have defined $ g_a^\infty$ as the asymptotic behaviour of $\p_a \left( \frac{f}{4\CK_L}\right)$ along the limit \eqref{limitL}. Notice that the second term asymptotes as $ \frac{\cK_{L a}}{\cK_L} \to 0$, and so the qualitative behaviour of the system depends on the functional behaviour of $g_a^\infty$. We have two possibilities:
\begin{itemize}
 
    \item[-] If $g_a^\infty \to 0$ for some $a$, then \eqref{Nfluxineqlin} will set an upper bound on this limit. If moreover the $K_L^{(3)}$ correction dominates over $g_a$, then  the bound will be similar to \eqref{boundIIB2}. Indeed, from \eqref{Nfluxineqlin2} we then obtain
  \begin{eqn}
 \frac{9}{8} |K^{(3)}_L|  \frac{\cK_{L a}}{\cK_L} \gtrsim N_{\rm flux}^{-2} \, .
\end{eqn}

    \item[-] If for all $a$, $g_a^\infty$ tends to a finite number bigger than $N_{\rm flux}^{-1}$, then \eqref{Nfluxineqlin} is automatically satisfied and no bound is imposed on the saxion vevs in order to find vacua in this region. This is for instance the case of the overall rescaling  $t^a  \to \lambda t^a, \lambda \to \infty$, since due to the homogeneity of $f$ and $\CK_L$ all the $g_a^\infty$ tend to quotients of intersection numbers. Therefore for sufficiently large values of $N_{\rm flux}$ \eqref{Nfluxineqlin} becomes trivial. Notice that in this case the monodromy-invariant flux combination $N_a$ only scans a finite number of values along the limit, and so the set of inequivalent flux vacua in this regime should be finite.
    
\end{itemize}
A potential third possibility would be that $g_a \to \infty, \forall a$, which would also imply that the bound \eqref{Nfluxineqlin} is automatically satisfied and that all possible values of $N_a$ are scanned along the limit, yielding an infinity of flux vacua. However, this scenario is not realised here. To see this, we note that the limit \eqref{limitL} requires that we have to blow up (some of) the saxions $t^a$, possibly at different rates. Now, there is always (at least) one saxion $t^*$ that grows the fastest. Due to the fibrations structure of the mirror $X_4$ the terms appearing in $f$ are determined by the intersection numbers $\cK_{Labc}$ of the fibre $X_3$ and the details of the twist of $X_3$ over $\mathbb{P}^1$. As a consequence in the limit \eqref{limitL} we can estimate 
\begin{align}
    f \lesssim t^* \cK_L\,.  
\end{align}
With this information, we can now evaluate the component $|g_*^\infty|$  as 
\begin{align}
    4|g_*^\infty|=\left|\partial_* \left(\frac{f}{\cK_L}\right)\right| \leq \left|\frac{\partial_* f}{\cK_L} \right| + \left|\frac{f \partial_* \cK_L}{\cK_L^2}\right| \lesssim \left|\frac{\cK_L}{\cK_L} \right|+ \left|\frac{t^* \cK_L \partial_* \cK_L}{\cK_L^2}\right| \,.
\end{align}
We can further use $t^* \partial_* \cK_L \lesssim \mathcal{O}(\cK_L)$ to see that the second term on the RHS is finite in the limit \eqref{limitL}. Given that the first term on the RHS is $\mathcal{O}(1)$ we find that at least $|g_*^\infty|$ can never diverge along \eqref{limitL}, and so the possibility $g_a\rightarrow \infty$, $\forall a$ cannot be realised. 

\subsubsection*{Beyond large complex structure}

As we have seen, the linear scenario is quite natural in the context of F-theory four-fold compactifications at large complex structure, and one may construct several explicit examples like the one discussed in section \ref{sec:counter}.  A natural question is then if the same setup can be realised along other limits of infinite distance within the complex structure field space. To address this question let us extract the key features and the underlying geometric picture that lies behind the linear scenario, in order to connect with the results of \cite{Grimm:2019ixq}, where techniques were developed to address the features of flux potentials along general infinite distance limits.

For this, notice that the leading-order saxion-dependent matrix \eqref{ZABlin} is of the form
\be
\label{ZABTC}
2{\cal V}_3^2 Z + \chi_0 = 
 \begin{pmatrix}
H & &  &  & & &  \\
& M & M \varepsilon_a  & &  &   &  \\ 
& M \varepsilon_a & H_{ab}   &  &  &  & \\
 & & &   g_{\mu\nu}  & & &  \\
& &    & &   H^{ab} &   -H^{ab}\varepsilon_a &  \\
&  &  & &  -H^{ab}\varepsilon_a & M^{-1} + H^{ab}\varepsilon_a \varepsilon_b&  \\
 & & &    & & & H^{-1} \\
\end{pmatrix} \, ,
\ee
where $\chi_0$ is defined in \eqref{eq:chi0}, and $H^{ac}H_{cb} = \delta^a_b$. From the results of appendix \ref{ap:georho}, we can interpret this matrix as the Hodge star action on the basis of four-forms $ \{ \tilde\a, \tilde\a_i, \tilde\sigma_\mu, \tilde\b^i, \tilde\b\}$ in which the component of $G_4$ are the flux-axion polynomials $\rho_A$, see eq.\eqref{ap:G4rho}. In \eqref{ZABTC} this action is block-diagonal, which is a general feature of the large complex structure regime, cf.\eqref{ZAB}. In fact, it follows from the results of \cite{Grimm:2019ixq} that the Hodge star action is approximately block-diagonal in any complex structure region in the vicinity of an infinite distance point.

Now, it is also a general result of \cite{Grimm:2019ixq} that as we approach an asymptotic region in complex structure field space, the different blocks in the Hodge star action behave differently. Some of them tend to infinity and some of them tend to zero, while the rest remain of finite order. In the linear scenario we have that $H, H_{ab} \to \infty$, $H^{-1}, H^{ab} \to 0$, and $M$ remains of order one. To find vacua one then needs to set $\tilde{\rho} = \tilde{\rho}^a =0$, which implies the flux constraint $m = m^a =0$. Finally, it is reasonable to assume that  it is consistent with the vacua equations to set to zero some of the fluxes $\hat{m}^\mu$, in such a way that $\eta_{\mu\nu} \hat{m}^\mu\hat{m}^\nu =0$. The only contribution to the tadpole is then
\be
N_{\rm flux} = - m^L\bar{e}_L\, , 
\ee
which is independent of $h^{3,1}(Y_4)$. This leads to a vector of flux-axion polynomials of the form $\vec{\rho}^{\, t} = \left(0, m^L, 0, \hat{\rho}^a, 0,  \bar{\rho}_a, \bar{e}_L, \bar{\rho} \right)$, from where the equations of motion follow
\begin{eqn}
\bar \rho = 0\, ,\qquad \hat{\rho}^a   =  0\, , \qquad \bar{\rho}_a = \varepsilon_a  \bar{e}_L \, , \qquad \bar{e}_L + M m^L  & = 0\, .
\label{vacuagen}
\end{eqn}
From here one obtains that $M \in (N_{\rm flux}^{-1}, N_{\rm flux})$, and the inequalities
\begin{eqn}
N_{\rm flux} |\varepsilon_a| \gtrsim 1\ \implies \ M|\varepsilon_a| \gtrsim N_{\rm flux}^{-2}\, .
\label{Nfluxineqgen}
\end{eqn}
The relevance of these bounds depends on the asymptotic behaviour of the $\varepsilon_a$ along each limit. By the results of \cite{Grimm:2019ixq} one would expect that $\varepsilon_a$ either tends to zero, increasing the number of blocks in which the Hodge star action is divided, or it remains finite. If all $\varepsilon_a$ tend to zero, then we recover a bound for the saxion vevs, just as in the \textbf{IIB2} scheme of section \ref{sec:moduliIIB}. If they do not, there is a priori no bound for the saxion vevs, but the values that the monodromy-invariant flux bilinear $N_a$ can take is limited, and so should be the number of inequivalent flux vacua. 

As we depart from the large complex structure region, some of the entries of \eqref{ZABTC} will stop being zero, and the above block-diagonal structure will be further broken. A clear example of this is the effect of $K^{(3)}$ corrections in the \textbf{IIB2} scheme, that besides generating a non-vanishing $\varepsilon_a$, induce additional non-vanishing off-diagonal entries in \eqref{ZABTC}. However, in that case such additional corrections do not deform significantly the set of vacua equations \eqref{vacuagen}, as can be appreciated from \eqref{eq:MinkcorrIIB2}. As a result, this moduli stabilisation scheme can be taken to be valid on a large region of complex structure field space. Whether this last feature is also present along limits outside of the large complex structure regime is yet to be seen, although the robustness of the equations in the \textbf{IIB2} setup suggests that this could well be the case.


\section{Examples}
\label{s:examples}

As stressed in section \ref{s:potential}, the most subtle part of the flux potential is the piece related to the four-forms $\sigma^\mu$, whose basis is not known in general. Exceptions to this  are four-folds $Y_4$ whose mirror dual $X_4$ is a smooth fibration, of which the setups in sections \ref{s:IIB} and \ref{s:linear} are particular subcases. In this section we provide explicit constructions that illustrate our previous results, by considering two types of fibrations for $X_4$. In section \ref{sec:elliptic} we apply our framework to the case in which $X_4$ is an elliptic fibration, which is a natural generalisation of the type IIB case. In section  \ref{sec:twofield} we study a concrete two-field model of this setup, and show how the bounds for the saxion vevs obtained in section \ref{sec:moduli} are realised in practice. Section \ref{sec:counter} considers a four-fold $X_4$ that is a fibration of a Calabi--Yau three-fold over $\mathbb{P}^1$, yielding a concrete realisation of the linear scenario of section \ref{s:linear}. This illustrates how our general formulas apply to specific geometries, while a more detailed  description of the set of flux vacua for each case is left for the future.

\subsection{Elliptically fibered mirror}
\label{sec:elliptic}

A natural generalisation of the type IIB limit is given by Calabi--Yau four-folds $Y_4$ whose mirror $X_4$ is a smooth, elliptically fibered four-fold with a section. In this case all the topological invariants of $X_4$ are determined by the three-fold base $B_3$ and so, as pointed out in \cite{Cota:2017aal}, one has explicit control over the set of four-forms $\sigma^\mu$. In our language, this allows us to determine the intersection numbers $ \zeta_{\mu, ij}$ explicitly, specify the form of the flux potential, and to carry out our analysis with the same degree of detail as in the type IIB limit. 

To see how this works, let us construct explicitly a basis of holomorphic $2p$-cycles classes in the mirror four-fold $X_4$, as done in the type IIB case. On the three-fold base $B_3$ of $X_4$, a basis of holomorphic $2p$-cycles is given by the point class $\mathcal{O}_{pt}$, the generators of the Mori cone  $[\mathcal{C}'^a]$, $a = 1, \dots, h^{1,1}(B_3) = h^{1,1}(X_4)-1$, the divisors classes $[D_a']$ that generate the K\"ahler cone, and the class of $B_3$. The relevant topological invariants for us will be the triple intersection numbers and the first Chern class of $B_3$:
\be
\kappa_{abc} = [D_a']\cdot [D_b']\cdot [D_c']\, , \quad \text{and} \quad c_1(B_3) = c_1^a  [D_a']\, .
\ee
We embed the holomorphic cycles of $B_3$ into $X_4$ by using the projection of the fibration $\pi$ and the divisor class of the section $[E]$. In particular, the  Mori cone of $X_4$ is generated by
\be
[\mathcal{C}^a] = [E . \pi^{-1} (\mathcal{C}'^a)] \, , \qquad [\mathcal{C}^0]\, ,
\ee
with $[\mathcal{C}^0]$ the class of the fibre. The K\"ahler cone is generated by the dual basis of divisor classes 
\be\label{kahlerconeelliptic}
[D_a] =\pi^*[D_a']\, , \qquad [D_0] = [E] + \pi^* c_1(B_3)\, .
\ee
Similarly to the type IIB case, we can construct a set of holomorphic four-cycle classes as 
\be
[\gamma_{ij}]  = [D_i . D_j] \, , \quad i = \{0, a\}\, .
\label{gammafib}
\ee
Again, all holomorphic four-cycle classes can be generated from linear combinations of $[\gamma_{ij}]$, but \eqref{gamma} does not form a basis because it is not a linearly independent set. For the case at hand, one can construct such a basis from
\be
  [H_a] = [D_0 . \pi^{-1}(D_a)] \,, \qquad [H_{\hat{a}}] = \pi^{*}[\mathcal{C}^a] \, ,
  \label{4formbasisfib}
\ee
which reduces to \eqref{4formbasis} when the fibration is trivial. This is a different choice of basis compared to the one taken in \cite{Cota:2017aal}, but more convenient for our purposes. The integral basis of four-form classes $[\sigma_\mu]$ that correspond to the period \eqref{period4} is then given by $\{[\sigma_\mu]\} = \{[H_a], [H_{\hat{a}}]\}$, and so $\mu = \{a, \hat{a}\}$, with $a, \hat{a} = 1, \dots, h^{1,1}(B_3)$. Notice that the number of elements of the basis \eqref{gammafib} is $2h^{1,1}(X_4) -2$, smaller than the $\half h^{1,1}(X_4) (h^{1,1}(X_4)+1)$ elements in \eqref{gammafib}. The tensor $\zeta^\mu_{ij}$ connecting both sets of four cycles as
\be
[\gamma_{ij}] = \zeta^\mu_{ij} [\sigma_\mu] = \zeta^a_{ij} [H_a] + \zeta^{\hat{a}}_{ij} [H_{\hat{a}}]\, ,
\ee
is specified by
\begin{align}
\zeta^a_{0b} = \zeta^a_{b0}  = \delta_{ab}\, , \qquad   \zeta_{a,bc}\equiv \zeta^{\hat{a}}_{bc}\eta_{\hat{a}a}= \kappa_{abc}\, ,\qquad \zeta_{00}^a = c_1^a \,,
    \label{Koszulfib}
\end{align}
with all remaining components vanishing. This clearly reduces to \eqref{Koszul} for $c_1^a =0$, and one can check that it satisfies the relation \eqref{interrel}. The intersection matrix for the basis \eqref{4formbasisfib} is given by 
\begin{align}\label{etafibration}
    \eta_{\hat{a}\hat{b}}=0\,,\qquad \eta_{a\hat{b}} = \delta_{a\hat{b}} \,,\qquad \eta_{ab}=  \kappa_{abc}c_1^c \equiv c_{ab} \,,
\end{align}
and so applying  \eqref{interrel} we recover
\begin{eqn}
    \cK_{0abc}&= \kappa_{abc}\,,\qquad \qquad\qquad \cK_{00ab}= \kappa_{abc} c_1^c\equiv c_{ab} \,,\\
    \cK_{000a}&= \kappa_{abc} c_1^b c_1^c\equiv c_a \,,\qquad \cK_{0000}=\kappa_{abc}c_1^a c_1^b c_1^c\equiv c\, ,
    \label{c1rel}
\end{eqn}
which indeed are  the quadruple intersection numbers of the elliptically fibered four-fold $X_4$. Furthermore, for a $X_4$ a smooth Weierstrass model the Euler characteristic $\chi(X_4)$ can be calculated from the adjunction formula as 
\begin{align}
    \chi(X_4) = \int_{B_3} \left[12 c_1(B_3)\wedge c_2(B_3) + 360 \, c_1(B_3)\wedge c_1(B_3) \wedge c_1(B_3) \right]\, ,
\end{align}
which also gives the Euler characteristic for the mirror $Y_4$. In the mirror four-fold $Y_4$ \eqref{Koszulfib} translates, via \eqref{hatm}, into the following dictionary for the set of $G_4$-flux quanta 
\be
{m}_a \equiv \delta_{a\hat{b}} \hat{m}^{\hat{b}} = \oh  \kappa_{abc} m^{bc}\, , \qquad \hat{m}^a = m^{0a} + \oh c_1^a m^{00}\, ,
\label{hatmfib}
\ee
which are the generalisation of the type IIB fluxes $m_a$, $\hat{m}^a$ to the present case and  the actual $G_4$-flux quanta,\footnote{Because the mirror manifold $X_4$ is a smooth elliptic fibration the quantisation condition for the $G_4$ flux \cite{Witten:1996md} is trivial, in the sense that $[G_4]$ must be an integer class \cite{Collinucci:2010gz}. In the present setup this implies that $m_a, \hat{m}^a \in \Z$. In fact, all flux quanta in \eqref{G4} should be integers when $X_4$ is a smooth elliptically fibered four-fold.} while $m^{ij}$ should be seen as auxiliary quanta. The superpotential then reads
\begin{align}
    W^{\rm corr}=&\, \bar{e}+\bar{e}_i T^{i}+\frac{1}{2}\kappa_{abc}\bar{m}^cT^bT^c+T^0 c_{ab}\bar{m}^a T^b+\frac{1}{2} (T^0)^2c_a\bar{m}^a+T^0T^a\bar{m}_a+\frac{1}{2}(T^0)^2 c_1^a\bar{m}_a\nonumber\\
    &+\frac{1}{6}m^0\kappa_{abc}T^aT^bT^c+\frac{1}{2}T^0\kappa_{abc}m^aT^bT^c+\frac{1}{2}m^0T^0c_{ab}T^aT^b+\frac{1}{2}(T^0)^2c_{ab}m^aT^b\nonumber\\
    &+\frac{1}{2}m^0 (T^0)^2c_{a}T^a+\frac{1}{6}(T^0)^3c_{a}m^a+\frac{1}{6}m^0(T^0)^3c  \nonumber\\
    &+\frac{m}{24}\left(c(T^0)^4+4(T^0)^3c_a T^a+6(T^0)^2c_{ab}T^aT^b+4T^0\kappa_{abc}T^aT^bT^c\right) \nonumber \\
    & -iK_{i}^{(3)}(m^i+mT^i)\, ,
\end{align}
where we have included the polynomial corrections of section \ref{sec:poly}, and in particular the flux redefinition \eqref{eq:fluxshift}. Similarly, the corrected K\"ahler potential is 
\begin{equation}
    K_{\rm cs}^{\rm corr}=-\log\left(\frac{2}{3}(4t^0\kappa+6(t^0)^2\kappa_ac^a_1+4(t^0)^3\kappa_{ab}c^a_1c^b_1+(t^0)^4c)+4K_i^{(3)}t^i\right)\, .
\end{equation}

It then follows from our general analysis that we recover a scalar potential of the form \eqref{bilinear}, where the flux-axion polynomials are given by 
\begin{align}
 \nonumber \bar{\rho} =&\, \bar{e} + \bar{e}_0 b^0+  \bar{e}_ab^a+ \bar{m}_a\left(b^a+c_1^a b^0\right) b^0 + \frac{1}{2}  \kappa_{abc} \hat{m}^{a} b^bb^c + \frac{1}{2} \kappa_{abc} \hat m^a c_1^bb^0 (b^c+ c_1^cb^0) \\
 \nonumber &+\frac{1}{6}\kappa_{abc}\left(3 m^ab^bb^cb^0 +3m^a c_1^b (b^0)^2 b^c+m^a c_1^b c_1^c (b^0)^3 +m^0(b^a+c_1^ab^0)(b^b+c_1^bb^0)(b^c+c_1^c b^0)\right)\\ 
 \nonumber&+\frac{m}{24} \kappa_{abc}\left(4b^ab^bb^c b^0+6b^a b^b c_1^c(b^0)^2 +4 b^ac_1^b c_1^c (b^0)^3 + c_1^a c_1^bc_1^c (b^0)^4\right)\,,\\
\nonumber \bar{\rho}_0=& \,\bar{e}_0+\bar{m}_a(b^a +c_1^ab^0)+\kappa_{abc}\hat m^ac_1^b(b^c+c_1^cb^0) +\frac{1}{2}\kappa_{abc} \left(m^a+c_1^am^0\right)(b^b+c_1^b b^0)(b^c+c_1^cb^0) \\
\nonumber &+\frac{m}{6}(b^a+c_1^a b^0)(b^b+c_1^b b^0)(b^c+c_1^cb^0) \,,\\
\nonumber \bar{\rho}_a=& \,\bar{e}_a + \bar{m}_ab^0 +  \kappa_{abc} \hat{m}^{b}\left(b^c + c_1^cb^0\right)  + \kappa_{abc} \left(m^b b^0 \left(b^c + \frac{1}{2} c_1^c b^0\right) +\frac{1}{2}m^0 (b^b+b^0c_1^b)(b^c+b^0c_1^c)\right)\\
 &+ \frac{m}{6}\kappa_{abc}\left( 3b^b b^c b^0+3(b^0)^2 c_1^b b^c +(b^0)^3 c_1^b c_1^c\right)\,,\\
\nonumber \bar{\rho}_a^\prime=& \,\bar{m}_a +  \kappa_{abc} \left(m^b b^c + m b^b b^c\right) \,,\\
\nonumber \bar{\rho}^{a}=&\, \bar{m}^{a} + m^a b^0 +m^0 b^a+ c_1^a m^0 b^0 + m \left(b^0 b^a +\frac{1}{2} c_1^a (b^0)^2\right)  \,,\\
\nonumber \tilde \rho^a=&\, m^a + m b^a \,, \\
\nonumber \tilde \rho^0=&\,m^0 + mb^0 \,, \\
\nonumber \tilde \rho=&\,m\, ,
\end{align}
and the saxion-dependent matrix reads, in the limit where the corrections $K^{(3)}_i$ can be neglected
\begin{align}
\label{ZABfib}
Z^{AB} =
\frac{e^K\cK}{3}
\begin{pmatrix}
\frac{\cK}{24} & & & & & -1 \\
&  \frac{\cK}{6} g_{ij} & & &  \d^i_j &  \\
& &   \tilde{B}_a{}^c\tilde{A}_{cd}\tilde{B}^d{}_b  & \tilde{B}_a{}^b & & \\
& & \tilde{B}^b{}_a & \tilde{A}^{ab} &  & \\
&  \d^i_j & & & \frac{6}{\cK} g^{ij} &  \\
-1 & & & & & \frac{24}{\cK} \\
\end{pmatrix} \, ,
\end{align}
with $\vec{\rho}^{\, t} = \left(\tilde{\rho},  \tilde{\rho}^i,   \bar{\rho}^{a},  \bar{\rho}_a^\prime, \bar{\rho}_i,    \bar\rho   \right)$. Here we have separated the tensor $g_{\mu\nu} -\eta_{\mu\nu}=  \frac{6}{\cK}  g^{ij}_{P}\zeta_{\mu i}\zeta_{\nu j}$   in \eqref{ZAB} into four blocks, reflecting the  splitting $\mu = \{a,\hat{a}\}$.  In particular, the matrices that appear in \eqref{ZABfib} are related to the metric $g_{\mu\nu}$ defined below \eqref{ZABdiag} by
\begin{equation}
   \tilde{A}^{ab}\delta_{a\hat{c}}\delta_{b\hat{d}}= {g}_{\hat{c}\hat{d}}\, , \qquad \tilde{B}_a{}^b\delta_{b\hat{c}}= {g}_{a\hat{c}} - {\delta}_{a\hat{c}}\, ,
   \label{eq: gmunu fib}
\end{equation}
and their explicit form is
\begin{align}
\label{tildeA}
  \tilde{A}^{ab}=&-2\left[\cK^{00}\left(t^at^b+t^0(t^ac_1^b+t^bc_1^a)+(t^0)^2c_1^ac_1^b\right)+\cK^{0a}t^0(t^b+t^0c_1^b)+\cK^{0b}t^0(t^a+t^0c_1^a)\right.\nonumber\\
    &\left.+\cK^{ab}(t^0)^2\right]+\frac{2(t^0)^2}{\cK}\left[4t^at^b+2t^0(t^ac_1^b+t^bc_1^a)+(t^0)^2c_1^ac_1^b\right]\, , \\
   \tilde{B}^b{}_a =&\   -2\left[\cK^{00}\left(\kappa_{ac}c_1^ct^b+t^0(c_at^b+\kappa_{ac}c_1^cc_1^b)+(t^0)^2c_ac_1^b\right)+\cK^{0b}t^0(\kappa_{ac}c_1^c+c_at^0)\right.\nonumber\\
    &\left.+\cK^{0c}\left(\kappa_{ac}t^b+t^0(\kappa_{ac}c_1^b+c_{ac}t^b)+(t^0)^2c_{ac}c_1^b\right)+\cK^{bc}t^0(\kappa_{ac}+t^0c_{ac})\right]\nonumber\\
    &+\frac{2t^0}{\cK}\left[2\kappa_a t^b+(t^0)\left(4\kappa_{ac}c_1^ct^b+\kappa_ac_1^b\right)+(t^0)^2\left(2\kappa_{ac}c_1^cc_1^b+2c_at^b\right)+(t^0)^3c_ac_1^b\right]\, .
  \end{align}
Finally, $\tilde{A}_{ac}\tilde{A}^{cb} = \delta_a^b$, from where the structure \eqref{blockZ} is manifest.

\subsubsection*{Moduli stabilisation}

Let us now write down the Minkowski vacuum equations for the case at hand, and study to what extent the results from the Type IIB orientifold limit generalise to this class of compactifications. In this setup the on-shell conditions \eqref{eq:Mink} become
\begin{subequations}
    \label{eq:Mink elliptic compact}
\begin{empheq}[box=\widefbox]{align}
 \bar\rho &= \frac{1}{24}\mathcal{K}\tilde{\rho}   \\
  \bar{\rho}_i&=-\frac{1}{6}\mathcal{K}g_{ij}\tilde{\rho}^j    \\
  \bar \rho_a'&= \Gamma_{ab} \bar \rho^b  \label{eq:Mink elliptic compact rho mu}
\end{empheq}
\end{subequations}
where we have defined $\Gamma_{ab} \equiv - \tilde{A}_{ac}\tilde{B}^c{}_b$. An explicit expression for this matrix is given in appendix \ref{ap:elliptic}, from where one can see that for vanishing $c_1^a$, $\Gamma_{ab}  \to \frac{2}{3} \frac{\kappa}{t^0} g_{ab}^\kappa$, and  we recover \eqref{eq:MinkIIB}.  Using the vacuum equations we can rewrite the  flux contribution to the tadpole as
\begin{align}
\label{eq: elliptic fibered flux}
    N_{\rm flux} = \bar{\rho} \tilde{\rho} - \bar{\rho}_i \tilde{\rho}^i + \frac{1}{2} \eta_{\mu\nu} \bar{\rho}^{\mu} \bar{\rho}^{\nu}
    \stackrel{\rm vac}{=} \frac{\mathcal{K}}{24}\left(\tilde{\rho}^2+4g_{ij}\tilde{\rho}^i\tilde{\rho}^j\right)+\frac{1}{2}\left(c_{ab} + \Gamma_{ab} + \Gamma_{ba} \right) \bar{\rho}^a\bar{\rho}^b\, ,
\end{align}
where the last term is positive definite by construction, as it equals $\oh g_{\mu\nu}\bar{\rho}^{\mu} \bar{\rho}^{\nu}$.  Hence, as in the type IIB case, in order not to overshoot the D3-brane tadpole we have to set $\tilde \rho=m=0$ and we further set $\tilde \rho^a=m^a=0$. However, unlike in the type IIB case the saxion $t^0$ now enters with a fourth power in $\cK$. Thus, based on our general discussion in section \ref{s:vacua}, we also need to demand $\tilde \rho^0=m^0=0$ to find vacua that do not violate the tadpole constraint at large complex structure.   

As before, including the corrections $K_i^{(3)}$ will modify the vacuum equations. At linear order in these corrections we have \eqref{eq:Minkcorr} adapted to this setup, which reads:
\begin{subequations}
\label{eq:Minkcorr elliptic}
\begin{align}
&\bar\rho- \frac{1}{24}\mathcal{K}\tilde{\rho}  =  - \frac{3}{8}\eps_it^i \left[ \frac{\cK}{18}\tilde{\rho} +  \varpi \right] \, , \\
& \bar{\rho}_i+\frac{1}{6}\mathcal{K}g_{ij}\tilde{\rho}^j   =   \frac{1}{3}  \CK_i  \left( \eps_j -  \epsilon_k t^k \frac{ \cK_j}{\cK} \right)\tilde{\rho}^j   - \frac{1}{6}\epsilon_i \cK_j \tilde \rho^j   \, , \\
 &\bar{\rho}_a'- \Gamma_{ab} \bar{\rho}^b  = \frac{E_a}{4t^0} \left[\frac{\cK}{2}\tilde{\rho}+ \varpi\right] \, ,
 \end{align}
 \end{subequations}
where we have defined $\varpi = (2t^at^0+(t^0)^2c_1^a)\bar{\rho}_a'+(\kappa_a+2\kappa_{ab}c_1^bt^0+c_a(t^0)^2)\bar{\rho}^a$ and
\begin{equation}
    E_a=\left[\epsilon_b-\frac{\cK_b}{(\cK-2\cK_0 t^0)}\left(2\epsilon_c t^c - \epsilon_i t^i\right)\right]\left[\delta_a^b-\frac{\cK_a c_1^b t^0}{\cK-2\cK_0 t^0+\cK_a c_1^a t^0}\right] \, .
\end{equation}
Let us now turn to the restricted flux scenario $m=m^i=0$ which yields $\tilde{\rho}=\tilde{\rho}^i=0$, $\bar{\rho}^a=\hat{m}^a$ and $\bar{\rho}_a'= {m}_a$. In this case \eqref{eq:Minkcorr elliptic} reduces to
\begin{subequations}
\label{eq: minkcorr elliptic simplify}
\begin{align}
 \bar\rho &=- \frac{3}{8}\eps_it^i \, \varpi \, , \label{eq: minkcorr elliptic simplify bar rho}\\
  \bar{\rho}_i& =0\, ,\\
 m_a - \Gamma_{ab}  \hat{m}^b &= \frac{1}{4} \left(\eps_a -\frac{\cK_a(\epsilon_ct^c-\epsilon_0t^0+t^0\epsilon_bc_1^b)}{\cK-2\cK_0t^0+\cK_ac_1^at^0}\right)   \varpi \, .
 \label{eq: minkcorr elliptic simplify rho mu}
 \end{align}
\end{subequations}
In order to stabilise all complex structure fields, we need to choose the flux quanta $(m_a, \hat m^a)$ such that the matrix $M$ defined in \eqref{Meq} is invertible. In the present case of a smooth elliptic fibration the matrix $M$ is given by 
\begin{align}\label{Mellipticfibration}
 M = \left(\begin{matrix} M_{00} & M_{0a} \\ M_{a0} & M_{ab} \end{matrix} \right) =  \left(\begin{matrix} c_1^a m_a + c_{a}\hat m^a & m_a + c_{ab}\hat m^b \\ m_a + c_{ab}\hat m^b  & \kappa_{abc} \hat m^b \end{matrix} \right) \,. 
\end{align}
To see whether this matrix is invertible, let us define the matrices $S_{ab} \equiv \kappa_{abc} \hat m^b$ and
\begin{align}
    \tilde S_{ab} = S_{ab} - \frac{\left(m_a + c_{ac}\hat m^c\right) \left(m_b + c_{bd}\hat m^d\right)}{c_1^c m_c +c_c\hat m^c}\,.
\end{align}
Now the block-matrix \eqref{Mellipticfibration} is invertible if one of the two is fulfilled 
\begin{eqn}\label{invertibleMelliptic}
    a)&:\qquad S_{ab}\; \;\text{invertible} \qquad\qquad \text{and} \qquad m_a S^{ab} m_b + c_1^a m_a \ne 0 \,,\\
    b)&: \qquad c_1^c m_c +c_c\hat m^c \ne 0 \,,\;\qquad \text{and} \qquad \tilde S_{ab}\; \;\text{invertible}\,. 
\end{eqn}

The solution \eqref{splitmu} now reads 
\begin{equation}
    m_a = A\kappa_a + C_a + \cO(\eps_i) \, , \qquad \hat{m}^a = A\left(2t^at^0+(t^0)^2c_1^a\right) + C^a + \cO(\eps_i)\, , \label{eq: ansatz elliptic fib}
\end{equation}
with the coefficients $C_a$ and $C^a$ satisfying
\begin{equation}
    C_at^0=-(\kappa_{ab}+c_{ab}t^0)C^b\, , \qquad (c^a\kappa_{ab}t^0+\kappa_b)C^b=0\, .
\end{equation}
Then \eqref{eq: minkcorr elliptic simplify bar rho} allows us to recover \eqref{astiA}
\begin{equation}
    A=-\frac{4\bar{\rho}}{9 K^{(3)}_i t^i}\, .
\end{equation}
In addition, \eqref{eq: minkcorr elliptic simplify rho mu} simplifies to
\begin{equation}
    m_a- \Gamma_{ab} \hat{m}^b=\frac{A}{4}\left(\eps_a -\frac{\cK_a(\epsilon_ct^c-\epsilon_0t^0+t^0\epsilon_bc_1^b)}{\cK-2\cK_0t^0+\cK_ac_1^at^0}\right)\, ,
\end{equation}
and \eqref{eq: elliptic fibered flux} becomes
\begin{equation}
    N_{\rm flux}=\frac{1}{2}A^2\cK-\frac{1}{t^0}(\kappa_{ab}+\frac{1}{2}c_{ab}t^0)C^aC^b+\mathcal{O}(\epsilon_i)\geq \frac{1}{2}A^2\cK +\mathcal{O}(\epsilon_i)\, .
\end{equation}
At this point we may apply the reasoning below \eqref{Meq} to obtain the inequality $N^p_{\rm flux}\bar{\rho}\gtrsim d^{2p-1}$ with $d=\textrm{gcd}(m_a,\hat{m}^a)$ and $p\leq h^{(3,1)}$. Hence we conclude that
\begin{equation}
\cK <   N_{\rm flux}^{2p+1}d^{2-4p}  (K_i^{(3)}t^i)^2 \, .
\end{equation}

\subsection{A two-field model}
\label{sec:twofield}

As a concrete example of a Calabi--Yau four-fold $Y_4$ for which the mirror $X_4$ is elliptically fibered, let us consider $X_4$ to be the degree 24 hypersurface in $\mathbb{P}^{5}_{(1,1,1,1,8,12)}$ and denote it by $X_{24}$. This manifold has been studied in the context of moduli stabilisation for instance in \cite{Cota:2017aal}. This hypersurface can be viewed as an elliptic fibration over $\mathbb{P}^3$ with intersection polynomial 
\begin{align}
    I(X_4) = 64 D_0^4 + 16 D_0^3 D_1 + 4 D_0^2 D_1^2 + D_0D_1^3\,,
\end{align}
where $D_0$ is the K\"ahler cone divisor associated to the zero section $E$, $[D_1]=\pi^*[H]$ the pull back of the hyperplane class in $\mathbb{P}^3$ and $c_1(\mathbb{P}^3)=4H$. For this four-fold we have the following basis of four-cycles 
\begin{align}
    [H_1] = [D_0 . D_1]\,,\qquad [H^1]= \pi^*[\mathcal{C}^1]\,,
\end{align}
with $\mathcal{C}^1$ the single Mori cone generator of $\mathbb{P}^3$. The non-vanishing components of the tensor $\zeta^\mu_{ij}$ as in \eqref{Koszulfib} are thus given by
\begin{align}
    \zeta^{1}_{01} = 1 \,,\qquad \zeta_{1,11}= \kappa_{111}=1\,,\qquad \zeta^1_{00}=c_1^1=4\,,
\end{align}
and the intersection matrix $\eta$ reduces to 
\begin{align}
    \eta^{11}=0\,,\qquad \eta^1_1=1\,,\qquad \eta_{11}=4\,.
\end{align}
Furthermore, the corrections $K^{(3)}_i$ for this example are given by 
\begin{align}
    K^{(3)}_0 = -3860\frac{\zeta(3)}{(2\pi)^3}\,,\qquad K^{(3)}_1=  -960\frac{\zeta(3)}{(2\pi)^3}\,. 
\end{align}
Finally, the Euler number of $X_{24}$ and its mirror $Y_4=X_{24}^*$ is $\chi(X_{24})=\chi(X_{24}^*)= 23328$. With this preparation we can now look at flux vacua for F-theory on  $X_{24}^*$ in the large complex structure regime. To find vacua for large values of the saxions $t^0, t^1$ we restrict to the  flux vector 
\begin{align}
    \vec{q}^t = (0,0,0, \hat m^1, m_1, \bar e_1, \bar e_0, \bar e)\,. 
\end{align}
The vacuum equations $\bar \rho_i=0$ then translate to 
\begin{align}
    \bar e_0 + m_1\left(b^1 +4b^0\right) +4\hat m^1\left(b^1 +4b^0\right)  =0 \,,\qquad \bar e_1 +\hat m^1\left(b^1+4b^0\right) +m_1 b^0=0\,, 
\end{align}
such that the matrix $M$ in \eqref{Mellipticfibration} is given by 
\begin{align}
    M = \left(\begin{matrix} 4m_1 +16\hat m^1 & m_1 + 4\hat m^1 \\ m_1 +4\hat m^1 & \hat m^1  \end{matrix}\right)\,,
\end{align}
which is invertible provided $m_1+4\hat m^1 \ne 0$ and $m_1\ne 0$. In case this is fulfilled we obtain
\begin{align}
    b^0&=\frac{\hat m^1 \bar e_0}{m_1 \left(4\hat m^1 +m_1\right)}-\frac{\bar e_1}{m_1}\,,\\
    b^1&=-4\frac{\hat m^1 \bar e_0}{m_1 \left(4\hat m^1 +m_1\right)} -\frac{\bar e_0}{4\hat m^1+m_1}+\frac{4\bar e_1}{m_1}\,. 
\end{align}
From here we can deduce 
\begin{align}\label{barrhoX24}
\bar \rho = \frac{2\left(4\hat m^1+m_1\right)m_1 \bar e - m_1 \bar e_1 \bar e_0 + \hat m^1\bar e_0\left(\bar e_0 -4\bar e_1\right)+3\bar e_0\bar e_1\left(4\hat m^1+m_1\right)}{2\left(4\hat m^1+m_1\right)m_1} \,, 
\end{align}
for which the numerator is a combination of integer fluxes and thus at least of $\mathcal{O}(1)$ if non-vanishing. 
We can further use \eqref{eq:Mink elliptic compact rho mu} to solve the vacuum equations for $\rho^\mu$ at leading order. For our particular two-modulus case we have
\begin{align}
    m_1 = \Gamma_{11} \hat m^1\,,
\end{align}
with
\begin{align}
    \Gamma_{11} = \frac{(t^1)^4+ 4t^0(t^1)^3}{2(t^1)^3t^0+12(t^1)^2 (t^0)^2+ 16 (t^0)^3t^1}\,. 
\end{align}
The corrected equations of motion for $\bar \rho$ now give
\begin{align}\label{eqofmbarrhoX24}
\frac{2\left(4\hat m^1+m_1\right)m_1 \bar e - m_1 \bar e_1 \bar e_0 + \hat m^1\bar e_0\left(\bar e_0 -4\bar e_1\right)+3\bar e_0\bar e_1\left(4\hat m^1+m_1\right)}{2\left(4\hat m^1+m_1\right)m_1} = -\frac{3}{8}\epsilon_i t^i \zeta_\mu m^\mu\,,
\end{align}
with 
\begin{align}\label{zetamummuX24}
    \zeta_\mu m^\mu = \left[(t^1)^2 + 4t^1 t^0 +16 (t^0)^2\right]\hat m^1 + \left(t^1 t^0 +4(t^0)^2\right) \Gamma_{11} \hat m^1\, .
\end{align}
Furthermore, in this model the contribution to the tadpole is then given by 
\begin{align}\label{NfluxX24}
        N_\text{flux}=\hat m^1m_1 + 4(\hat m^1)^2= \left(\Gamma_{11}+4\right)(\hat m^1)^2\,.
\end{align}
We now want to find the bound on $\cK$ and $t^i$ for which we expect solutions to the vacuum equations similar to \eqref{eq: K3 saxion bound}. To that end, let us distinguish three different cases depending on the hierarchy between $t^1$ and $t^0$: 
\begin{itemize}
    \item[$i)$] {\em Vacua with the hierarchy $t^1\gg t^0$}. In this case we can approximate
    \begin{align}
        m_1 = \left[\frac{t^1}{2t^0}\left(1+ 4 \frac{t^0}{t^1} + \mathcal{O}\left(\frac{t^0}{t^1}\right)^2 \right)\right]\hat m^1= \left(\frac{t^1}{2t^0}+2\right)\hat m^1+ \mathcal{O}\left(\frac{t^0}{t^1}\right)\,. 
    \end{align}
    Thus in order to have the required hierarchy we need $m_1 \gg \hat m^1$ such that the contribution to the tadpole goes essentially as $N_\text{flux}\gtrsim \Gamma_{11}$. From \eqref{barrhoX24} we then find
    \begin{align}\label{NfluxbarrhoX24i}
        \qquad N_\text{flux}^2 \bar \rho\gtrsim 1\,,
    \end{align} 
    i.e. we would expect \eqref{eq: K3 saxion bound} to hold for $p=2$. The RHS of \eqref{eqofmbarrhoX24} to leading order is then given by
    \begin{align}
        -\frac{9}{16} \frac{K_0^{(3)} t^0 + K_1^{(3)}t^1}{t^0 (t^1)^3} \left((t^1)^2 \hat m^1+t^1 t^0 m_1\right) = \frac{27}{32} \frac{K_1^{(3)}}{t^0} \hat m^1+\mathcal{O}\left(\frac{1}{t^1}\right) \,. 
    \end{align}
    Using the bound \eqref{NfluxbarrhoX24i} we can derive 
    \begin{align}
        t^0 \lesssim \left(4\hat m^1+m_1\right)\left(m_1 \hat m^1\right)|K_1^{(3)}| = \Gamma_{11}^2(\hat m^1)^3 |K_1^{(3)}|\lesssim |K_1^{(3)}| N_\text{flux}^{2}\,.
    \end{align}
    Accordingly, $t^1$ is bounded by
    \begin{align}
        t^1\sim \frac{m^1}{\hat m^1} t^0 \lesssim \left(4\hat m^1+m_1\right)\left(m_1 m_1\right)|K_1^{(3)}|\lesssim N_\text{flux}^{3}|K_1^{(3)}|\,.
    \end{align}
    Combining the bound for $t^0$ and $t^1$ we find 
    \begin{align}
         \cK \lesssim \left(K^{(3)}_i t^i\right)^2 N_\text{flux}^5\,,
    \end{align}
    in accordance with \eqref{eq: K3 saxion bound} for $p=2$. 
    \item[$ii)$] {\em Vacua with both saxions  of the same order, i.e. $t^0/t^1=\gamma$ with $\gamma\sim \mathcal{O}(1)$}. In this case
    \begin{align}
         m_1 = \left(\gamma^{-1} \frac{1+4\gamma }{2+12\gamma+16\gamma^2}\right)\hat m^1\,,
    \end{align}
    such that in order for $\gamma$ to be $\mathcal{O}(1)$ we also need $\hat m^1$ and $m_1$ to be of the same order. Accordingly, from \eqref{barrhoX24} and \eqref{NfluxX24} we find 
    \begin{align}
        \qquad N_\text{flux} \bar \rho\gtrsim 1\,,
    \end{align}
    such that we expect the bound \eqref{eq: K3 saxion bound} with $p=1$. We can now set a bound on the overall saxion $t^1$. Using \eqref{zetamummuX24} we have that 
    \begin{align}
        -\frac{3}{8}\epsilon_i t^i \zeta_\mu m^\mu \sim \frac{1}{t^1} \hat m^1 f(\gamma)\,,
    \end{align}
    with $f$ a function of $\gamma$. From here, we derive the bound 
    \begin{align}
        t^1 \lesssim \hat m^1 \left(4\hat m^1+m_1\right)m_1 |K_1^{(3)} + \gamma K_0^{(3)}| \lesssim N_\text{flux}^{3/2} |K_1^{(3)} + \gamma K_0^{(3)}|\,, 
    \end{align}
    and similar for $t^0$. Combining the scaling of $t^0$ and $t^1$ we find the bound 
    \begin{align}
         \cK \lesssim \left(K^{(3)}_i t^i\right)^2 N_\text{flux}^3\,,
    \end{align}
    in accordance with \eqref{eq: K3 saxion bound} with $p=1$. 
    \item[$iii)$] {\em Vacua with the hierarchy $t^0\gg t^1$}. Here we find 
    \begin{align}
        m_1 = \left[ \frac{1}{4}\left(\frac{t^1}{t^0}\right)^2 + \mathcal{O}\left(\frac{t^1}{t^0}\right)^3 \right] \hat m^1   \,,
    \end{align}
    such that we need to impose $\hat m^1\gg m_1$ to achieve the required hierarchy. In view of \eqref{NfluxX24} and \eqref{barrhoX24} we then find the bound 
    \begin{align}\label{NfluxbarrhoX24iii}
       \qquad N_\text{flux}^{1/2}\bar \rho \gtrsim 1\,,
    \end{align}
    which should lead to \eqref{eq: K3 saxion bound} with $p=1/2$. In this regime we have that 
    \begin{align}
        -\frac{3}{8} \epsilon_i t^i \zeta_\mu m^\mu \sim -K_0^{(3)}\frac{\hat m^1}{t^0}\,.
    \end{align}
    From here we can then derive the bounds
    \begin{align}
        t^0 \lesssim 4 (\hat m^1)^2 m_1 |K_0^{(3)}| \lesssim  N_\text{flux} |K_0^{(3)}|\,, \qquad t^1\lesssim N_\text{flux}^{1/2} |K_0^{(3)}|\,.
    \end{align}
    Putting things together we then find 
    \begin{align}
         \cK \lesssim \left(K^{(3)}_i t^i\right)^2 N_\text{flux}^2\,,
    \end{align}
    in accordance with \eqref{eq: K3 saxion bound} for $p=1/2$.
\end{itemize}
\subsubsection*{Type IIB limit}
F-theory compactified on the four-fold $X_{24}^*$ can be viewed as the F-theory lift of type IIB compactified on the mirror of the $\mathbb{Z}_5$ orbifold of the quintic, which has a single complex structure modulus $T^1$. The intersection number and Euler characteristic of the mirror, i.e. in the quintic itself, are 
\begin{align}
    \kappa_{111}=1\,,\qquad \chi_E=-40\,.  
\end{align}
The main difference to the case of $X_{24}^*$ discussed before is that now $t^0$ only appears linearly in the K\"ahler potential. In this case, the set of vacuum equations simplifies considerably. For instance at the classical level \eqref{eq:MinkIIB mu} reduces to
\begin{align}
     \bar{\rho}'_1 = \frac{1}{2} \frac{t^1}{t^0} \hat \rho^1 \,.
\end{align}
Focusing on the restricted flux case $\vec{q}^{\, t} = (0,0,0, \hat{m}^a, \bar{m}_a, \bar{e}_a, \bar{e}_0, \bar{e})$ this translates into 
\begin{align}
    \frac{t^1}{t^0} = \frac{2m_1}{\hat m^1}\,. 
\end{align}
In this case the equation for $\rho_i$ read 
\begin{align}
    \bar \rho_0 = \bar e_0 + m_1 b^1 =0 \,, \qquad \bar \rho_1 = \bar e_1 + m_1 b^0 + \hat m^1 b^1 =0\,,
\end{align}
which are solved by 
\begin{align}
    b^1 = -\frac{\bar e_0}{m_1} \,,\qquad b^0 = -\frac{\bar e_1}{m_1} + \frac{\hat m^1}{m_1^2} \bar e_0\,. 
\end{align}
This can be inserted into $\bar \rho$ to find 
\begin{align}
    \bar \rho = \bar e +\frac{1}{2} \bar e_i b^i = \frac{1}{m_1^2} \left((m_1)^2\bar e - \bar e_1\bar e_0 m_1 +\frac{1}{2} \bar e_0^2 \hat m^1\right)\,. 
\end{align}
We can now give an estimate for the range where the moduli $t^0$, $t^1$ can be fixed, based on \eqref{eq:MinkcorrIIB1 0}: \begin{align}
     (m_1)^2\bar e - \bar e_1\bar e_0 m_1 +\frac{1}{2}\bar e_0^2 \hat m^1 = \frac{9}{8} K^{(3)} \frac{\hat m^1}{t^1} (m_1)^2.
\end{align}
As in the case of $X_{24}^*$ we distinguish three cases: 
\begin{itemize}
    \item[$i)$] {\em At the vacuum we have the hierarchy $t^1\gg t^0$}. In this case we have $m_1 \gg \hat m^1$ such that the flux contribution to the tadpole is determined by $m_1$. As a consequence
    \begin{align}
        t^1 \lesssim |K^{(3)}|\, m_1 \left(m_1 \hat m^1\right) < |K^{(3)}| N^{2}_\text{flux}\,,
    \end{align}
    and accordingly 
    \begin{align}
        \frac{\kappa}{t^0} \lesssim (K^{(3)})^2 N_\text{flux}^5\,, 
    \end{align}
    which agrees with \eqref{boundIIB1} for $p=2$. 
    \item[$ii)$] {\em At the vacuum $t^0\sim t^1$}.  For this we need $m_1\sim \hat m^1$. In this case we find 
    \begin{align}
        t^1 \lesssim |K^{(3)}|\, m_1 \left(m_1 \hat m^1\right) < |K^{(3)}| N^{3/2}_\text{flux}\,,
    \end{align}
    where we used $N_\text{flux}^{1/2} \gtrsim \hat m^1 \sim m_1$. Hence 
    \begin{align}
        \frac{\kappa}{t^0}\lesssim (K^{(3)})^2 N_\text{flux}^3\,, 
    \end{align}
    which corresponds to \eqref{boundIIB1} for $p=1$.
    \item[$iii)$] {\em At the vacuum we have the hierarchy $t^0\gg t^1$}. In this case we have $N_\text{flux}\gtrsim \hat m^1\gg m_1$ such that our bound becomes 
    \begin{align}
        t^1 \lesssim |K^{(3)}|\, m_1 \left(m_1 \hat m^1\right) < |K^{(3)}| N_\text{flux}\,,
    \end{align} 
    and 
    \begin{align}
        \frac{\kappa}{t^0}\lesssim (K^{(3)})^2 N_\text{flux}\,, 
    \end{align}
    reproducing \eqref{boundIIB1} for $p=1/2$. 
\end{itemize}
 We see that compared to the $X_{24}^*$ discussion the bound on $t^1$ in the case $i)$ is stronger whereas it is the same in case $ii)$ and even less constraining in case $iii)$. In the present example we further have 
\begin{align}
    |K^{(3)}| = \frac{\zeta(3)}{8\pi^3} |\chi_E| \simeq 1\,, 
\end{align}
such that $|K^{(3)}|\, m_1 \left(m_1 \hat m^1\right)$ can be made moderately larger than 1 to ensure that we are always in the regime where the perturbations $\epsilon \ll 1$.  

\subsection{A realisation of the linear scenario}
\label{sec:counter}

In section \ref{s:linear} we discussed a linear scenario that resembles certain features of the \textbf{IIB2} scheme  and in particular allows for full moduli stabilisation for a flux choice with only one contribution to the D3-brane tadpole $N_\text{flux}$. In the following we would like to give an explicit example of an F-theory construction that realises this linear scenario. In our concrete model the number of complex structure moduli is four, but as discussed in section \ref{s:linear} the construction can be easily generalised to an arbitrary $h^{3,1}(Y_4)$. 

As pointed out in section \ref{s:linear} we can realise the linear scenario in case the mirror manifold $X_4$ admits a fibration of a Calabi--Yau three-fold $X_3$ over a $\mathbb{P}^1$. As the example in this section, we take the mirror manifold $X_4$ to be a triple fibration $\mathbb{T}^2 \rightarrow \mathbb{P}^1 \rightarrow \mathbb{P}^1 \rightarrow \mathbb{P}^1$, which can either be seen as an elliptic fibration over a base $B_3=\mathbb{P}^1\rightarrow \mathbb{F}_2$ or as a fibration of a Calabi--Yau $X_3=\mathbb{T}^2 \rightarrow \mathbb{F}_1$ over $\mathbb{P}^1$. Here, $\mathbb{F}_n$ the $n$-th Hirzebruch surfaces. Such a manifold can be constructed using toric methods -- the toric data for this manifold is given e.g. in \cite{Mayr:1996sh}. For this model we have four generators of the K\"ahler cone $D_0, \, D_1, \,  D_2$ and $D_L$ with intersection polynomial 
\begin{eqn}\label{intersectionpolyIII}
    I(Y_4) =& \left(8D_0^3+D_0 D_1 D_2 + D_0D_2^2 + 2 D_0^2 D_1 + 3 D_0^2 D_2\right)D_L + 6 D_0^2 D_2 D_1 + 2D_0 D_2 D_1^2 \\
    &+ 2 D_0D_2^2D_1 + 16 D_0^3 D_1 + 2 D_0 D_2^3 +4D_0^2 D_1^2 + 6 D_0^2 D_2^2 + 18 D_0^3 D_2 +52 D_0^4\,.
\end{eqn}
We can identify $D_0$ as the K\"ahler cone generator related to the zero section of the elliptic fibration as in \eqref{kahlerconeelliptic}. Furthermore $D_L$, satisfying $D_L. D_L=0$, denotes the class of the generic Calabi--Yau three-fold fibre $X_3$ and $D_1$ and $D_2$ are the divisors dual to the curves inside the base $\mathbb{F}_1$ of $X_3$. 
From \eqref{intersectionpolyIII} we can read off
\begin{eqn}
    \cK =& \left[8(t^0)^3+t^0t^1t^2 + t^0(t^2)^2 + 2 (t^0)^2 t^1 + 3 (t^0)^2 t^2\right]t_L + 6 (t^0)^2 t^2t^1 + 2t^0t^2(t^1)^2 \\
    &+ 2 t^0(t^2)^2t^1 + 16 (t^0)^3 t^1 + 2 t^0(t^2)^3 +4(t^0)^2 (t^1)^2 + 6 (t^0)^2 (t^2)^2 + 18 (t^0)^3 t^2 +52 (t^0)^4\,.
\end{eqn}
In the following we will use the indices $a,b,\dots$ to refer to $i=0,1,2$ and $\alpha,\beta, \dots$ to refer just to $i=1,2$. The first Chern class of the base $B_3$ is given by 
\begin{align}
    c_1(\mathbb{F}_1\rightarrow \mathbb{P}^1) = 2D_2 + D_1\,, 
\end{align}
and the corrections $K_i^{(3)}$ can be read off from \cite{Mayr:1996sh}
\begin{align}\label{c_3ExampleIII}
    c_3(Y_4) D_i = -3136 D_0-960 D_1 -1080D_2 - 480 D_L\,. 
\end{align}
Since $X_4$ can be seen as an elliptic fibration, a basis of four-cycles is given as in \eqref{4formbasisfib}. However, here we choose a different basis of four-cycles that is better suited for the study of the linear scenario that is given by \eqref{linearscenarioH1} and \eqref{linearscenarioH2}. The first set of four-cycles is given by divisors of the generic three-fold fibre $X_3$: 
\begin{align}
    H_0= D_0.D_L \,,\qquad  H_{\alpha} = D_\alpha . D_L\, , \qquad \alpha = 1,2\,. 
\end{align}
The second set of four-cycles is obtained by fibering the Mori-cone generators $\mathcal{C}^\alpha$ of $X_3$ over the base $\mathbb{P}^1$. The so-obtained four-cycles $H$ satisfy 
\begin{align}
    D_\alpha.D_\beta = \lambda_{\alpha \beta} H_{\hat 0}\,,\;\, D_0.D_\alpha = \lambda_{\alpha\beta}\left(\delta^{\beta \hat{\beta}}H_{\hat \beta} +c_1^\beta H_{\hat 0}\right)\,,\;\, D_0.D_0 =\lambda_{\alpha\beta }c_1^\alpha \left(\delta^{\beta \hat \beta} H_{\hat \beta} +c_1^\beta H_{\hat 0}\right)\,,
\end{align}
where $\lambda_{\alpha\beta}=\cK_{L 0 \alpha \beta}$ is the intersection on the two-fold base of $X_3$. From here we can read off the non-vanishing components of the $\zeta$ tensor 
\begin{eqn}
    \zeta^0_{L 0}&=1\,,\qquad \zeta^\alpha_{L\beta} =\delta^\alpha_\beta \,,\qquad \zeta^{\hat 0}_{\alpha \beta}=\lambda_{\alpha \beta}\,,\qquad \zeta^{\hat \alpha}_{\beta 0}=\delta^{\hat \alpha \alpha} \lambda_{\alpha \beta}\,,\\
    \zeta^{\hat \alpha}_{00}&=\delta^{\hat \alpha \alpha} \lambda_{\alpha \beta}\,,\qquad \zeta^{\hat 0}_{\alpha 0}=\lambda_{\alpha\beta}c_1^\beta\,,\qquad \zeta^{\hat 0}_{00}=\lambda_{\alpha \beta}c_1^\alpha c_1^\beta. 
\end{eqn}
The non-vanishing components of the intersection matrix $\eta_{\mu \nu}$ in the four-cycle sector are
\begin{eqn}\label{intersectionexampleIII}
     \eta_{a\hat b} &= \delta_{a\hat b} \,,\qquad \eta_{\hat 0\hat \alpha}=\delta_{\hat \alpha \alpha} \lambda^{\alpha \gamma} \lambda^{\delta \rho} D_0 D_\gamma D_\delta D_\rho\\
     \qquad \eta_{\hat \alpha \hat \beta} &= \delta_{\hat \alpha \alpha} \delta_{\hat \beta \beta}\left[\lambda^{\alpha \gamma}\lambda^{\beta \delta} D_0^2 D_\gamma D_\delta-(c_1^\alpha \lambda^{\beta \delta}+c_1^\beta \lambda^{\alpha \delta})\lambda^{\gamma\rho} D_0 D_\delta D_\gamma D_\rho\right]\,. 
\end{eqn}
In the following, we use the notation $ m_a = \delta_{a\hat a} \hat m^{\hat{a}}$ for the fluxes associated to $H_{\hat a}$. With this information, we can now look for solutions to the vacuum equations. We are interested in vacua that realise the linear scenario of section \ref{s:linear} and hence look at the limit \eqref{limitL}, which in the present case 
 can be viewed as some sort of Sen's limit. As before, to find vacua in the region probed by this limit we must set $\tilde \rho=m=0$, and since $\cK g_{ab}$ will generically diverge we also set $\tilde\rho^a=m^a=0$ in order not to violate the tadpole constraint. However, we can have $m^L\ne 0$ since \eqref{limgLL} is finite. If we further set $\bar \rho_a'=m_a=0$ by \eqref{intersectionexampleIII} we have a single pair of fluxes contributing to the D3-brane tadpole as $ N_\text{flux} = - m^L \bar{e}_L$. This results  in  the following restricted flux vector \eqref{eq:fluxchoiceL}:
\begin{align}\label{eq:fluxchoiceexampleIII}
    \vec q^{\, t} = (0, m^L, 0,0, \hat m^a,0,   \bar e_\alpha, \bar e_0, \bar e_L,
    \bar e)\,,
\end{align}
and the following flux-axion polynomials: 
\begin{align}
    \nonumber \bar \rho &= \bar e +\bar e_i b^i + \frac{ \cK_{L 0 \alpha \beta} }{2}\left(\hat m^0 (b^\alpha+c_1^\alpha b^0)(b^\beta +c_1^\beta b^0)+\hat m^\alpha b^0 (2b^\beta +c_1^\beta b^0)\right)+\frac{1}{6} \cK_{L abc} m^L b^a b^b b^c \,, \\
    \nonumber \bar \rho_0 &= \bar e_0 +  \cK_{L 0 \alpha \beta}\left(\hat m^\alpha (b^\beta+c_1^\beta b^0)+ c_1^\alpha \hat m^0(b^\beta +c_1^\beta b^0)\right) + \frac{1}{2} \cK_{L 0ab} m^L  b^0 b^a b^b\,,\\
    \nonumber \bar \rho_\alpha &= \bar e_\alpha + \hat m^0 \cK_{L 0 \alpha \beta}\left(b^\beta +c_1^\beta b^0 \right) + \hat m^\beta \cK_{L 0 \alpha \beta}b^0  + \frac{1}{2} \cK_{\alpha L ab} m^L b^\beta b^a b^b\,, \\
    \bar \rho_L  &= \bar e_L  \,,\\
    \nonumber\bar \rho_a^\prime &= 0 \,,\\
    \nonumber\hat \rho^a &= \hat m^a + m^L  b^a \,,\\
    \nonumber\tilde \rho^a &=0 \,, \\ \nonumber \tilde \rho^L  &= m^L \,, \\ \nonumber \tilde\rho&=0\, ,
\end{align}
where we used that the intersection numbers $\cK_{L abc}$ are related to $\lambda_{\alpha \beta}$ and $c_1^\alpha$ via
\begin{align}\label{kappaIII}
    \cK_{L0\alpha \beta} = \lambda_{\alpha \beta}\,,\qquad \cK_{L00\alpha}= \lambda_{\alpha \beta}c_1^\beta \,,\qquad \cK_{L000}= \lambda_{\alpha \beta}c_1^\alpha c_1^\beta\, .
\end{align}
One can check that the polynomials in \eqref{rhoAexample} correspond to those obtained in the general linear scenario in section \ref{s:linear}. The axions are stabilised as in \eqref{axionsL}, and also the stabilisation of the saxions works as in the general case. For concreteness, let us focus on the overall rescaling
\begin{align}\label{limitlambda}
   t^a = v^a \lambda \,,\qquad v^a\sim \mathcal{O}(1)\,,\qquad \lambda\rightarrow \infty\,. 
\end{align}
together with $t_L  \sim \lambda^3 \to \infty$. We can thus write $\cK= 4 t_L   \kappa(v)\lambda^3 + f(v) \lambda^4$ and $\cK_a =3 t_L \kappa_a(v)\lambda^2 + f_a(v)\lambda^3$. The values for the parameters $v^a$ can then be inferred  from the equation of motion for $\bar \rho_a$ as in \eqref{vaclin2} where $\varepsilon_a$ in the present example is given by 
\begin{align}
    \varepsilon_a = \frac{g_{L a}}{g_{LL}}   - 6\eps_L\cK_L\frac{\cK_a}{g_{L L}\cK^2}\quad \stackrel{\eqref{limitL}}{\to} \quad \frac{\kappa(v) f_a(v) -\frac{3}{4} \kappa_a(v) f(v)}{\kappa(v)^2}- \frac{27}{4} K^{(3)}_L \frac{\kappa_a(v)}{\kappa(v)^2} \frac{t_L}{\lambda^4}\,. 
\end{align}
Then the equation of motion \eqref{Nfluxineqlin}  fixes the $v^a$.  Since by assumption the $v^a$ are of order one, we also find $\varepsilon_a\sim \mathcal{O}(1)$, $\forall a$, such that the bound $N_\text{flux} |\varepsilon_a| \ge 1$
is trivially satisfied. As a result there is no upper bound for the value of $\lambda$, in accordance with the general discussion in section \ref{s:linear}. Still, since $N_a$ is a monodromy-invariant flux combination there should only be  a finite number of inequivalent vacua along the limit. Finally, the ratio $t_L/\lambda^3$ is fixed by \eqref{vaclin1}: 
\begin{align}
     \bar e_L = -\frac{\cK}{6} g_{LL}  m^L \rightarrow -\frac{\lambda^3}{t_L} \frac{m^L}{6} \ \implies \ \frac{t_L}{\lambda^3} \lesssim N_\text{flux}\,. 
\end{align}
We thus conclude that the present example indeed captures all the key features discussed for the general linear scenario in section \ref{s:linear}.

\section{Conclusions}
\label{s:conclu}

In this paper we analysed flux potentials and their vacua for F-theory compactifications on smooth elliptically fibered Calabi--Yau four-folds. We restricted our analysis to the regime of moderate to large complex structure,  where the complex structure moduli split into an axionic and a saxionic component and the periods of the holomorphic four-form $\Omega$ can be well approximated by polynomial expressions, neglecting exponentially suppressed terms. 
In this regime we provided an explicit expression for the scalar potential that allows for a systematic study of its vacua. To arrive at this result, we used that  the periods of the four-fold in the large complex structure regime are captured, through homological mirror symmetry, by the central charges of B-branes wrapping the holomorphic $2p$-cycles in the mirror four-fold. This strategy was promoted in \cite{CaboBizet:2014ovf,Cota:2017aal} to calculate the Gukov-Vafa-Witten superpotential. 

Since in our limit of consideration exponential corrections to the periods can be ignored, the resulting axionic shift symmetry allows us to separate the scalar potential into a saxion-dependent matrix $Z^{AB}$ and a set of flux-axion polynomials $\rho_A$ that depend on the axions and the $G_4$-flux quanta. This structure is in fact a general feature of the scalar potential close to generic large distance singularities, as argued in \cite{Grimm:2019ixq}. In terms of the $\rho_A$ the vacua conditions, i.e. the self-duality constraint for the $G_4$-flux, take the particularly simple form \eqref{eq:Mink} and can be analysed systematically. Using this form of the self-duality condition allowed us to directly compute the flux contribution to the D3-brane tadpole $N_{\rm flux}$ in terms of the $\rho_A$ on-shell values.  

Our analysis shows that for generic Calabi--Yau four-folds we have to restrict the choice of fluxes in order not to violate tadpole cancellation parametrically. This led us to consider the generic flux choice \eqref{truncflux}. In fact this constraint on the possible fluxes can be viewed as a generalisation of the result of \cite{Brodie:2015kza,Marsh:2015zoa}, where it was shown that in 4d type IIB/F-theory compactifications switching on the flux associated to the top period is inconsistent with tadpole cancellation and moduli stabilisation at large complex structure. 

As it turns out, our generic choice of fluxes compatible with the tadpole cancellation is too constrained in order for the leading vacua equations to stabilise all complex structure fields. In particular, the analysis of the set of leading order vacua equations revealed that at least one saxionic direction necessarily remains flat. This problem is circumvented when polynomial corrections to the periods are included. While most of these polynomial corrections can be treated as a re-definition of the flux quanta, the correction $K^{(3)}_i$, that is related to the third Chern class of the mirror four-fold, has important consequences for the vacua equations as it gives a correction to the action of the Hodge $*$ operator on $Y_4$. Including this correction allows us to generically stabilise all the complex structure fields. Still, to achieve full moduli stabilisation the fluxes need to be chosen in such a way that the matrix $M$ appearing in \eqref{Meq} is invertible. Invertibility of this matrix should be read as a constraint on the fluxes $\hat{m}^\mu$ contributing to the tadpole $N_{\rm flux}$. In the light of the recent conjecture put forward in \cite{Bena:2020xrh, Bena:2021wyr} it would be very interesting to translate this constraint into a precise relation between $N_{\rm flux}$ and the number of fields that need to be stabilised, which a priori could exist for this particular family of vacua.  

In any event, we observed that in this generic flux scenario the regime for the saxion vevs in which we can find vacua without violating tadpole cancellation is bounded from above by $|K^{(3)}| N_\text{flux}^{p+\oh}$.\footnote{We stress that even taking into account this upper bound, we can find vacua consistent with our approximation of neglecting exponentially suppressed terms, since the saxion vevs are still allowed to be moderately large depending on the precise value of $K^{(3)}$. } As discussed in section \ref{s:vacua}, the exponent $p$ is bounded by the number of complex structure fields in the system, and the upper bound on the saxionic vevs can be understood as arising due to the full stabilisation of the complex structure moduli by means of perturbatively suppressed terms. This bound on the saxion vevs nicely parallels the prediction for the total number of flux vacua based on statistical methods \cite{Ashok:2003gk,Denef:2004ze, Denef:2004cf}. Indeed, it was found that the number of vacua in type IIB flux compactifications grows like $N_{\rm flux}^{Q/2}$, with $Q$ the number of flux quanta. Since in type IIB the number of flux quanta is twice the number of complex structure plus dilaton fields, our bound on the saxion vevs is indeed in line with the expected number of flux vacua in type IIB. It would be interesting to make this link more precise, also by adding the D7-brane flux contribution as in \cite{Gomis:2005wc}.

Reducing our general F-theory setup to type IIB, we connected with several existing results in the literature. We realised that the flux choice made in \cite{Blanco-Pillado:2020hbw} is one of the simplest that guarantee that the matrix $M$ is invertible, implying that all complex structure moduli and the dilaton are fixed. In our scheme, the mass spectrum clearly depends on the correction $K^{(3)}$, as one of the fields is only stabilised when they are taken into account. This is also consistent with the results of \cite{Blanco-Pillado:2020wjn,Blanco-Pillado:2020hbw}, since the parameter $\xi$ that controls their mass spectrum is a simple function of $K^{(3)}$. Furthermore,  we also showed that in one particular case in which the matrix $M$ is not invertible, we recover the residual flat direction found in \cite{Demirtas:2019sip} for the same flux choices. In that reference it was shown that this flat direction can be stabilised by including non-perturbative corrections, possibly yielding to an exponentially small superpotential. Our analysis of section \ref{sec:moduli} provide an F-theory generalisation of both of these type IIB constrained flux scenarios, and we expect them to display similar features, see e.g. \cite{Honma:2021klo}. In particular, notice that the vacuum obtained in \cite{Demirtas:2019sip} after including exponential corrections is located at $\mathcal{O}(1)$ values for the saxionic fields. This is analogous to our observation that the small corrections which yield full stabilisation of all complex structure fields set an upper bound for the regime in which we expect to find vacua. Based on our analysis presented in this paper, it would be interesting to investigate whether also in general F-theory models non-perturbative corrections can lift the perturbatively flat direction of the potential when $M$ is not invertible. Finally, it would also be interesting to connect our results with the type IIB analysis made in \cite{Dimofte:2008jg}.

Besides the class of vacua associated to the flux choice \eqref{truncflux}, which is present in generic F-theory models, we found a second class of vacua arising for a different pattern of flux quanta when at least one of the complex structure fields only enters linearly in $e^{-K}$ and the superpotential. In this case there exists a region in field space where we can fix all complex structure moduli with the flux choice \eqref{eq:fluxchoiceL}, without violating the tadpole constraint. Most importantly, for this flux choice there is only a pair of  flux quanta that contribute to the tadpole. As we argued in section \ref{s:linear}, in the linear scenario the full moduli stabilisation can be achieved provided the matrix $Z^{AB}$ entering the scalar potential has enough off-diagonal components. In the type IIB limit these off-diagonal components are again related to the $K^{(3)}$ correction and reproduce the mirror dual of the Minkowski vacua studied in \cite{Escobar:2018rna}. However, as discussed in section \ref{s:linear} in the generic F-theory setup we do not necessarily need to rely on the $K^{(3)}$ corrections, and full moduli stabilisation can be already achieved just on the level of the classical contributions to the periods of the four-fold. Notice that in this case the off-diagonal terms of $Z^{AB}$ are not necessarily suppressed in the large complex structure limit. As a consequence there is in general no bound on the value of the saxion vevs for which we can find these kind of vacua. Still, as argued in section \ref{s:linear}, we expect the number of vacua in this class to be finite. This follows from inequivalent vacua being characterised by a monodromy-invariant integer which can only take values in a finite range. 

The analysis presented in this paper offers some avenues to follow in the future: first of all it would be interesting to calculate the precise mass spectra for the F-theory vacua obtained here as done in \cite{Blanco-Pillado:2020wjn,Blanco-Pillado:2020hbw} for the type IIB case. In this way one could verify whether there is indeed a hierarchy between the masses of the fields in the complex structure sector. More precisely, given our analysis one would expect that the mass of at least one of the fields that is stabilised only through the effect of the $K^{(3)}$ corrections is  smaller than the masses of the fields that are already stabilised using the leading order vacua equations. 

Second, one could try to generalise our explicit expressions to other asymptotic regions in the four-fold complex structure moduli space, as classified in \cite{Grimm:2019ixq}. Our analysis shows that in order to achieve full moduli stabilisation it is often necessary to include contributions that are sub-leading in the asymptotic region. In the particular case of the large complex structure regime, we were fortunate to get general expressions for these contributions via mirror symmetry. However, for limits other than the large complex structure limit, the subleading contributions to the periods are not known in general, though for three-folds there has been recent progress in that direction  \cite{Bastian:2021eom}. As a first step one could consider infinite distance limits that involve intersections with divisors corresponding to conifold-like singularities. Moduli stabilisation in type IIB string theory around such regions has been considered in \cite{Demirtas:2020ffz, Blumenhagen:2020ire}. Furthermore, for three-folds such limits have been classified in the dual K\"ahler moduli space as emergent string/decompactification limits in \cite{Lee:2019oct}, see also \cite{Lee:2019jan, Klaewer:2020lfg} for a discussion of similar emergent string limits in CY four-folds. Since such limits are defined in the large volume phase of the mirror dual four-fold, it should be easier to relate them to the discussion presented in here. It would be interesting to see whether or not in these more general limits one also sees the two classes of vacua discussed in this work, or even if one encounters additional classes of vacua that enhance the flux landscape.

\bigskip

\centerline{\bf  Acknowledgments}

\vspace*{.5cm}

We would like to thank Pierre Corvilain for collaboration at the initial stages of the project, and Alberto Castellano, Alvaro Herr\'aez, Severin L\" ust, Raffaele Savelli and Thorsten Schimannek for useful discussions.  This work is supported by the Spanish Research Agency (Agencia Estatal de Investigaci\'on) through the grant IFT Centro de Excelencia Severo Ochoa SEV-2016-0597, and by the grant PGC2018-095976-B-C21 from MCIU/AEI/FEDER, UE. The work of DP was supported through the FPU grant No. FPU19/04298. The work of MW received the support of a fellowship from 'la Caixa' Foundation (ID 100010434) with fellowship code LCF/BQ/DI18/11660033 and funding from the European Union's Horizon 2020 research and innovation programme under the Marie Sklodowska-Curie grant agreement No. 713673.


\appendix


\section{Geometric interpretation of the $\rho_A$}
\label{ap:georho}

In this appendix we provide a geometric interpretation of the flux-axion polynomials $\rho_A$, introduced in section \ref{s:potential} to describe the scalar potential in regions of large complex structure, as well as of the saxion-dependent matrix $Z^{AB}$ that appears in \eqref{ZAB}. While the discussion below is focused in the large complex structure region without polynomial corrections for simplicity, our reasoning can be easily extended to include them and can also be generalised to other limits in which approximate axionic shift symmetries appear in the moduli space metric.

To understand the flux-axion polynomials geometrically, one may first realise that they can be seen as the components of the  flux $G_4$ in a particular basis of four-forms. More precisely we have that
\begin{align}
    G_4 =  \rho \tilde \alpha -  \rho^i \tilde \alpha_i + \rho^\mu \tilde \sigma_\mu - \rho_i \tilde \beta^i + \rho \tilde \beta\, ,
    \label{ap:G4rho}
\end{align}
where we have defined
\begin{eqn}\label{apeq:tildealpha}
    \tilde \alpha &= \alpha + b^i \alpha_i +\frac{1}{2} b^i b^j \zeta^\mu_{ij} \sigma_\nu^Y +\frac{1}{6} \cK_{ijkl}b^ib^j b^k \beta^l +\frac{1}{24} \cK_{ijkl}b^ib^j b^k b^l\beta\,,\\
    \tilde \alpha_i &=\alpha_i + \zeta^{\mu}_{ij}b^j \sigma_\mu^Y + \frac{1}{2} \cK_{ijkl}b^j b^k \beta^l + \frac{1}{6} \cK_{ijkl}b^j b^k b^l \beta \,,\\
    \tilde \sigma_\mu &=\sigma_\mu^Y + \zeta_{\mu kl} b^k \beta^l +\frac{1}{2}\zeta_{\mu kl} b^kb^l \beta\,,\\
    \tilde \beta^i &= \beta^i + b^i \beta\,,\\
    \tilde \beta &= \beta \,. 
\end{eqn} 
The geometric interpretation of the $\rho$'s then boils down to the geometric significance of this tilded set of four-forms, in comparison with the basis of integer four-forms $\{ \a, \a_i, \sigma_\mu^Y, \b^i, \b\}$, that span the horizontal subspace  $H_H^4(Y_4)$. As one can check, two key properties properties of this new basis are that:
\begin{itemize}

\item[{\it i)}] It has the same intersection numbers as the initial basis $\{ \a, \a_i, \sigma_\mu^Y, \b^i, \b\}$.

\item[{\it ii)}] Their elements are invariant under monodromies around the large complex structure point.  

\end{itemize}

The first property can be easily checked by direct computation, and it implies the tadpole identity \eqref{Nfluxrho}. The second one follows from the characterisation of the large complex structure monodromies as \eqref{eq: monodromy P}, given that the monodromy generators $P_i$  also specify the change of basis $\{ \a, \a_i, \sigma_\mu^Y, \b^i, \b\} \to \{ \tilde\a, \tilde\a_i, \tilde\sigma_\mu, \tilde\b^i, \tilde\b\}$.  Combined, these two properties also allows us to relate the saxion-dependent matrix $Z^{AB}$ with  the action of the Hodge star operator on the basis $\{ \tilde\a, \tilde\a_i, \tilde\sigma_\mu, \tilde\b^i, \tilde\b\}$.

Indeed, this tilded basis is particularly suitable to express monodromy-invariant quantities like the holomorphic four-form $\Omega$ and its derivatives. To simplify the discussion, let us ignore the contribution of the corrections $K_i^{(3)}$ to the expression of $\Omega$. That is, we consider the expression \eqref{Omega}, from where we find
\begin{align}\label{eqap:Omega}
    \Omega & = \tilde \alpha + i t^i \tilde \alpha_i - \frac{1}{2}\zeta^\mu \tilde \sigma_\mu - \frac{i}{6}\cK_i \tilde \beta^i + \frac{\cK}{24} \tilde \beta \,, \\
    \label{eqap:derOmega}
    D_i \Omega & =   \tilde \alpha_i + i \zeta^\mu_i \tilde \sigma_\mu - \frac{1}{2}\cK_{ik} \tilde \beta^k - \frac{i}{6}\cK_i \tilde \beta +\frac{2 i \cK_i}{\cK}\left[\tilde \alpha + i t^i\tilde \alpha_i-\frac{1}{2}\zeta^\mu \tilde \sigma_\mu- \frac{i}{6}\cK_i \tilde \beta^i+\frac{\cK}{24}\tilde \beta \right]\,, \\
    \label{eqap:derderOmega}
     D_iD_j \Omega & = \zeta_{ij}^\mu \tilde\sigma_\mu + i \cK_{ijk}\tilde \beta^k -\frac{1}{2} \cK_{ij} \tilde \beta -\left(g_{ij} + \frac{4 \cK_i \cK_j}{\cK^2}\right) \Omega + \frac{2i\cK_i}{\cK}\partial_{T^j} \Omega + \frac{2i\cK_j}{\cK}\partial_{T^i} \Omega\\
     \nonumber &+\left(\frac{2i \cK_{ij} t^k}{\cK} -\frac{2i}{\cK}\left(\delta_i^k\cK_j + \delta_j^k \cK_i) \right)+i \cK^{kl}\cK_{ijl}\right) D_k\Omega \, .
\end{align}
We now use the fact that the Hodge star operator has a simple action on each of these four-forms
\be
* \Omega = \Omega\, ,\quad * D_i \Omega = - D_i\Omega\, , \quad * D_iD_j \Omega = D_i D_j\Omega\, ,
\ee
and in particular that 
\be
\frac{1}{3}t^it^j D_iD_j \Omega + \Omega = i t^i \tilde \alpha_i - \frac{2}{3} \zeta^\mu \tilde\sigma_\mu - \frac{i}{6} \cK_{i} \tilde \beta^i
\ee
is self dual. From the real part of this expression we obtain that 
\begin{align}\label{eqap:starsigma1}
    * \left(\zeta^\mu \tilde \sigma_\mu\right)=\zeta^\mu \tilde \sigma_\mu \,,
\end{align}
and from its imaginary part that
\begin{align}
    * \left(t^i \tilde \alpha_i -\frac{\cK_i}{6} \tilde \beta^i \right)&= t^i \tilde \alpha_i -\frac{\cK_i}{6} \tilde \beta^i\,.\label{eqap:alphaistar0}
\end{align}
In addition, using that $*\Omega = \Omega$ and the above relations we obtain
\begin{align}
    * \left(\tilde \alpha + \frac{\cK}{24} \tilde \beta\right) &= \tilde \alpha  + \frac{\cK}{24} \tilde \beta\, .
\end{align}
Moreover, from $* D_i \Omega = - D_i\Omega$ we obtain the following two conditions
\begin{align}\label{eqap:alphaistar1}
     *\left(\tilde \alpha_i -\frac{\cK_{ik}}{2}\tilde \beta^k\right) & = - \tilde \alpha_i + \frac{\cK_{ik}}{2}\tilde \beta^k -\frac{2}{3} \frac{\cK_i\cK_k}{\cK} \tilde \beta^k + \frac{4\cK_i}{\cK}\tilde \alpha_k t^k\,,\\
* \left(\zeta^\mu_{i}\tilde \sigma_\mu-\frac{1}{6}\cK_i \tilde \beta \right) & =  - \zeta^{\mu}_i\tilde \sigma_\mu+ \frac{1}{6} \cK_i \tilde \beta -\frac{4\cK_i}{\cK} \left(\tilde \alpha -\frac{1}{2} \zeta^\mu \tilde \sigma_\mu +\frac{\cK}{24} \tilde \beta\right)\,,
\end{align}
where we used \eqref{eqap:alphaistar0}. Taking this into account as well as the above relations, one finds that the action of the Hodge star operator on the basis $\{ \tilde\a, \tilde\a_i, \tilde\sigma_\mu, \tilde\b^i, \tilde\b\}$ must be given by
\begin{eqn}\label{eqap:starrest}
    * \tilde \alpha &= \frac{\cK}{24} \tilde \beta \,,\qquad\qquad  * \tilde \beta= \frac{24}{\cK} \tilde \alpha\,,\\
    * \tilde \alpha_i& = -\frac{1}{6} \cK g_{ij} \tilde \beta^i \,,\qquad  * \tilde \beta^i = - \frac{6}{\cK}g^{ij} \tilde \alpha_j\,,
\end{eqn}
together with \eqref{eqap:starsigma1} and
\begin{align}\label{eqap:starsigma2}
    * \left(\zeta^\mu_i -\frac{\cK_i}{\cK}\zeta^\mu \right)\tilde\sigma_\mu= -\left( \zeta^\mu_i - \frac{\cK_i}{\cK}\zeta^\mu\right) \tilde \sigma_\mu \,.
\end{align}

It is now easy to identify the action of the Hodge star with the diagonal entries of the saxion-dependent matrix \eqref{ZAB}. More precisely, we have that the matrix $2{\cal V}_3^2 Z + \chi_0$ defined in there corresponds to the entries of the standard four-form metric
\be
G^{AB} = \int_{Y^4} \omega^A \wedge * \omega^B\, ,
\label{Hmetric}
\ee
with $\{\omega^A\} = \{ \tilde\a, \tilde\a_i, \tilde\sigma_\mu, \tilde\b^i, \tilde\b\}$, computed to the same level of approximation. In fact, to fully show this statement one must verify that
\begin{align}
    g_{\mu \nu} = \int_{Y_4} \tilde \sigma_\mu \wedge * \tilde \sigma_\nu\,,
    \label{ap:sigmetric}
\end{align}
with $g_{\mu\nu}$ as defined below \eqref{ZABdiag}. This is easy to argue from the results above. For this, let us perform the decomposition
\be
\rho^\mu \tilde{\sigma}_\mu = \left(A\zeta^\mu + B^\mu +C^\mu\right) \tilde{\sigma}_\mu\, ,
\ee
with components such that
\begin{align}
    B^\mu=  \left(\zeta_i^\mu-\frac{\cK_i}{\cK} \zeta^\mu \right)\xi^i\,, \qquad \zeta_{\mu i}C^\mu=0\ \forall i\, ,
    \label{ap:sigmadec}
\end{align}
for some arbitrary vector $\xi^i$. This splitting is directly related to the decomposition introduced in \eqref{splitmu}, to which one can give a geometric meaning in terms of self-duality properties. Indeed, it follows from \eqref{eqap:starsigma1} and \eqref{eqap:starsigma2} that the first and second components are Hodge self-dual and anti-self-dual, respectively, and it is easy to convince oneself (either using mirror symmetry or \eqref{eqap:derderOmega}) that $*C^\mu \tilde\sigma_\mu = C^\mu \tilde\sigma_\mu$. Putting all these together we have 
\begin{align}
\rho^\mu \rho^\nu \int_{Y_4} \tilde \sigma_\mu \wedge * \tilde \sigma_\nu & = \rho^\mu \eta_{\mu\nu}\rho^\nu  - 2 B^\mu \eta_{\mu\nu} B^\nu = \rho^\mu \eta_{\mu \nu}\rho^\nu - 2 \xi^i \left(\cK_{ij} - \frac{\cK_i \cK_j}{\cK} \right) \xi^j \nonumber \\
& = \rho^\mu \eta_{\mu \nu}\rho^\nu-2 \rho^\mu \left(\cK^{ij} - \cK^{-1} t^i t^j \right)\zeta_{\mu i}\zeta_{\nu j} \rho^\nu\,,
\end{align}
where we have used that $ \xi^i (\cK_{ij} - \frac{\cK_i \cK_j}{\cK})= (\zeta_{\mu j} -\frac{\cK_j}{\cK} \zeta_\mu)\rho^\mu$, and so \eqref{ap:sigmetric} follows.

Notice that our results imply a prescription to construct the flux-axion polynomials $\rho_A$, without the knowledge of \eqref{apeq:tildealpha}, and that one can apply it to any other field space region with approximate axionic symmetries. Indeed, given a real integral basis of horizontal four-forms $\{\omega^A\}$ one may construct an alternative basis $\{\tilde \omega^A\}$ from axion-independent linear combinations of the real and imaginary parts of $\Omega$, $D_i\Omega$ and $D_iD_j\Omega$, so that the elements of the new basis are automatically monodromy-invariant. One must moreover choose the new basis such that $\chi^{AB} \equiv \int \tilde \omega^A \wedge\tilde \omega^B = \int \omega^A \wedge \omega^B$. We then define the flux-axion polynomials $\rho_A$ as the coefficients of the four-form flux in this basis, and the saxion-dependent matrix in terms of its Hodge and intersection products:
\be
G_4 = \rho_A \tilde{\omega}^A\, , \qquad Z^{AB} = \frac{1}{2{\cal V}_3^2}  \left(G^{AB} - \chi^{AB}\right)\, ,
\ee
with $G^{AB}$ defined as in \eqref{Hmetric}.


\section{Curvature corrections on four-folds}
\label{ap:curvature}

In this appendix we cover several technical details regarding the polynomial corrections discussed in section \ref{sec:poly}. In \ref{sap:corrper} we elaborate on the computation of the corrections to the  periods and the intersection matrix, both seen as curvature corrections in the dual Calabi--Yau four-fold $X_4$. In \ref{sap:corrka} we provide an alternative derivation of the corrected K\"ahler potential \eqref{Kcscorr}. In \ref{sap:corrFterm} we provide the flux potential including all the polynomial corrections. In \ref{sap:corrvac} we focus on the corrections to the F-terms, which we use to provide the corrected vacuum equations. 

\subsection{Corrected periods and intersection matrix}
\label{sap:corrper}

Section \ref{sec:poly} discusses the polynomial corrections to the four-fold periods in the large complex structure regime. These can be obtained via mirror symmetry from the central charges of B-branes wrapped on holomorphic $(2p)$-cycles in the mirror four-fold $X_4$. In the large volume regime the leading polynomial form of the central charge of a $(2p)$-brane that corresponds to a complex $\mathcal{E}$ is given by 
\begin{align}\label{eqap:centralcharge}
    Z(\mathcal{E})= \int_{X_4} e^J \Gamma_\mathbb{C}(X_4) \left(\text{ch}(\mathcal{E})\right)^\vee \,,
\end{align}
where $J$ is the complexified K\"ahler class. The Calabi--Yau $n$-fold complex $\Gamma$-class is given by 
\begin{align}\label{eqap:gammaclass}
    \Gamma_\mathbb{C} (X_n)= \sqrt{\text{Td}(X_n)} \exp(i\Lambda_{X_n})\,,
\end{align}
with $\text{Td}(X_n)$ the Todd class of $X_n$ and
\begin{align}
    \Lambda_{X_n} = -\frac{\zeta(3)}{(2\pi)^3}c_3 + \frac{\zeta(5)}{(2\pi)^5}\left(c_5-c_2c_3\right) + \dots 
\end{align}
To evaluate these central charges one needs a basis of $(2p)$-branes, which we take as type IIA D$(2p)$-branes on a four-fold $X_4$. For $p\ne2$ such a basis was constructed in \cite{Gerhardus:2016iot}: the D8-brane wrapped on $X_4$ is associated with the structure sheaf $\mathcal{O}_{X_4}$ with Chern character $\text{ch}(\mathcal{O}_{X_4}) = 1$. 
A basis of D6-branes is given by the sheaves $\mathcal{O}_{D_i}$ with $D_i$ the generators of the K\"ahler cone. For these sheaves the Chern character is given by 
\begin{align}
    \text{ch}(\mathcal{O}_{D_i})=D_i -\frac{1}{2} D_i^2 + \frac{1}{6} D_i^3 - \frac{1}{24} D_i^4\,. 
\end{align}
A basis for D2-branes is obtained from the Mori cone generators $C^i$ via $\mathcal{C}^i= \iota_!\mathcal{O}_{C^i}\left(K^{1/2}_{C^i}\right)$ for which the Chern character is simply
\begin{align}
    \text{ch}(\mathcal{C}^i)= C^i \,. 
\end{align}
Finally, as shown in \cite{Cota:2017aal} in many cases a basis of D4-branes can be constructed from the intersection of two divisors $D_i . D_j$. The Chern character of the associated sheaf $\mathcal{O}_{D_i. D_j}$ is then 
\begin{align}
    \text{ch}(\mathcal{O}_{D_i . D_j}) = D_i . D_j -\frac{1}{2} D_i.D_j. \left(D_i + D_j\right) + \frac{1}{12}D_i. D_j \left(2 D_i^2 + 3 D_iD_j + 2 D_j^2\right)\,.
\end{align}
Using these expressions for the Chern characters, the central charges in \eqref{eqap:centralcharge} can be explicitly evaluated yielding the periods \eqref{eq:corrper}. Let us stress that these expressions for the central charges are valid in the large volume regime. Away from these limits in principle exponential corrections need to be taken into account that do not necessarily converge in the entire classical K\"ahler cone. In order to ensure that we are in the regime of validity of the polynomial approximation to the central charges we impose that the classical contribution to the central charges of $8$-, $6$- and $4$-branes is suitably large. We will in particular assume that the curvature corrections due to $c_i(X_4)$ are small compared to the leading polynomial expression in \eqref{eq:corrper}. As an example of what this constraint entails, let us consider the central charge of a D6-brane on a divisor $D_i$ that satisfies $D_i.D_i.D_i.D_j=0$, $\forall D_j$.   In the limit of large $t^i$ we find that
\begin{eqn}
    Z(\mathcal{O}_{D_i}) = - \frac{1}{6} \cK_{iijk} T^i T^j T^k - \frac{1}{24} T^i \int c_2 \wedge D_i\wedge D_i + \dots = -T^i\left(\frac{1}{6} \cK_{iijk} T^j T^k + K_{ii}^{(2)}\right) +\dots
\end{eqn}
Since the term in the brackets is constant for large values of $t^i$ we see that for the curvature correction to be subleading we need to impose
\begin{align}\label{eqap:constraintK2}
     \frac{1}{6} \cK_{iijk} t^j t^k > |K_{ii}^{(2)}|\,, 
\end{align} 
which is a condition on the other saxions. For a related discussion of the role of the second Chern class for the validity of the perturbative expansion in  type IIA on CY three-folds, see \cite{Lee:2019oct}.

Besides the periods, to extract the form of the flux potential we also need the corrected intersection matrix $\chi$ associated to the integer basis of $2p$-cycles on the Calabi--Yau four-fold $X_4$. As reviewed in the main text, this intersection matrix is given by the open string index
\begin{align}\label{eqapp:chi}
    \chi(\mathcal{E}, \mathcal{F}) = \int_{X_4} \text{Td}(X_4) (\text{ch} \mathcal{E})^\vee (\text{ch}\mathcal{F})\,,
\end{align}
where the Todd class is given by \eqref{eq: Todd} and $\mathcal{E}$ and $\mathcal{F}$ are complexes corresponding to the branes wrapped on the $2p$-cycles. 
Using the Chern characters of the associated complexes reviewed above, we can calculate the intersection matrix to be
\begin{align}
    \hat{\chi} = \left( \begin{matrix} \frac{1}{720} \int 3c_2^2 -c_4 & -K_{ii}^{(2)} -\frac{1}{24} \cK_{iiii} & \chi(\cO_{D_i. D_j}, \cO_Y)&0 &1 \\ 
    -K_{kk}^{(2)} -\frac{1}{24} \cK_{kkkk} &\chi(\cO_{D_k}, \cO_{D_i})  & -\frac{1}{2} \cK_{kkij} +\frac{1}{2}(\cK_{kiij}+\cK_{kijj})&-\delta_k^i &0\\
    \chi(\cO_{D_k. D_l}, \cO_Y) & -\frac{1}{2} \cK_{iikl} + \frac{1}{2} (\cK_{klli} + \cK_{kkli}) & \cK_{klij}&0&0\\
    0&-\delta_i^k &0&0&0 \\ 1 &0&0&0&0\end{matrix} \right) \,,
\end{align}
where 
\begin{align}
         \chi(\cO_{D_i. D_j}, \cO_Y)=\frac{1}{12} (2\cK_{iiij} + 3\cK_{iijj}+ 2 \cK_{ijjj} )+2 K^{(2)}_{ij}\,,
     \\
     \chi(\cO_{D_i}, \cO_{D_k})=-2 K_{ik}^{(2)} + \frac{1}{4} \cK_{iikk} -\frac{1}{6} (\cK_{iiik}+ \cK_{ikkk})\,. 
\end{align}
This matrix can now be rewritten as a product of three matrices 
\begin{align}
    \hat{\chi} = \hat \Lambda^t \hat \chi_0 \hat \Lambda\,, 
\end{align}
where 
\begin{align}
\label{hatLambda}
    \hat \Lambda = \left(\begin{matrix} 1&0&0&0&0 \\ 0&\delta_i^j & 0&0&0 \\ \frac{1}{24} c_2^{jl} & - \frac{1}{2}\delta^{j}_{i} \delta^{l}_{i} & \delta^j_i \delta^l_k & 0&0 \\ 0 & \frac{1}{6}\cK_{jiii} +K_{ji}^{(2)} & -\frac{1}{2}\left(\cK_{jiik} +\cK_{jikk}\right) &\delta_j^i&0 \\ 
    K^{(0)}& -\frac{1}{24}\cK_{iiii} - \half K_{ii}^{(2)} & \lambda_{ik} &0&1 \end{matrix}\right)\,,
\end{align}
with $\lambda_{ik}= \frac{1}{12} \left(2\cK_{iiik} + 3\cK_{iikk} + 2\cK_{ikkk}\right) + K_{ik}^{(2)}$, and we have 
\begin{align}
    \hat \chi_0 = \left(\begin{matrix} 0&0&0&0&1 \\ 0&0&0&-\delta_j^i &0 \\ 0&0&\cK_{ijkl} &0&0\\ 0&-\delta_i^j &0&0&0 \\ 1&0&0&0&0   \end{matrix}\right)\,.
\end{align}

As emphasized in the main text, to describe the potential in terms of physical fluxes we need to rewrite the intersection matrix so that it describes the intersections on the actual basis of four-cycles $\sigma_\mu$. Starting with the strict large complex structure limit, we can do this by defining
\begin{align}
    \hat \chi_0 = \Theta \chi_0 \Theta^t \,,
\end{align}
with 
\begin{align}
    \Theta = \left(\begin{matrix}1&0&0&0&0 \\0& \delta_i^j &0&0&0 \\ 0&0& \zeta_{ij}^\mu &0&0\\ 0&0&0&\delta_j^i &0 \\ 0&0&0&0&1 \end{matrix}\right)\,,
    \label{Theta}
\end{align}
and $\chi_0$ given in \eqref{eq:chi0}. Note $\Theta$ is the matrix $\hat{\mathcal{M}}$ introduced in \eqref{eq: Piij transformation} in the limit in which the induced lower $Dp$-brane charges go to zero, so $\hat{\cM}_{ij}^0=\hat{\cM}_{ij,k}=0$.
To account for polynomial corrections and the full intersection matrix $\chi$, which keeps track of these induced lower $Dp$-brane charges, we need to carefully consider the relation between the central charges of B-branes over the basis $\sigma_\mu$ versus the intersections $D_i.D_j$ discussed in  \eqref{eq: Pimu transformation} and \eqref{eq: Piij transformation}. Writing the transformation matrices explicitly as
\begin{eqn}
    \mathcal{M}=\left(\begin{matrix}1&0&0&0&0 \\0& \delta_i^j &0&0&0 \\ 0&0& \mathcal{M}_\mu^{ij} & \mathcal{M}_{\mu,i}&\mathcal{M}_\mu^0\\ 0&0&0&\delta_j^i &0 \\ 0&0&0&0&1 \end{matrix}\right)\,,\qquad 
    \hat{\mathcal{M}}=\left(\begin{matrix}1&0&0&0&0 \\0& \delta_i^j &0&0&0 \\ 0&0& \zeta^\mu_{ij} & \hat{\mathcal{M}}_{ij,k}&\hat{\mathcal{M}}_{ij}^0\\ 0&0&0&\delta_j^i &0 \\ 0&0&0&0&1 \end{matrix}\right)\,,
    \label{eq: M explicit}
\end{eqn}
and defining 
\begin{eqn}
    \Lambda = \Theta^t \hat \Lambda \mathcal{M}^t\,,
\end{eqn} 
 we arrive at the expression \eqref{eq:Lambda} and
\begin{align}
    \chi=\Lambda^t \chi_0 \Lambda\,.
\end{align}

\subsection{Corrections to the K\"ahler potential}
\label{sap:corrka}

In the main text, we derived the polynomial corrections to the Kähler potential \eqref{Kcs} via the correction to the periods of $\Omega$ and the intersection numbers. We noted that the resulting K\"ahler potential \eqref{Kcscorr} remains of the classical form up to a term proportional to the third Chern class of the mirror. In the following we will review a more direct way to arrive at the same result, based on the results of \cite{Halverson:2013qca}. 

In \cite{Halverson:2013qca} the Kähler potential on the complexified Kähler moduli space of general Calabi--Yau $n$-fold $X_n$ was argued to be of the form 
\begin{align}\label{eqapp:KahlerXn}
 e^{-K}= \int_{X_n} \exp\left(2i\sum\limits_{i=1}^{h^{(1,1)}(X_n)} t^i D_i\right)\left(\frac{\hat\Gamma_{\mathbb{C}}(X_n)}{\bar{\hat\Gamma}_{\mathbb{C}}(X_n)}\right)+ \mathcal{O}(e^{2\pi i T})\,,
\end{align}
based on calculating the perturbative corrections to the $S^2$ partition function of the associated gauged linear sigma model. Here $t^i = \text{Im}\,T^i$ is the saxionic part of the complexified Kähler moduli of $X_n$ and $\hat \Gamma_\mathbb{C} (X_n)$ is the complex $\Gamma$-class \eqref{eqap:gammaclass} that also appears in the calculation of the central charges \eqref{eqap:centralcharge}. Since the Todd class is real, its contribution to the Kähler potential drops out and we are left only with contributions from the term $\exp(i\Lambda_{X_n})$. For Calabi--Yau four-folds there is only one term in $\Lambda_{X_4}$ proportional to the third Chern class, indicating that only the third Chern class gives a correction to the K\"ahler potential. Evaluating \eqref{eqapp:KahlerXn} for a four-fold thus yields 
\begin{align}
    e^{-K}= \frac{2}{3} \cK_{ijkl}t^i t^j t^k t^l + \frac{4\zeta(3)}{(2\pi)^3} \int_{X_4} c_3(X_4) . D_i t^i = \frac{2}{3} \cK_{ijkl}t^it^j t^k t^l + 4 K_i^{(3)} t^i\,,
\end{align}
up to exponentially-suppressed corrections, with $K_i^{(3)}$ defined as in \eqref{K23}. This polynomial structure was previously conjectured in \cite{Honma:2013hma}, and one can easily check that it agrees with \eqref{Kcscorr}. 

\subsection{Corrected F-term potential}
\label{sap:corrFterm}

To compute the F-term potential we use the standard Cremmer el al. formula \cite{Cremmer:1982en}

\begin{equation}
    e^{-K}V_F=g^{m\bar{n}}D_m W D_{\bar{n}}\bar{W}-3|W|^2\, ,
\end{equation}
where $D_m W=\partial_{T^m}W+(\partial_{T^m}K) W$, $g^{m\bar{n}}$ is the inverse field space metric and $m, n$ run over all moduli. Ignoring corrections to the K\"ahler sector of the compactification we recover the standard cancellation of no-scale structure models and the above expression simplifies to
\begin{align}
    e^{-K}V_F&=g^{i\bar{j}}D_{i}W D_{\bar{j}} \bar{W}\\
    &=g^{i\bar{j}}\left[\Re W_i \Re W_{\bar{j}} +\Im W_i \Im W_{\bar{j}}+\left((\Re W)^2+\Im(W)^2\right)K_i\bar{K}_{\bar{j}}+K_i W\bar{W}_{\bar{j}}+\bar{K}_{\bar{j}}W_i \bar{W}\right]\, ,\nonumber
\end{align}
where $W_i=\partial_i W$ and now $i,j$ only run over complex structure moduli.  We proceed to consider the version of the superpotential \eqref{eq:supo and der corr} and the Kähler potential \eqref{Kcscorr} that include the polynomial corrections:
\begin{align}
     W&=\, \bar{\rho}_0+i\bar{\rho}_it^i-\frac{1}{4}\mathcal{K}_{ij}\bar{\rho}^{ij}-i \left(\frac{1}{6}\mathcal{K}_i + K_i^{(3)}\right)\tilde{\rho}^i + \left(\frac{\mathcal{K}}{24} +  K_i^{(3)}t^i \right)\tilde{\rho} \, ,
\end{align}
\be
K_{\rm cs}=-\log\left(\frac{2}{3}\mathcal{K}_{ijkl}t^it^jt^kt^l + 4 K_i^{(3)} t^i \right)\, ,
\ee
where the $\rho$'s are given by \eqref{corrhos}. From the Kähler potential we can derive the corrected version of the metric of the complex structure moduli space. We have
\begin{align}\label{apeq:KTi}
    K_{T^i}&\equiv \partial_{T^i} K_{\rm cs}=\frac{i\left(2\mathcal{K}_i+3K^{(3)}_i\right)}{\mathcal{K}+6K^{(3)}_kt^k}=\frac{i}{2\cK}\frac{4\cK_i+\cK\epsilon_i}{1+\epsilon_k t^k}\, ,\\
    g_{ij}&\equiv \partial_{T^i}\partial_{\bar{T^j}}  K_{\rm cs}\nonumber\\
    &=\frac{1}{\left(1+\epsilon_k t^k\right)^2}\left[\frac{4\cK_i\cK_j}{\cK^2}-\frac{3\cK_{ij}}{\cK}+\frac{1}{\cK}\left(\cK_i\epsilon_j+\cK_j\epsilon_i-3\cK_{ij}\epsilon_kt^k\right)+\frac{1}{4}\epsilon_i\epsilon_j\right]\, ,
\end{align}
where we have defined $\epsilon_i\equiv 6K^{(3)}_i/\cK$. The inverse metric can be computed as a series in powers in $\epsilon_i$, whose first terms are given by
\begin{align}
    g^{ij}=&(1+\epsilon_kt^k)^2 \left[\frac{4}{3}t^it^j-\frac{1}{3}\cK\cK^{ij}+\frac{\epsilon_k\cK}{3}\left(\cK^{ij} t^k+\cK^{ik} t^j+\cK^{jk}t^i\right)\right.\nonumber\\
    &\left.+\left[\frac{\cK^2}{12}\cK^{ik}\cK^{jl}-\frac{\cK}{3}\left(\cK^{ij} t^k t^l+t^it^j\cK^{kl}\right)+\frac{4}{3}t^it^j t^kt^l\right]\epsilon_k\epsilon_l+\cO(\epsilon_k^3)\right]\, .
\end{align}
Working with the inverse metric in its full extension would be extremely cumbersome. We take a different approach with the final aim of obtaining an expression for the scalar potential where the uncorrected part can be easily identified. To do so we make use of the following relation:
\begin{equation}
    g^{ik}K_{T^{i}}=2it^i-\frac{3i}{2}\frac{\tilde{\epsilon}^i}{(1+\epsilon)^2}\, ,
\end{equation}
with $\epsilon=\epsilon_it^i$ and $\tilde{\epsilon}^i=g^{ij}(\epsilon_j-4\epsilon\cK_j/\cK)$. Then $V_F$ becomes
\begin{align}
    e^{-K}V_F=&g^{ij}(\Re W_i\Re W_j+\Im W_i\Im W_j)+4\Re W(\Re W+t^i\Im W_i)\nonumber\\
    &+4\Im W(\Im W -t^i\Re W_i)+\frac{3\tilde{\epsilon}^i}{(1+\epsilon)^2}\left(\Im W \Re W_i -\Re W\Im W_i\right)\nonumber\\
    &-\left((\Re W)^2+(\Im W)^2\right)L\, ,
\end{align}
where
\begin{equation}
    L=\frac{3\epsilon}{1+\epsilon}+\frac{3\tilde{\epsilon}^i}{4(1+\epsilon)^3\cK}(4\cK_i+\cK\epsilon_i)\, .
\end{equation}
Substituting the superpotential and its derivatives in terms of the flux polynomials and denoting the uncorrected metric and its inverse \eqref{metric} by $g^0_{ij}$ and $g_0^{ij}$, respectively, we arrive to
\begin{align}
   e^{-K}V_F=&\; 4\left(\bar{\rho}-\frac{\mathcal{K}}{24}\tilde{\rho}\right)^2+g_0^{ij}\left(\rho_i+\frac{\mathcal{K}}{6}g^0_{ik}\tilde{\rho}^k\right)\left(\rho_j+\frac{\mathcal{K}}{6}g^0_{jl}\tilde{\rho}^l\right)+
    (g^{ij}-t^it^j)\zeta_{\mu i}\zeta_{\nu j} \bar{\rho}^\mu  \bar{\rho}^\nu\nonumber\\
    &+\frac{1}{36}\left(g^{ij}-g_0^{ij}\right)\cK_i\cK_j\tilde{\rho}^2+\frac{1}{36}g^{ij}(2\cK_i\cK\epsilon_j+\cK^2\epsilon_i\epsilon_j)\tilde{\rho}^2-\frac{1}{12}\epsilon\cK^2\tilde{\rho}^2+\frac{2\epsilon}{3}\cK\bar{\rho}\tilde{\rho}\nonumber\\
    &+\frac{1}{3}\left(g_0^{ij}-g^{ij}\right)\zeta_{\mu i}\cK_j \bar{\rho}^\mu\tilde{\rho}-\frac{\cK}{3}g^{ij}\zeta_{\mu i}\epsilon_j\bar{\rho}^\mu\tilde{\rho}+\frac{\epsilon}{3}\cK\zeta_\mu\bar{\rho}^\mu\tilde{\rho}+(g^{ij}-g_0^{ij})\bar{\rho}_i\bar{\rho}_j\nonumber\\
    &-\left(g^{ij}-g^{ij}_0\right)\cK_{jk}\bar{\rho}_i\tilde{\rho}^k+\frac{1}{4}\left(g^{ij}-g^{ij}_0\right)\cK_{ik}\cK_{jl}\tilde{\rho}^k\tilde{\rho}^l-\frac{2\cK}{3}\epsilon_j\tilde{\rho}^j\bar{\rho}_it^i-\frac{\cK}{9}\cK_i\epsilon_j\tilde{\rho}^i\tilde{\rho}^j+\frac{\cK^2}{9}(\epsilon_i\tilde{\rho}^i)^2\nonumber\\
    &+\frac{3\tilde{\epsilon}^i}{(1+\epsilon)^2}\left[\bar{\rho}_i\bar{\rho}_k t^k-\frac{1}{6}(\cK_k\delta^j_i+3\cK_{ik}t^j)\bar{\rho}_j\tilde{\rho}^k-\frac{1}{6}\cK\epsilon_k\tilde{\rho}^k\bar{\rho}_i+\frac{1}{12}(\cK_j+\cK\epsilon_j)\cK_{ik}\tilde{\rho}^j\tilde{\rho}^k\right.\nonumber\\
    &-\bar{\rho}\zeta_{\mu i}\bar{\rho}^\mu+\frac{1}{6}(\cK_i+\cK\epsilon_i)\tilde{\rho}\bar{\rho}+\frac{1}{2}\zeta_\mu\bar{\rho}^\mu\zeta_{\nu i}\bar{\rho}^\nu-\frac{1}{12}(\cK_i+\cK\epsilon_i)\tilde{\rho}\zeta_\mu\bar{\rho}^\mu\nonumber\\
    &\left.-\left(\frac{1}{24}+\frac{\epsilon}{6}\right)\cK \tilde{\rho}\zeta_{\mu i}\bar{\rho}^\mu+\frac{\cK}{6}\left(\frac{1}{24}+\frac{\epsilon}{6}\right)(\cK_i+\cK\epsilon_i)\tilde{\rho}^2\right]\nonumber\\
    &-L\left[\bar{\rho}^2+\frac{1}{4}(\zeta_\mu\bar{\rho}^\mu)^2+\left(\frac{1}{24}+\frac{\epsilon}{6}\right)^2\cK^2\tilde{\rho}^2-\zeta_\mu\bar{\rho}^\mu\bar{\rho}+\left(\frac{1}{12}+\frac{\epsilon}{3}\right)\cK\tilde{\rho}\bar{\rho}\right.\nonumber\\
    &\left.-\zeta_\mu\bar{\rho}^\mu\left(\frac{1}{24}+\frac{\epsilon}{6}\right)\cK\tilde{\rho}+(\bar{\rho}_it^i)^2+\frac{1}{36}[(\cK_i+\cK\epsilon_i)\tilde{\rho}^i]^2-\frac{1}{3}\bar{\rho}_it^i(\cK_j+\cK\epsilon_j)\tilde{\rho}^j\right]\, . \label{scalarpotcorr}
\end{align}
One can then see that in the limit $\eps_i \to 0$ we recover the leading form of the potential \eqref{scalarpot} from the first line of this expression. Notice that as expected all terms are quadratic on the flux-axion polynomials $\rho_A$, and so one has a potential of the form \eqref{bilinear}. The expression for the matrix $Z$ is, however, much more complicated than \eqref{ZAB}, with several new non-vanishing entries that destroy its block-diagonal structure. 

We can use the result in \eqref{scalarpotcorr} to generalise \eqref{ZAB} to account for the presence of linear order corrections in $\epsilon_i$. The new matrix will be given by
\begin{eqn}
Z = Z_0 + \eps_k Z^k + \cO(\eps_k^2) \, ,
\end{eqn}
where $Z_0$ is the uncorrected matrix from \eqref{ZAB} and $Z^k$ is given by
\begin{align}
&2{\cal V}_3^2 Z^{k}=\begin{pmatrix}
\frac{\cK t^k}{48} & &-\frac{t^k}{2}\zeta_\mu+\frac{\cK}{8}\cK^{ik}\zeta_{\mu i} & &  -t^k\\
& \mathcal{X}^k_{ij} & &  \mathcal{Y}^{ki}_j& &  \\
-\frac{t^k}{2}\zeta_\nu+\frac{\cK}{8}\cK^{ik}\zeta_{\nu i}& & \mathcal{P}^{k}_{\mu\nu} & & 3\cK^{ik}\zeta_{\nu i}+\frac{6t^k}{\cK}\zeta_\nu  \\
&  \mathcal{Y}^{kj}_i& &  \mathcal{Z}^{kij} &  \\
-t^k & &3\cK^{ik}\zeta_{\mu i}+\frac{6t^k}{\cK}\zeta_\mu & &  \frac{36t^k}{\cK} \\
\end{pmatrix} \, ,
\end{align}
where we have arranged the flux-axion polynomials in a vector of the form
$\vec{\rho}^{\, t} = \left(\tilde{\rho}, \tilde{\rho}^i,   \bar{\rho}^{\mu}, \bar{\rho}_i,   \bar{\rho}   \right)$ and we have defined
\begin{align}
    \mathcal{X}_{ij}^k\equiv&\frac{t^k}{\cK}\cK_i\cK_j-\frac{t^k}{2}\cK_{ij}-\frac{1}{12}\delta_i^k\cK_j-\frac{1}{12}\delta_j^k\cK_i\, ,\\
    \mathcal{Y}_j^{ki}\equiv& \frac{2t^k}{\cK}t^i\cK_j+t^k\delta^i_j-\frac{1}{2}\cK_j\cK^{ik}-\frac{3t^i}{2}\delta_j^k\, ,\\
    \mathcal{Z}^{kij}\equiv&\frac{4t^k}{\cK}t^it^j-2\cK^{ij}t^k-\cK^{ik}t^j-\cK^{jk}t^i\, ,\\
    \mathcal{P}_{\mu\nu}^k\equiv& \frac{t^k}{\cK}\zeta_\mu\zeta_\nu-2t^k\cK^{ij}\zeta_{\mu i}\zeta_{\nu j}+\frac{1}{2}\cK^{ik}\zeta_{\mu i}\zeta_{\nu}+\frac{1}{2}\cK^{ik}\zeta_{\nu i}\zeta_{\mu}\, .
\end{align}
In general, given the complicated form of the potential, it is easier to characterise the corrected vacuum equations in terms of the corrected F-terms, as we now turn to discuss.

\subsection{Corrected vacuum equations}
\label{sap:corrvac}

The polynomial corrections to the superpotential \eqref{supofinal} and K\"ahler potential \eqref{Kcscorr} modify the  on-shell conditions \eqref{eq:Mink} at leading order. In the following we would like to compute such a modification which, as pointed out in the main text, essentially depends on $K_i^{(3)}$. Using 
 \eqref{eq:supo and der corr} and \eqref{apeq:KTi} we find the F-term condition $D_i W=0$ to be equivalent to 
\begin{eqn}\label{apeq:DiW=0}
        &\left(\cK + 6 K_j^{(3)} t^j\right)\left[\bar \rho_i +i \zeta_{\mu,i}\bar \rho^{\mu} -\frac{1}{2} \cK_{ij} \tilde \rho^j -\frac{i}{6} \cK_i \tilde \rho- i K_i^{(3)} \tilde \rho\right]\\
    &=-2i \left(\cK_i +\frac{3}{2} K_i^{(3)}\right)\left[\bar \rho + i\bar \rho_j t^j - \frac{1}{2}\zeta_\mu \bar \rho^{\mu} -\frac{i}{6}\cK_j \tilde \rho^j -i K_j^{(3)} \tilde \rho^j+\frac{\cK}{24}\tilde \rho + K_j^{(3)} t^j \tilde \rho \right]\,. 
\end{eqn}
Contracting this expression with $t^i$ yields 
\begin{eqn}\label{apeq:tiDiW=0}
    &\left(\cK + 6 K_i^{(3)} t^i\right)\left[\bar \rho_jt^j +i \zeta_\mu \bar \rho^\mu -\frac{1}{2} \cK_j \tilde \rho^j -\frac{i}{6} \cK \tilde \rho-  i K_j^{(3)}t^j \tilde \rho\right]\\
    &=-2i \left(\cK +\frac{3}{2} K_i^{(3)}t^i\right)\left[\bar \rho + i\bar \rho_j t^j - \frac{1}{2}\zeta_\mu \bar \rho^\mu -\frac{i}{6}\cK_j \tilde \rho^j -i K_j^{(3)} \tilde \rho^j+\frac{\cK}{24}\tilde \rho + K_j^{(3)} t^j \tilde \rho \right]\,.
\end{eqn}
We now split this equation into real and imaginary part. The real part gives 
\begin{align}
   \left(1 +\half \epsilon_it^i\right) \bar \rho_i t^i = -\left(1 +\frac{5}{2} \epsilon_i t^i\right) \frac{\cK_j\tilde \rho^j }{6} + \left(4 + \epsilon_i t^i\right)\frac{\cK\epsilon_j  \tilde \rho^j}{12}\,, 
\end{align}
and the imaginary part 
\begin{align}\label{apeq:vacuumrho}
   (1 + \frac{1}{4} \epsilon_it^i) \bar \rho = \frac{\cK \tilde \rho}{24} - \frac{\cK \epsilon_i t^i\tilde \rho}{96}  - \frac{3  \epsilon_i t^i \zeta_\mu \bar \rho^{\mu}}{8} +\frac{\cK(\epsilon_i t^i)^2 \tilde \rho}{24}\, ,
\end{align}
where again $\epsilon_i \equiv 6 K_i^{(3)}/\cK$. Inserting the above expressions back into \eqref{apeq:DiW=0} we obtain 
\begin{align}\label{apeq:vacuumrho_i}
    \bar \rho_i = \frac{1}{2} \cK_{ij}\tilde \rho^j -\frac{2\left(\frac{\cK_i}{\cK} +\frac{1}{4} \epsilon_i \right)}{1+ \half \epsilon_j t^j - \half (\epsilon_k t^k)^2}\left[\frac{1}{3} \cK_j \tilde \rho^j + \frac{1}{3} \epsilon_k t ^k \cK_j \tilde \rho^j - \frac{1}{6} \cK \epsilon_j \tilde \rho^j \left(1 + \epsilon_k t ^k \right)\right]\,,
\end{align}
and 
\begin{eqn}\label{apeq:vacuumrhomu}
    \left(\zeta_{\mu, i} - \frac{\cK_i}{\cK} \zeta_\mu \right) \bar \rho^\mu =& -\frac{1}{8}\left(1+\epsilon_k t^k\right)\left( \cK_i \epsilon_k t^k - \epsilon_i \cK \right)\tilde \rho - \frac{5}{4}\epsilon_k t^k \zeta_{\mu, i}\hat \rho^\mu \\
    &+ \left(\frac{\cK_i}{\cK}\epsilon_k t^k +\frac{1}{4} \epsilon_i\right) \zeta_\mu \bar \rho^\mu  + \frac{1}{4} \left(\epsilon_i \epsilon_k t^k \zeta_\mu - (\epsilon_k t^k )^2 \zeta_{\mu, i}\right) \bar \rho^\mu\,.
\end{eqn}
As expected, in the limit $\epsilon_i \to 0$ equations \eqref{apeq:vacuumrho}, \eqref{apeq:vacuumrho_i} and \eqref{apeq:vacuumrhomu} reduce to the classical vacuum equations \eqref{eq:Mink}. To capture the leading effect of the corrections, we can also expand to linear order in $\epsilon_i$ to find 
\begin{subequations}\label{apeq:vaclinear}
\begin{align}
   \bar \rho- \frac{1}{24}\cK \tilde \rho  &= - \frac{1}{48} \epsilon_i t^i \left[\cK \tilde \rho + 18 \zeta_\mu \bar \rho^\mu \right] + \cO(\eps_i^2) \,,\\
    \bar \rho_i + \frac{1}{6} \cK g_{ij} \tilde \rho^j &=  -\frac{1}{6} \epsilon_i \cK_j \tilde \rho^j -\frac{1}{3} \cK_i\left(\epsilon_j t^j \frac{\cK_k}{\cK} - \epsilon_k\right) \tilde \rho^k + \cO(\eps_i^2)  \,,\\
    \left(\zeta_{\mu, i} - \frac{\cK_i}{\cK} \zeta_\mu \right)\bar \rho^\mu &= \frac{1}{8} \left(\eps_i - \epsilon_k t^k \frac{\cK_i }{\cK} \right)   \left( \cK \tilde \rho +  2 \zeta_\mu \bar\rho^{\mu}\right)+ \cO(\eps_i^2) \, ,
\end{align}
\end{subequations}
which gives \eqref{eq:Minkcorr} in the main text. If we further impose the condition for supersymmetric vacua $W=0$ we get the additional constraints
\begin{subequations}
\begin{align}
    \bar \rho_i t^i &= \frac{1}{4} \left(\cK \epsilon_i \tilde \rho^i -\epsilon_i t^i \cK_j \tilde \rho^j \right) + \cO(\eps_i^2) \,,\\
    \zeta_\mu \bar{\rho}^\mu &=\frac{\cK}{6}\left(1 + \eps_it^i \right) \tilde{\rho} + \cO(\eps_i^2) \, .
\end{align}
\end{subequations}
where we have also made use of the linearised equations \eqref{apeq:vaclinear}.


\section{Flux invariants and moduli fixing}
\label{ap:invariants}

The introduction of the flux-axion polynomials $\rho_{A}$ is a powerful technique that allows for the study of moduli stabilisation in a clear and systematic way. Since the flux polynomials depend on the axions $b^i$, fixing the moduli amounts to solve the system of algebraic equations in the saxions $t^i$ and the flux polynomials $\rho_{A}$ that arises from the vanishing derivatives of the scalar potential with respect to the set of moduli.

As discussed in \cite{Marchesano:2020uqz}, using the $\rho_{A}$ as a stepping stone to stabilise the $b^i$ may lead to some questions regarding whether it is actually possible to accomplish this task, since in most examples the number of polynomials will exceed the rank of the system of equations. The solution to this problem comes through the fact that the $\rho_{A}$ are not a set of fully independent variables. There are many constraints that arise from their definition and they can be expressed by the set of combinations of flux polynomials which are invariant under shifts of the axions. More precisely, we look for invariant multilinear combinations of $\rho_{A}$ under the transformation $\vec\rho \rightarrow R(b) \vec\rho$, with $R(b)$ given by \eqref{eq: axion matrix}. After some algebra we find that these invariants are
\begin{subequations}
\label{eq: invariants}
\begin{align}
    \tilde{\rho}^2\rho_i-\tilde{\rho}\zeta_{\mu,ij}\tilde{\rho}^j\bar{\rho}^\mu+\frac{1}{3}\mathcal{K}_{ijkl}\tilde{\rho}^j\tilde{\rho}^k\tilde{\rho}^l & = m^2e_i - m \zeta_{\mu,ij} m^j \bar{m}^\mu + \frac{1}{3} \mathcal{K}_{ijkl} m^jm^km^l  \, ,\\
    \bar{\rho} \tilde{\rho} - \bar{\rho}_i \tilde{\rho}^i + \frac{1}{2} \eta_{\mu\nu} \bar{\rho}^{\mu} \bar{\rho}^{\nu} & = \bar{e} m - \bar{e}_i m^i + \frac{1}{2} \eta_{\mu\nu} \bar{m}^\mu\bar{m}^\nu\, , \label{eq: invariants tadpole} \\
    \bar{\rho}^\mu\tilde{\rho}-\frac{1}{2}\zeta^\mu_{ij}\tilde{\rho}^i\tilde{\rho}^j & = \bar{m}^\mu m -\frac{1}{2}\zeta^\mu_{ij} m^im^j \, ,\\
    \tilde{\rho} & = m\, .
\end{align}
\end{subequations}

Looking at \eqref{eq:Mink}, we could think the system is composed of $2h^{(3,1)}+1$ linearly independent equations but note that the last family of equations has an additional constraint, since $\left(  \cK \zeta_{\mu i}- \mathcal{K}_i\zeta_{\mu}\right)t^i $ is trivially zero.  Therefore we actually have $2h^{(3,1)}$ equations in the variables $\{t^i, \rho_{A}\}$, which amount to  $3h^{(3,1)}+h^{(2,2)}+2$ unknowns. If it were not for the invariants this would imply that we have an extremely unconstrained system. However, the existence of invariant combinations of axion polynomials greatly reduces the number of degrees of freedom. From \eqref{eq: invariants} we see that we have  $2+h^{(3,1)}+h^{(2,2)}$ constraints. Consequently, the $\rho$'s move in an orbit of dimension $h^{(3,1)}$ which is just enough to fix all the axions using half of the vacua equations. The remaining $h^{(3,1)}$ vacua equations can be used to fix the saxions $t^a$.

 Notice that, by construction, the multilinear combinations of flux quanta in the rhs of \eqref{eq: invariants} are invariant under the monodromies ${\cal T}_i$ around the complex structure point, see  \eqref{eq: monodromy P}. This implies that they label flux-inequivalent vacua, and therefore that the saxion vevs should only depend on such invariants, simply because the value of the invariants $\rho_A$ in the vacuum also must depend on them.  Finally, in some specific scenarios where some flux quanta vanish, like in sections \ref{sec:moduli},  \ref{sec:moduliIIB} and \ref{s:linear}, 
the flux-axion polynomials will simplify and some other combinations of fluxes may play the role of those in  \eqref{eq: invariants}. For instance, only   \eqref{eq: invariants tadpole} remains non-vanishing in the moduli stabilisation scheme of section \ref{sec:moduli}, but other invariants like $\bar{m}^\mu$ appear in this case.


\section{Vacua equations for elliptic fibered mirrors}
\label{ap:elliptic}

In this appendix we analyse the vacua equations for the particular case in which the mirror manifold $X_4$ is elliptically fibered, as considered in section \ref{sec:elliptic}. In particular we want to provide an explicit expression for $\Gamma_{ab} \equiv - \tilde{A}_{ac}\tilde{B}^c{}_b$ in \eqref{eq:Mink elliptic compact rho mu}. While one could simply compute the inverse of \eqref{tildeA} and apply the definition, in the following we would like to obtain an expression for $\Gamma$ directly from  \eqref{eq:Mink mu}, in the same spirit as in \eqref{vacua mu strategy}. This strategy should be useful in cases where $X_4$ is not a fibration, and so the index splitting $\mu = \{ a, \hat{a}\}$ does not occur. Then, in general $g_{\mu\nu} -\eta_{\mu\nu}$ will be a singular matrix, and we cannot have an expression of the form \eqref{ZABfib}, because $\tilde{A}$ does not have an inverse.

To proceed one may expand the vacua equations  \eqref{eq:Mink mu} in the basis \eqref{4formbasisfib}. This is equivalent to consider the equations
\begin{align}
    \zeta^b_0\left[ \tilde{B}_b{}^c\tilde{A}_{cd}\tilde{B}^d{}_e  \bar{\rho}^e + \tilde{B}_b{}^c \bar{\rho}_c'\right]=0 \, ,\\
     \zeta^b_a \left[ \tilde{B}_b{}^c\tilde{A}_{cd}\tilde{B}^d{}_e  \bar{\rho}^e + \tilde{B}_b{}^c \bar{\rho}_c'\right] +\zeta_{ab} \left[\tilde{B}^b{}_c\bar{\rho}^c+ \tilde{A}^{bc}\bar{\rho}_c'\right]=0\, ,
\end{align}
which are in turn equivalent to \eqref{eq:Mink elliptic compact rho mu}. Expanding \eqref{eq:Mink mu} using \eqref{Koszulfib} and after some algebra we obtain:
\begin{subequations}\label{eqap:mink a}
\begin{align}
    \cK \left(t^a + c_1^a t^0\right)(\bar \rho_a + c_{ab}\bar \rho^b) &=\cK_0\left[t^0(2t^c + t^0 c_1^c)(\bar \rho_c + c_{cb} \bar \rho^c) + \kappa_b \bar \rho^b \right]\,, \\
    \cK\left(\kappa_{ab} \bar \rho^b + t^0 (\bar \rho_a + c_{ab} \bar \rho^b)\right)&= \cK_a \left[ t^0(2 t^b + t^0 c_1^b) (\bar \rho_b + c_{bc} \bar \rho^c) +\kappa_b \bar \rho^b\right] \, ,
\end{align}
\end{subequations}
which can be simplified with the following change of basis
\begin{align}
    \varrho_a = \bar \rho_a + c_{ab}\bar \rho^b\,,\qquad \varrho^a = \bar \rho^a \, ,
\end{align}
in terms of which \eqref{eqap:mink a} read 
\begin{subequations}\label{eqap:varmink a}
\begin{align}
    \cK \left(t^a + c_1^a t^0\right)\varrho_a &=\cK_0\left[t^0(2t^c + t^0 c_1^c)\varrho_c+ \kappa_b \varrho^b \right]\,, \label{eqap:varrho0}\\
    \cK\left(\kappa_{ab} \varrho^b + t^0 \varrho_a \right)&= \cK_a \left[ t^0(2 t^b + t^0 c_1^b) \varrho_b+\kappa_b \varrho^b\right] \,.\label{eqap:varrhoa}
\end{align}
\end{subequations}

Note that there is some redundancy among this set of equations, inherited from the fact that the contraction of \eqref{eq:Mink mu} with $t^i$ vanishes identically. To extract the information contained in \eqref{eqap:varrhoa} that is independent of \eqref{eqap:varrho0} we introduce  two projection operators 
\begin{align}
    \left(\mathbb{P}_p\right)^a_b = \delta^a_b - \frac{\cK_a t^b}{\cK - \cK_0t^0} \,,\qquad  \left(\mathbb{P}_{np}\right)^a_b=\frac{\cK_a t^b}{\cK - \cK_0t^0}\,. 
\end{align}
Then applying $\mathbb{P}_p$ to \eqref{eqap:varrhoa} we obtain 
\begin{align}
    t^0\left(\varrho_{a}-\frac{\cK_a}{\cK -\cK_0t^0} t^c\varrho_{c}\right)=\frac{\cK_a}{\cK - \cK_0t^0}\kappa_{b} \varrho^{b} 
- \kappa_{ab}\varrho^{b} \,,
\end{align}
which is solved by 
\begin{align}\label{ansatzPp}
    \varrho_{a}=\mathcal{G}_{ab} \varrho^{b}\, , \quad \text{with} \quad  \mathcal{G}_{ab} \equiv \left[\frac{\cK_a v_b}{\cK-\cK_0t^0} + \frac{1}{t^0}\left(\frac{\cK_a}{\cK-\cK_0t^0} \kappa_{b} -\kappa_{ab} \right)\right]\,,
\end{align}
where $v_b$ is a vector that still needs to be determined. Projecting \eqref{eqap:varrhoa} with 
 $\mathbb{P}_{np}$ is equivalent to \eqref{eqap:varrho0}, which can be rewritten as 
\begin{align}
    (\cK -\cK_0t^0)\left(t^a \varrho_{a} + t^0 c_1^a\varrho_{0a} \right)=\cK_0\left(t^0 t^a\varrho_{a}+\kappa_b \varrho^{b}\right)\,.
\end{align}
From this equation we can determine $v_b$ to be 
\begin{align}
    v_b= \frac{\left(\cK -\cK_0 t^0 \right)c_1^c \kappa_{cb} + (\cK_0-c_1^c \cK_c) \kappa_b}{\cK-2\cK_0t^0 +t^0c_1^a \cK_a}\,, 
\end{align}
such that the matrix $\Gamma_{ab} =  \mathcal{G}_{ab} - c_{ab}$ is given by 
\begin{align}
    \Gamma_{ab} =&\, \frac{1}{\cK - \cK_0 t^0}\left[ \frac{ \left(\cK -\cK_0 t^0 \right)c_1^c \kappa_{cb} \cK_a+ (\cK_0-c_1^c \cK_c) \cK_a \kappa_b}{\cK-2\cK_0t^0 +t^0c_1^a \cK_a} + \frac{1}{t^0}\cK_a\kappa_{b}   \right] -\frac{1}{t^0} \kappa_{ab} - c_{ab}\\
    =&\, \frac{\cK_a(\kappa_b+t^0\kappa_{bc}c_1^c)}{t^0(\cK-2\cK_0 t^0+t^0c_1^a\cK_a)}-\frac{1}{t^0}\kappa_{ab} - c_{ab}\, .
\end{align}
Finally, we may rewrite $\Gamma_{ab}$ in terms of base quantities by expanding it in $t^0 c_1^b$. The result is:
\begin{eqn}\label{eqap:expansionAab}
     t^0\left(2\kappa+3c_1^c\kappa_c t^0 + c_1^c c_1^d\kappa_{cd}(t^0)^2 \right)& \Gamma_{ab} =3 \kappa_a \kappa_b  -2 \kappa  \kappa_{ab}  \\ 
     &+ t^0\left(3\kappa_{ac}  \kappa_{b}  + 3 \kappa_{a}  \kappa_{bc}  - 3\kappa_{ab} \kappa_c  - 2 \kappa \kappa_{abc}  \right)c_1^c\\ 
    & +(t^0)^2\left(3\kappa_{ac} \kappa_{bd} -\kappa_{cd} \kappa_{ab}  + \kappa_{acd} \kappa_b -3 \kappa_{abc}  \kappa_d  \right)c_1^cc_1^d \\
     &+  (t^0)^3\left(\kappa_{acd} \kappa_{be}  - \kappa_{cd} \kappa_{abe} \right)c_1^c c_1^d c_1^e\, .
\end{eqn}


\bibliographystyle{JHEP2015}
\bibliography{papers}

\end{document}